\documentclass{aastex63}
\usepackage{appendix}

\usepackage{CJKutf8}
\usepackage{longtable}

\usepackage{float}
\usepackage{hyperref}
\usepackage{algorithm}

\usepackage[caption=false]{subfig}
\usepackage{graphicx}
\usepackage{supertabular}
\usepackage{longtable}
\usepackage{multirow}
\usepackage{booktabs}
\usepackage{amsmath,amssymb}
\usepackage[T1]{fontenc}
\graphicspath{{./}{figures/}}
\newcommand{\ra}[4]{$#1^{\rm h}#2^{\rm m}#3^{\rm s}.#4$}
\newcommand{\dec}[4]{$#1^{\circ}2'#3''.#4$}

\begin{document}

\title{Using the Optical--NIR Spectral Energy Distributions To Search for the Evidence of Dust Formation of 66 Supernovae}

\correspondingauthor{Shan-Qin Wang \begin{CJK*}{UTF8}{gbsn}(王善钦)\end{CJK*}}
\email{shanqinwang@gxu.edu.cn}

\author{Jing-Yao Li \begin{CJK*}{UTF8}{gbsn}(李京谣)\end{CJK*}}
\affiliation{Guangxi Key Laboratory for Relativistic Astrophysics,
School of Physical Science and Technology, Guangxi University, Nanning 530004,
China}

\author{Shan-Qin Wang \begin{CJK*}{UTF8}{gbsn}(王善钦)\end{CJK*}}
\affiliation{Guangxi Key Laboratory for Relativistic Astrophysics,
School of Physical Science and Technology, Guangxi University, Nanning 530004,
China}

\author{Wen-Pei Gan \begin{CJK*}{UTF8}{gbsn}(甘文沛)\end{CJK*}}
\affiliation{Guangxi Key Laboratory for Relativistic Astrophysics,
School of Physical Science and Technology, Guangxi University, Nanning 530004,
China}

\author{Tao Wang \begin{CJK*}{UTF8}{gbsn}(王涛)\end{CJK*}}
\affiliation{Guangxi Key Laboratory for Relativistic Astrophysics,
School of Physical Science and Technology, Guangxi University, Nanning 530004,
China}

\author{Ji-Shun Lian \begin{CJK*}{UTF8}{gbsn}(连纪顺)\end{CJK*}}
\affiliation{Guangxi Key Laboratory for Relativistic Astrophysics,
School of Physical Science and Technology, Guangxi University, Nanning 530004,
China}

\author{Song-Yao Bai \begin{CJK*}{UTF8}{gbsn}(白松瑶)\end{CJK*}}
\affiliation{Guangxi Key Laboratory for Relativistic Astrophysics,
School of Physical Science and Technology, Guangxi University, Nanning 530004,
China}

\author{En-Wei Liang \begin{CJK*}{UTF8}{gbsn}(梁恩维)\end{CJK*}}
\affiliation{Guangxi Key Laboratory for Relativistic Astrophysics,
School of Physical Science and Technology, Guangxi University, Nanning 530004,
China}

\begin{abstract}

In this paper, we searched for the dust formation evidence of 66 supernovae (SNe)
by using the blackbody model and the blackbody plus dust {emission} model to fit their
early$-$time optical$-$near infrared (NIR) spectral energy distributions (SEDs).
We find that, while the blackbody model can fit most SEDs of the SNe in our sample,
the model cannot fit the SEDs of some SNe, in which the SEDs of 2 SNe
(SNe~2010bq and 2012ca) show NIR excesses which can be attributed to the
emission from the heated dust.
We use blackbody plus dust {emission} model to fit the SEDs showing NIR excesses,
finding that both graphite and silicate dust model{s} can fit the SEDs, and the graphite model
get reasonable temperatures or better fits. Assuming that the dust is graphite,
the best-fitting temperatures (masses) of the dust of the SNe~2010bq and 2012ca
are {$\sim 1300-1800$ K ($\sim 0.1-3.4 \times 10^{-4}$~M$_\odot$) and
$\sim 600-1000$ K ($\sim 0.6-7.5 \times 10^{-3}$~M$_\odot$),}
respectively. We compare the vaporization radii and the blackbody radii of the dust shells of
the 2 SNe with the upper limits of the ejecta radii of the SNe at the first epochs, and
demonstrate that the NIR excesses of the SEDs of {the 2 SNe} might be caused
by the pre-existing dust.

\end{abstract}


\section{Introduction}
\label{sec:intro}

It is believed that the radiation of supernovae (SNe) would heat the dust grains surrounding
the progenitors of the SNe or produced in the ejecta. The dust grains heated to a temperature
higher than their evaporation (or sublim) temperature which is typically $1100-2500$ K would
be evaporated and the SNe radiation would produce dust-free cavities \citep{Dwek1983}.
The evaporation radii of the SNe depend on the peak luminosities of the SNe, the chemical
composition and the radii of the dust grains. Outside the evaporation radii, the dust heated
to a temperature lower than the evaporation temperature would produce blackbody-like emission
{that} peak{s} at near infrared (NIR) or middle infrared (MIR). The IR emission from the dust would
contaminate the spectral energy distributions (SEDs) of SNe and produce IR excesses. Additionally,
the late-time SEDs of some SNe are dominated by the MIR emission from relatively cold dust
(a few hundred K).

Up to 2010, the IR excesses/detections of SEDs of at least 35 SNe had been confirmed and
explained by the dust emission model (see Table 1 of \citealt{Fox2011} and the references therein)
\cite{Fox2011} performed $Spitzer$/IRAC survey for 68 known SNe IIn discovered during 1999 to 2008,
finding that 10 of them show late-time IR emission that can be attributed to the heated dust
radiation. \cite{Fox2013} presents follow-up study of the 10 SNe that had been confirmed to have
evidence of dust formation.

Several similar studies for the dust formation of SNe in different samples were performed
in the past years. \cite{Tinyanont2016} use $Spitzer$/IRAC to get the MIR bands
$I_1$ {(3.6 $\mu$m)} and $I_2$ {(4.5 $\mu$m)} photometry of 141 nearby SNe,
and detect 8 SNe Ia and 36 core-collapse SNe (CCSNe); the
masses and temperatures of the dust they derived are $10^{-6}-10^{-2}$~M$_\odot$ and $200-1280$ K,
respectively. \cite{Jencson2019} present the MIR ($I_1$ and $I_2$) photometry of 9
luminous IR transients discovered by the $Spitzer$/IRAC between 2014 and 2018, and suggest
that 5 are obscured CCSNe, 2 are erupting massive stars, 1 is luminous red nova, and
1 is intermediate-luminosity red transient. They conclude that most of CCSNe may be heavily
obscured by dust. Based on archival Spitzer/IRAC images of the positions of 1142 SNe,
\cite{Szalai2019b} perform a systematic study and find mid-infrared detections of 119 SNe,
45 of which hadn't been previously published. By modeling the MIR photometry, they reveal
that some SNe have significant amounts ($10^{-3}$~M$_\odot$) of warm dust.
\cite{Gan2021} model the optical-NIR SEDs of 4 SNe Ibn and find that, while the
SEDs of 3 of the 4 SNe can be well explained by the blackbody model, the SEDs of
OGLE-2012-SN-006 {show significant NIR excesses that can} be explained by
the blackbody plus dust {emission} model.
Recently, \cite{Szalai2021} published the MIR data of 31 SNe and studied the origin
of the dust and the possible connection between the dust and the circumstellar matter (CSM).

Besides the large sample studies, the IR excesses/detections of individual SNe
have also been studied comprehensively (e.g., \citealt{Smith2008,Mattila2008,Fox2009,Fox2010,Szalai2013,Gall2014,Kankare2014,
Johansson2017,Sarangi2018,Bhirombhakdi2019,Szalai2019a,Tinyanont2019,Tartaglia2020,Chen2021,Dwek2021}).

To date, however, there are a few dozen SNe that have optical--NIR ($J$, $H$, $K$) SEDs 
haven't been studied for the purpose of searching for the evidence of the dust formation.
In this paper, we collect their optical--NIR data and construct the
optical--NIR SEDs that can be used to confirm the dust formation evidence.

In Section \ref{sec:SED}, we list the basic information of the SNe in our sample, and
use the blackbody model and the blackbody plus dust {emission} model to fit the optical--NIR SEDs
of all SNe in the sample and the SEDs showing NIR excesses, respectively
\footnote{{It should be noted that, while the dust emission can account for the IR excesses
of the SEDs of SNe, the IR excesses can also be attributed to extended atmosphere effects
\citep{Vlemmings2019}. Nevertheless, we assume that the IR excesses of the SEDs are produced by the
dust emission in this paper, though the dust formation scenario is not the unique one to explain
the IR excesses.}}. 
We discuss our results Section \ref{sec:discussion} and draw some conclusions in Section \ref{sec:Con}.
Throughout the paper, we assume $\Omega_m = 0.3$, $\Omega_\Lambda = 0.7$,
and H$_0 = 68.3$\,km\,s$^{-1}$\,Mpc$^{-1}$. The values of the Milky Way
extinction ($E_\mathrm{B-V}$) of all events are from \cite{Schlafly2011}.

\section{The Optical and NIR SEDs and the Modeling}
\label{sec:SED}

The first step of the study is collecting the optical-NIR data of the SNe
that hadn't been studied for the aim of searching for the evidence of dust
formation via optical-NIR SEDs.
We don't included SNe Ia since their SEDs are usually polluted by
NIR emission produced by the process of Fe III to Fe II transition and
deviate from the blackbody SEDs.
Moreover, the data of PTF12dam and SN~2008am are also excluded
since the NIR flux of their host galaxies hadn't been removed
\footnote{\cite{Chen2015} remove the NIR flux of the host galaxy of PTF12dam
and corrected the early-time NIR photometry of PTF12dam presented in \cite{Nich2013},
but they do not list the corrected NIR data and the most NIR data of PTF12dam in
the {\it Open Supernova Catalog} (https://sne.space) \citep{Guillochon2017} are from \cite{Nich2013}.}.

We perform interpolation and/or extrapolation for some SNe (SNe~1996al, 2005em, 2005gj, 2007av,
2009dt, 2012ca, and LSQ13zm), for which the time intervals between the optical data and NIR data
are small enough. On the other hand, we exclude some SNe that the time intervals between the
optical and NIR data are {too} large so that the interpolation or extrapolation method
cannot be used to obtain the optical-NIR SEDs.

The basic information of the SNe in the sample we selected is listed in
in Table \ref{table:details}. The optical--NIR photometry of almost all SNe in the
sample are from the {\it Open Supernova Catalog},
while the data of SN~2005ek (optical), SN~2012ca, and SN~2013dn are from \cite{Drout2013}, \cite{Inserra2016},
and \cite{Fox2015}, respectively \footnote{The {\it Open Supernova Catalog} collects
the optical and MIR ($I_1$ and $I_2$) data of SN~2012ca and MIR data of SN~2013dn,
but doesn't present NIR ($J$, $H$, $K$) photometry of SN~2012ca and optical and NIR photometry of
SN~2013dn.}. The references providing the data are also listed in Table \ref{table:details}.

To search for the evidence of NIR excess, we use the blackbody model to fit the optical-NIR SEDs of the SNe.
The model description can be found in \cite{Gan2021}. The Markov Chain Monte Carlo (MCMC) method
by using the \texttt{emcee} Python package \citep{Foreman-Mackey2013}
are also adopted here. The free parameters of the blackbody model are
the the radius ($R_{\rm ph}$) and the temperature ($T_{\rm ph}$) of the SN photosphere.
The best-fitting parameters and the values of $\chi^2$/dof for all fits
using the blackbody model are presented in Table \ref{table:SED_BB} (see the Appendix).

The fits can be split into 3 subclasses. (1). The optical-NIR SEDs of most SNe
can be well fitted by the blackbody model {or show slight NIR excesses}, see Figure \ref{fig:SED-BB-1}.
(2). The optical-NIR SEDs of some SNe (SNe~2005kl, 2006fo, 2007ce, 2008ax and 2009K)
cannot be fitted by the same model but don't show
NIR excesses, see Figure \ref{fig:SED-BB-2}; in contrast, the optical-NIR SEDs of
those SNe show NIR absorption features.
(3). The SEDs of 2 SNe (SNe~2010bq and 2012ca)
show {significant} NIR excesses, see Figure \ref{fig:SED-BB-3}.

In principle, all deviation from the blackbody model curves can be
attribute to the undulation features of the spectra and SEDs of SNe.
For the SEDs showing NIR excesses, however, the NIR excesses might be
due to the emission from the dust surrounding the SNe.
Especially, the {significant} NIR excesses cannot be explained by the
SED undulation scenario, and the dust emission assumption might be
the best for explaining the NIR excesses.

Therefore, we use the blackbody plus dust {emission} model which assumes the flux is
produced by the SN photosphere and the circum-stellar or newly produced
dust grains that had been heated by the SN radiation.
The model description for the blackbody plus dust {emission} model can also be found in
\cite{Gan2021} and references therein, and the MCMC method is also employed.
We assume that the mean value of the dust grains is 0.10 $\mu$m.
The free parameters of the blackbody plus dust {emission} model are $R_{\rm ph}$,
$T_{\rm ph}$, the dust temperature $T_{\rm d}$, and the dust mass $M_{\rm d}$.

The fits of the blackbody plus dust {emission} model for the SEDs of SNe~2010bq and
2012ca show that the SEDs of the SNe at all epochs can be well fitted by the model
(see Figure \ref{fig:SED-double}).
The best-fitting parameters of the blackbody plus dust {emission} model are listed in
Table \ref{table:SED_PARAM-double}.

The derived masses of the dust of the SNe are respectively
$\sim 0.1-3.4 \times 10^{-4}$~M$_\odot$ ($\sim 0.2-2.6 \times 10^{-3}$~M$_\odot$) and
{$\sim 0.6-7.5 \times 10^{-3}$~M$_\odot$ ($\sim 2.5-23 \times 10^{-3}$~M$_\odot$)}
if the the dust grains are graphite (silicate).
We plotted the evolution of the masses of the graphite dust
in the left panel of Figure \ref{fig:evo_T_M}.
{We note that the derived masses of the dust are lower
limits, because the dust with lower temperatures emit radiation
peaks at MIR and the NIR emission might be lower than the detection limit.
The dust masses mentioned throughout the paper are also the lower limits
for the same reason.} 

We find that the derived temperatures of the dust of SNe~2010bq and
2012ca are respectively $\sim 1300-1800$ K ($\sim 1400-2000$ K) and
$\sim 600-1000$ K ($\sim 700-1200$ K) for graphite (silicate) grains.
By comparing the derived temperatures of the dust of 2 SNe and the evaporation (sublim)
temperatures of graphite ($\sim$ 1900 K, e.g., \citealt{Stritzinger2012}) and
silicate ($\sim 1100-1500$ K, e.g., \citealt{Laor1993,Mattila2008,Gall2014}),
we find that the dust of SNe~2010bq might be graphite,
while the dust of SN~2012ca favors both the two cases.
However, the values of $\chi^2$/dof of the graphite model
are smaller than that of the silicate model, indicating that
the best model fitting the SEDs of SN~2012ca is the graphite model.
The evolution of the temperature of the graphite dust is shown in the right panel of
Figure \ref{fig:evo_T_M}.

For completeness, we also use the same model to fit the SEDs can be fitted by the
blackbody model {or show slight NIR excesses (the best-fitting parameters and the
corresponding values of $\chi^2$/dof are listed in Table \ref{table:SED_PARAM-double-uppperlimit})}.
{We find that 52 SNe disfavor the blackbody plus dust emission model, because
the values of $\chi^2$/dof of the model for all SEDs of 44 SNe are larger than that of the blackbody model,
the derived masses and/or temperatures of 6 SNe (SNe~2005bf, 2005cs, 2005kj, 2006bo, 2009kr, and 2016bkv)
are unreasonable or have the error bars comparable or significantly larger than the best-fit values;
2 SNe (SNe~2004gk and 2007I) have only 1 SED favoring the model, but the SEDs prior and after the
corresponding SEDs disfavor the model.
The SEDs of the rest 7 SNe (SNe~1999el, 2006aa, 2008D, 2008S, 2008in, 2009ay, LSQ13zm) favor
the blackbody plus dust emission model at some epochs (for SNe having 2 or more SEDs)
or the unique epoch (for SNe having only 1 SED). It should be noted that, however,
the slight deviation from the blackbody model might be due to the undulation feature of
the SEDs, rather than the existence of the putative dust.}
{Hence, t}he derived masses of the dust {of SNe~1999el, 2006aa, 2008D, 2008S,
2008in, 2009ay, LSQ13zm should} be regarded as the upper limits of the masses of the dust
surrounding the SNe.

\section{Discussion}
\label{sec:discussion}

In this section, we discuss some issues associated with the dust of the SNe in
our sample. We focus on the graphite case since the blackbody plus graphite model
can obtain reasonable parameters or better fits for the SEDs of {both the 2} SNe.

\subsection{The Luminosities of the Dust Shells of the SNe}

According to the best-fit parameters of the blackbody plus dust {emission} model,
we derive the luminosities of the photospheres ($L_{\rm ph}$) of SNe
and the dust shells ($L_{\rm d}$) surrounding them using the equations
$L_{\rm ph} = 4 \pi \sigma T_{\rm ph}^4R_{\rm ph}^2$ and
$L_{\rm d} = 4\pi M_{\rm d} \int_0^{\infty} B_\nu(T_{\rm d}) \kappa_\nu(a)d\nu$.
Here, $\sigma = 5.67 \times 10^{-5} \,\mathrm{erg~cm^{-2}~K^{-4}}$ is the Stefan-Boltzmann constant,
$T_{\rm ph}$ and $R_{\rm ph}$ are the temperature and radius of the photosphere;
$M_{\rm d}$, $T_{\rm d}$, $B_\nu(T_{\rm d})$, $\kappa_\nu(a)$
are the mass of the dust shell, temperature of the dust grains,
the intensity of the blackbody radiation of a dust grain, the
mass absorption coefficient, respectively.

The derived values of $L_{\rm ph}$ and $L_{\rm d}$ are presented
in Table \ref{table:SED_L}.
We find that the derived luminosities of the dust shells of
SNe~2010bq and 2012ca are respectively
$\sim 4.2-29.3 \times 10^{41}$ erg s$^{-1}$ and
{$\sim 4.6-8.9 \times 10^{41}$ erg s$^{-1}$} for graphite grains.
{The optical depth ($\tau$) of the dust shells
calculated by using $\tau = {L_{\rm d}}/({L_{\rm ph}+L_{\rm d}})$ \citep{Fox2009}
is also listed in Table \ref{table:SED_L}.}

In Figure \ref{fig:evo_L}, we plot the luminosities of the dust and the photospheres of SNe~2010bq
and 2012ca {as well as the the optical depth of the dust shell}.
We find that the photosphere luminosities of SNe~2010bq ($\sim 10^{42}$ erg s$^{-1}-\sim 10^{43}$ erg s$^{-1}$)
are higher than the luminosities of the respective dust shells ($\sim 10^{41}$ erg s$^{-1}-\sim 10^{42}$ erg s$^{-1}$)
throughout the evolution. The luminosity of the photosphere of SN~2012ca is higher than that
of the dust at the first 2 epochs, and rapidly declined to be lower than the dust luminosity
which rebrightened at the last 2 epochs (>500 d).

\subsection{The Vaporization Radii of the SNe}
\label{subsec:vap}

The extremely high temperature {produced by the SN radiation} will vaporize the
dust surrounding SNe and form a cavity inside the dust shell.
The radius of the cavity reached a maximal value when the luminosity
peaked. The maximal radius of the dust-free cavity is the vaporization
radius ($R_{\rm vap}$).

Here we use Equation (22) of \cite{Corsi2014} to calculate the vaporization radii
of SNe~2010bq and 2012ca. Assuming that the radius of the dust grains is
0.1 $\mu$m and the vaporization temperature ($T_{\rm vap}$) of dust is $\sim$1900 K,
and adopting the derived values of the peak luminosities and the effective SN temperatures
{at the epochs} around the SN peaks
of the 2 SNe \footnote{The inferred date of the peaks of SNe~2010bq and 2012ca
are MJD 55311.35 and MJD 56047.88, respectively.},
we find that the respective $R_{\rm vap}$ of SNe~2010bq and 2012ca are
$\sim 4.1 \times 10^{16}$ cm and $\sim 4.6 \times 10^{16}$ cm.

\subsection{The Blackbody Radii of the Dust Shells of the SNe}
\label{subsec:rbb}

Supposing that the dust shells are optical thick, the blackbody radii
($R_{\rm bb}$) of the dust shells can be derived using
the equation $R_{\rm bb} = ({L_{\rm d}}/({4 \pi \sigma T_{\rm d}^4}))^{1/2}$.
The derived $R_{\rm bb}$ of the dust shells of
SNe~2010bq and 2012ca are respectively
{$\sim 7.7-30 \times 10^{15}$ cm and
$\sim 27-70 \times 10^{15}$ cm} for graphite grains
(see also Table \ref{table:SED_L}).

It should be noted that $R_{\rm bb}$ are always smaller than the radii of the dust shell
($R_{\rm d}$) since the dust shells are optical thin. In other words,
$R_{\rm bb}$ is the lower limits of $R_{\rm d}$ \citep{Fox2011}.

\subsection{The Origin of the Dust}

The NIR excesses of the SEDs can be caused {thermal IR echo from} pre-existing dust and/or
{newly formed dust in the cooling ejecta of SNe or the cooling shock-CSM interaction region
\citep{Gall2014,Sarangi2018,Dwek2021}}.
To determine the nature of the dust of the 2 SNe,
we must compare $R_{\rm vap}$ and $R_{\rm bb}$ with
the the radii of the SN ejecta ($R_{\rm ej}$) at the first epochs of the
SEDs of the SNe.

The method of determining the origin of the dust is comparing the radii of the
dust shells of the SNe and the radii of the ejecta of the SNe. The radii of the
shells of the dust must be larger than both of $R_{\rm vap}$ and $R_{\rm bb}$ which
have been calculated in subsections \ref{subsec:vap} and \ref{subsec:rbb}.
Here, we estimate the values of $R_{\rm ej}$ at the first epochs by determining
the duration ($\Delta t$) between the explosion date and the dates of the first
SEDs and supposing the ejecta velocity.

For SN~2010bq, \cite{Ofek2014} suggest that its rise time
is $\sim 15$ days (see their Table 1); it should be noted that, however,
the rise time \cite{Ofek2014} defined is the duration between the
first detection and the peak. We suggest that the real rise time of
SN~2010bq can be set to be $\lesssim 30$ days.
For SN~2012ca, the inferred explosion date is MJD 55998.2$\pm$20 \citep{Inserra2016},
the date of the first epoch of the optical--NIR SED is 56357.37, $\Delta t$
is $\lesssim 372$ days (the redshift of SN~2012ca is 0.019).
Assuming that the ejecta velocity of the {expanding metal-rich material of the}
SNe is {$\sim 3\times 10^8$ cm s$^{-1}$ (see, e.g., \citealt{Dwek2021})}, 
we find that the respective values of $R_{\rm ej}$ are {$\sim 7.8 \times 10^{14}$ cm
and $\sim 9.6 \times 10^{15}$ cm.}

For comparison, we plot $R_{\rm vap}$, $R_{\rm bb}$, and $R_{\rm ej}$ at the first epochs
in Figure \ref{fig:evo_R}. By comparing the values of $R_{\rm vap}$, $R_{\rm bb}$, and $R_{\rm ej}$,
we find that the radii of the dust shells of {SNe~2010bq and 2012ca are}
significantly larger than their ejecta radii at the first epochs, indicating that
the dust were produced before the explosions of the SNe, {rather than}
in the ejecta {or shocked CSM}.

\subsection{Comparison with Other Study}

\cite{Szalai2019b} demonstrate that the 3 SEDs of SN~2012ca
showed MIR excesses by modeling the $I_1$ and $I_2$ data.
They find that the best-fitting temperatures and the masses of the
dust at 3 epochs are 680 K and $\textless 5.37 \times 10^{-3}$~M$_\odot$ (MJD 56446),
610 K and $\textless 13.02 \times 10^{-3}$~M$_\odot$ (MJD 56619),
and 450 K and $\textless 44.45 \times 10^{-3}$~M$_\odot$ (MJD 56820).

The temperature values we derived {($659-952$ K)} are roughly
consistent with the values derived by \cite{Szalai2019b} ($450-680$ K).
Moreover, our derived dust mass of SN~2012ca {($0.6-7.5\times 10^{-3}$~M$_\odot$)}
is smaller than the upper limits derived by \cite{Szalai2019b} ($\textless 5.37-44.45\times 10^{-3}$~M$_\odot$),
and poses more stringent constraints for the mass of the dust of SN~2012ca.
We plot Figure \ref{fig:12ca_MT} to illustrate the consistency between our
parameters and the parameters or upper limits derived by \cite{Szalai2019b}.

\subsection{{The Energy Source of SNe~2010bq and 2012ca}}
{SN~2012ca is an interacting SN whose luminosity might be powered by the cascade
decay of $^{56}$Ni plus the interaction between the ejecta and the CSM. SN~2010bq might also
be an interacting SN. By using $^{56}$Ni cascade decay model (see \citealt{Wang2021} and the
references therein) to fit the multi-band light
curves of SNe~2010bq and 2012ca (see Figures \ref{fig:SN2010bq_lc} and \ref{fig:SN2012ca_lc}),
we find that the model cannot account for the light curves of the 2 SNe, because the derived
ejecta mass ($\sim 1.13$~M$_\odot$, see Figure \ref{fig:SN2010bq_corner}) of SN~2010bq is
too low for an SN IIn/II, and the derived $^{56}$Ni mass ($\sim 4-5$~M$_\odot$, Figure
\ref{fig:SN2012ca_corner}) of SN 2012ca is unreasonably large for an SN Ia-CSM.}

{It can be expected that the $^{56}$Ni plus the ejecta-CSM interaction model can get
more reasonable results for the 2 SNe. However, modeling the light curves of the 2
SNe using the $^{56}$Ni plus the ejecta-CSM interaction model is beyond the scope of the
paper, and we do not perform the fits.
\footnote{\cite{Inserra2016} use the ejecta-CSM interaction model as well as the $^{56}$Ni plus the ejecta-CSM interaction model
to fit the bolometric light curve of SN~2012ca and got reasonable parameters (see their Figure 13 and Table 2).}}

\section{Conclusion}
\label{sec:Con}

In this paper, we collect the early$-$time optical and NIR photometry of 66 SNe and
construct their optical--NIR SEDs. We use the blackbody model to search for
the evidence of NIR excesses. The SEDs of major fraction of SNe in the sample can
be fitted by the blackbody model, indicating that the SNe do not show significant
dust formation. On the other hand, the blackbody model cannot fit the SEDs of some SNe
, 2 (SNe~2010bq and 2012ca) of which show evident NIR excesses which might
be due to the emission from the dust heated by the radiation of the SN photospheres.

We use the blackbody plus dust {emission} model to fit the SEDs showing
{significant} NIR excesses and find that the model can fit the SEDs of the 2 SNe.
The best-fitting parameters suggest that the masses of the dust are respectively
$\sim 0.1-3.4 \times 10^{-4}$~M$_\odot$ ($\sim 0.2-2.6 \times 10^{-3}$~M$_\odot$)
and {$\sim 0.6-7.5 \times 10^{-3}$~M$_\odot$ ($\sim 2.5-23 \times 10^{-3}$~M$_\odot$)}
for graphite (silicate) grains; the respective best-fitting temperatures of the dust of the 2 SNe are
$\sim 1300-1800$ K ($\sim 1400-2000$ K) and $\sim 600-1000$ K ($\sim 700-1200$ K)
for graphite (silicate) grains.
{Moreover, we use the same model to fit the SEDs that can be fitted by the blackbody model
or show slight NIR excesses, finding that 52 SNe disfavor the model and 7 SNe (SNe~1999el,
2006aa, 2008D, 2008S, 2008in, 2009ay, LSQ13zm) favor the model. Due to the fact that
the SN SEDs have the undulation feature, any model more complicated than the blackbody
model can get better fits for some SEDs. Therefore, the derived masses of the dust of the
7 SNe should be regarded as the upper limits of the masses of the dust surrounding the SNe.}

By comparing the derived temperatures of the dust of the 2 SNe and the evaporation
temperatures of graphite and silicate, we find that the best model explaining the
NIR excesses of the SEDs of  SN~2010bq might be
graphite dust plus blackbody model. For graphite grains, the derived luminosities
of the respective dust shells of SNe~2010bq and 2012ca are
$\sim 4.2-29.3 \times 10^{41}$ erg s$^{-1}$ and
{$\sim 4.6-8.9 \times 10^{41}$ erg s$^{-1}$}.

We compare the vaporization radii, the blackbody radii, and the
estimated upper limits of ejecta radii of the 2 SNe, finding that the NIR excesses of
SNe~2010bq {and 2012ca might be} produced by the pre-existing dust.

According to the results we obtained, one might infer that the percentage of SNe
having early-time NIR excesses produced by the dust is {3.0\%.}
Previous studies for the dust formation of 6 SNe Ibn having optical-NIR SEDs
\cite{Mattila2008,Sanders2013,Gan2021} indicates that the percentage of
SNe Ibn that showed early-time NIR excesses is about 1/3.
While the latter is very rough estimate since the sample is very small,
we can suggest that the percentage of NIR-excess SNe is significantly
smaller than that of NIR-excess interacting SNe.

\acknowledgments
{We thank the anonymous referee for helpful comments and
suggestions that have allowed us to improve this manuscript.}
This work is supported by National Natural Science Foundation of China
(grants 11963001, 11673006, 11851304, 11973020 (C0035736), and U1938201),
the Bagui Scholars Program (LEW), and the Bagui Young Scholars Program (LHJ).

\clearpage

\begin{center}
\setlength{\tabcolsep}{0.05mm}{
\begin{longtable}{ccccccccc}
\caption{The information of the SNe in the sample}
\label{table:details} \\
\hline
\hline
\toprule
Name & R.A.     & Decl.      & ~~$z$       &  ~~type  & Filters & References$^a$ \\
         & (J2000)  &  (J2000)   &           &         &         &   \\
\hline
\hline
SN1996al   & \ra{23}{33}{16}{1}& \dec{-54}{05}{2}{0}  & ~~0.006571 &   ~~II/IIL  & $B$, $H$, $I$, $J$, $R$, $U$, $V$  &1 \\
SN1999el   & \ra{20}{37}{18}{03}& \dec{66}{6}{11}{9}  & ~~0.0047 &   ~~IIn/II  & $B$, $H$, $I$, $J$, $K$, $R$  &2 \\
SN2004ex   & \ra{0}{38}{10}{19}& \dec{2}{43}{17}{2}  & ~~0.018 &   ~~IIb/II  & $B$, $g$, $H$, $i$, $J$, $r$, $u$, $V$, $Y$   &3 \\
SN2004ff   & \ra{4}{58}{46}{19}& \dec{-21}{34}{12}{0}  & ~~0.023 &   ~~IIb/Ic  & $B$, $g$, $H$, $i$, $J$, $R$, $r$, $u$, $V$, $Y$   &3 \\
SN2004gk   & \ra{12}{25}{33}{21}& \dec{12}{15}{39}{9}  & ~~-0.00055 &   ~~Ic  & $H$, $i'$, $J$, $Ks$, $R$, $r'$, $V$  &4,5 \\
SN2004gq   & \ra{5}{12}{4}{81}& \dec{-15}{40}{54}{2}  & ~~0.006468 &   ~~Ib  & $B$, $g$, $H$, $i$, $J$, $R$, $r$, $u$,$V$, $Y$  &3,4 \\
SN2004gt   & \ra{12}{1}{50}{37}& \dec{-18}{52}{12}{7}  & ~~0.005477 &   ~~Ic  & $B$, $g$, $H$, $i'$, $J$, $Ks$, $r$, $r'$, $u$, $V$, $Y$  &3,5 \\
SN2004gv   & \ra{2}{13}{37}{42}& \dec{-0}{43}{5}{8}  & ~~0.02 &   ~~Ib/Ic  & $B$, $g$, $H$, $i$, $J$, $r$, $u$, $V$, $Y$  &3 \\
SN2005aw   & \ra{19}{15}{17}{44}& \dec{-54}{8}{24}{9}  & ~~0.0095 &   ~~Ic  & $B$, $g$, $H$, $i$, $J$, $r$, $u$, $V$, $Y$ &3 \\
SN2005ay   & \ra{11}{52}{48}{07}& \dec{44}{6}{18}{4}  & ~~0.002699 &   ~~II  & $B$, $H$, $I$, $J$, $K$, $R$, $V$ &6,7 \\
SN2005az   & \ra{13}{5}{46}{97}& \dec{27}{44}{8}{4}  & ~~0.0085 &   ~~Ic/Ib  & $B$, $H$, $i'$, $J$, $Ks$, $R$, $r'$, $V$ &4,5 \\
SN2005bf   & \ra{10}{23}{56}{99}& \dec{-3}{11}{29}{3}  & ~~0.018913 &   ~~Ib/Ic  & $B$, $H$, $i$, $i'$, $J$, $Ks$, $r$, $r'$, $U$, $V$ &5,8 \\
SN2005cs   & \ra{13}{29}{53}{37}& \dec{47}{10}{28}{2}  & ~~0.00137 &   ~~II/IIP  & $B$, $H$, $I$, $J$, $K$, $R$, $U$, $V$, $z$ &9 \\
SN2005ek   & \ra{3}{5}{48}{96}& \dec{36}{46}{10}{6}  & ~~0.016551 &   ~~Ic  & $B$, $H$, $I$, $J$, $K$, $R$, $V$  &5,10 \\
SN2005em   & \ra{3}{13}{45}{74}& \dec{-0}{14}{37}{0}  & ~~0.02517 &   ~~II/IIb/Ic  & $g$, $g'$, $H$, $i$, $i'$, $J$, $r$, $r'$, $Y$, $z'$  &3,11 \\
SN2005gj   & \ra{3}{1}{11}{96}& \dec{-0}{33}{14}{0}  & ~~0.0616 &   ~~Ia-CSM/Ia  & $H$, $i'$, $J$, $r'$, $Y$, $z'$  &11,12 \\
SN2005hg   & \ra{1}{55}{41}{87}& \dec{46}{47}{47}{4}  & ~~0.021 &   ~~Ib/Ic  & $B$, $H$, $i'$, $J$, $Ks$, $R$, $r'$, $U$, $V$  &13 \\
SN2005kj   & \ra{8}{40}{9}{18}& \dec{-5}{36}{2}{2}  & ~~0.016 &   ~~IIn/II  & $B$, $g'$, $H$, $i'$, $J$, $Ks$, $r'$, $u'$, $V$, $Y$  &14 \\
SN2005kl   & \ra{12}{24}{35}{68}& \dec{39}{23}{3}{5}  & ~~0.003486 &   ~~Ic  & $B$, $H$, $i'$, $J$, $Ks$, $r'$, $V$  &5 \\
SN2005mf   & \ra{9}{8}{42}{33}& \dec{44}{48}{51}{4}  & ~~0.02675 &   ~~Ic  & $B$, $H$, $i'$, $J$, $Ks$, $R$, $r'$, $V$  &4,5 \\
SN2006T   & \ra{9}{54}{30}{21}& \dec{-25}{42}{29}{3}  & ~~0.0081 &   ~~IIb/II  & $B$, $g$, $H$, $i$, $i'$, $J$, $r$, $r'$, $U$, $u$, $V$, $Y$   &3,5 \\
SN2006aa   & \ra{11}{53}{19}{89}& \dec{20}{45}{18}{2}  & ~~0.0207 &   ~~IIn/II  & $B$, $g'$, $H$, $i'$, $J$, $r'$, $u'$, $V$, $Y$   &14 \\
SN2006aj   & \ra{3}{21}{39}{67}& \dec{16}{52}{2}{27}  & ~~0.033023 &   ~~Ic/Ib  & $B$, $H$, $i'$, $J$, $Ks$, $r'$, $U$, $V$  &5,15 \\
SN2006au   & \ra{17}{57}{13}{56}& \dec{12}{11}{3}{2}  & ~~0.00958 &   ~~II  & $B$, $g'$, $H$, $i'$, $J$, $r'$, $V$, $Y$  &16 \\
SN2006ba   & \ra{9}{43}{13}{4}& \dec{-9}{36}{53}{0}  & ~~0.019 &   ~~II/IIb  & $B$, $g$, $H$, $i$, $i'$, $J$, $r$, $r'$, $V$, $Y$  &3,5 \\
SN2006bf   & \ra{12}{58}{50}{68}& \dec{9}{39}{30}{1}  & ~~0.024 &   ~~IIb/Ib  & $g$, $H$, $i$, $i'$, $J$, $r$, $r'$, $V$, $Y$  &3,5 \\
SN2006bo   & \ra{20}{30}{41}{9}& \dec{9}{11}{40}{8}  & ~~0.0153 &   ~~IIn/II  & $B$, $g'$, $H$, $i'$, $J$, $r'$, $u'$, $V$, $Y$   &14 \\
SN2006ep   & \ra{0}{41}{24}{88}& \dec{25}{29}{46}{7}  & ~~0.015 &   ~~Ib/Ic  & $B$, $H$, $i'$, $J$, $r'$, $V$, $Y$  &3,5 \\
SN2006fo   & \ra{2}{32}{38}{89}& \dec{0}{37}{3}{0}  & ~~0.020709 &   ~~Ib/Ic  & $B$, $g$, $g'$, $H$, $i$, $i'$, $J$, $Ks$, $r$, $r'$, $u$, $u'$, $V$, $z'$  &3,5,11 \\
SN2006ir   & \ra{23}{4}{35}{68}& \dec{7}{36}{21}{5}  & ~~0.02 &   ~~Ic/Ib  & $g$, $H$, $i$, $i'$, $J$, $r$, $r'$, $V$, $Y$  &3,5 \\
SN2006lc   & \ra{22}{44}{24}{48}& \dec{-0}{9}{53}{5}  & ~~0.016104 &   ~~Ib/Ic  & $B$, $g$, $H$, $J$, $r$, $r'$, $u$, $V$, $Y$  &3 \\
SN2006ld   & \ra{0}{35}{27}{81}& \dec{2}{55}{50}{7}  & ~~0.0139 &   ~~Ib  & $B$, $H$, $i'$, $J$, $Ks$, $r'$, $U$, $V$  &5 \\
SN2007C   & \ra{13}{8}{49}{3}& \dec{-6}{47}{1}{0}  & ~~0.005611 &   ~~Ib  & $B$, $g$, $H$, $i$, $i'$, $J$, $Ks$, $R$, $r$, $r'$, $V$, $Y$  &3,4,5 \\
SN2007I   & \ra{11}{59}{13}{15}& \dec{-1}{36}{18}{9}  & ~~0.021638 &   ~~Ic  & $B$, $H$, $i'$, $J$, $Ks$, $r'$, $V$  &5 \\
SN2007Y   & \ra{3}{2}{35}{92}& \dec{-22}{53}{50}{1}  & ~~0.0046 &   ~~Ib  & $B$, $g$, $H$, $i$, $J$, $r$, $u$, $V$, $Y$, $UVW1$, $UVW2$  &3,17 \\
SN2007aa   & \ra{12}{0}{27}{69}& \dec{-1}{4}{51}{6}  & ~~0.004887 &   ~~IIP/II  & $B$, $H$, $i'$, $J$, $K$, $r'$, $V$, $UVW1$  &6,17,18 \\
SN2007ag   & \ra{10}{1}{35}{99}& \dec{21}{36}{42}{0}  & ~~0.0207 &   ~~Ib  & $H$, $i'$, $J$, $r'$, $Y$  &3,5 \\
SN2007av   & \ra{10}{34}{43}{18}& \dec{11}{11}{39}{1}  & ~~0.00464 &   ~~II/IIP  & $H$, $i'$, $J$, $K$, $r'$, $V$  &6,18 \\
SN2007ce   & \ra{12}{10}{17}{96}& \dec{48}{43}{31}{5}  & ~~0.046 &   ~~Ic  & $B$, $H$, $i'$, $J$, $Ks$, $r'$, $V$  &5 \\
SN2007hn   & \ra{21}{2}{46}{85}& \dec{-4}{5}{25}{2}  & ~~0.03 &   ~~Ib/Ic  & $B$, $g$, $H$, $i$, $J$, $r$, $V$, $Y$  &3 \\
SN2007kj   & \ra{0}{1}{19}{58}& \dec{13}{6}{30}{6}  & ~~0.0179 &   ~~Ib/Ic  & $B$,  $H$, $i'$, $J$, $r'$, $V$, $Y$ &3,5 \\
SN2007rz   & \ra{04}{31}{10}{84}& \dec{07}{37}{51}{49}  & ~~0.013 &   ~~Ic  & $B$, $g$, $H$, $i$, $J$, $r$, $V$, $Y$  &3 \\
SN2007uy   & \ra{9}{9}{35}{4}& \dec{33}{7}{9}{9}  & ~~0.0065 &   ~~Ib  & $B$, $H$, $i'$, $J$, $Ks$, $r'$, $U$, $V$, $UVW1$, $UVW2$, $UVM2$  &5,17 \\
SN2008D   & \ra{9}{9}{30}{625}& \dec{33}{8}{20}{16}  & ~~0.006521 &   ~~Ib  & $B$, $H$, $i'$, $J$, $Ks$, $r'$, $U$, $V$, $UVW1$ &5,17,19 \\
SN2008S   & \ra{20}{34}{45}{35}& \dec{60}{5}{57}{8}  & ~~0.00016 &   ~~IIn  & $B$, $H$, $I$, $J$, $K$, $R$, $U$, $V$  &13 \\
SN2008aq   & \ra{12}{50}{30}{36}& \dec{-10}{52}{1}{3}  & ~~0.008 &   ~~IIb/II  & $B$, $g$, $H$, $i$, $i'$, $J$, $r$, $r'$, $U$, $V$, $Y$  &3,5 \\
SN2008ax   & \ra{12}{30}{40}{8}& \dec{41}{38}{16}{1}  & ~~0.001931 &   ~~IIb/II/Ib  & $B$, $g'$, $H$, $i'$, $J$, $Ks$, $r'$, $U$, $u'$, $V$, $z'$, $UVW1$, $UVW2$  &17,20 \\
SN2008hh   & \ra{1}{26}{3}{65}& \dec{11}{26}{26}{5}  & ~~0.0194 &   ~~Ic  & $B$, $g$, $H$, $i$, $J$, $Ks$, $r$, $u$, $V$  &3,5 \\
SN2008if   & \ra{9}{20}{23}{47}& \dec{-7}{52}{33}{1}  & ~~0.0115 &   ~~II  & $H$, $J$, $K$, $V$  &6,18 \\
SN2008in   & \ra{12}{22}{1}{77}& \dec{4}{28}{47}{5}  & ~~0.005224 &   ~~IIP/II  & $B$, $H$, $I$, $i'$, $J$, $K$, $R$, $r'$, $V$, $UVW1$  &6,18,21 \\
SN2009K   & \ra{4}{36}{36}{77}& \dec{-0}{8}{35}{6}  & ~~0.0117 &   ~~II/IIb  & $B$, $g$, $H$, $i$, $i'$, $J$, $K$, $r$, $r'$, $u$, $V$, $Y$  &3,5 \\
SN2009ay   & \ra{17}{48}{22}{97}& \dec{54}{8}{54}{7}  & ~~0.0222 &   ~~II  & $B$, $H$, $I$, $i'$, $J$, $K$, $R$, $r'$, $V$  &6,21 \\
SN2009bb   & \ra{10}{31}{33}{92}& \dec{-39}{57}{28}{2}  & ~~0.0104 &   ~~Ic  & $B$, $g$, $H$,  $I$, $i$, $J$, $R$, $r$, $u$, $V$, $Y$  &3,22 \\
SN2009ca   & \ra{21}{26}{22}{2}& \dec{-40}{51}{48}{6}  & ~~0.0899 &   ~~Ic  & $B$, $g$, $H$, $i$, $J$, $r$, $V$, $Y$  &3 \\
SN2009dt   & \ra{22}{10}{09}{18}& \dec{-36}{05}{44}{2}  & ~~0.0104 &   ~~Ic  & $B$, $g$, $H$, $i$, $J$, $r$, $V$, $Y$  &3 \\
SN2009er   & \ra{15}{39}{29}{84}& \dec{24}{26}{5}{3}  & ~~0.035 &   ~~Ib  & $B$, $H$, $i'$, $J$, $Ks$, $r'$, $V$  &5 \\
SN2009ib   & \ra{4}{17}{40}{09}& \dec{-62}{46}{40}{3}  & ~~0.00448 &   ~~IIP/II  & $B$, $g'$, $H$, $I$, $i'$, $J$, $R$, $r'$, $V$, $z'$  &23 \\
SN2009iz   & \ra{2}{42}{15}{41}& \dec{42}{23}{50}{1}  & ~~0.014 &   ~~Ib  & $B$, $H$, $i'$, $J$, $Ks$, $r'$, $u'$, $V$ &5 \\
SN2009kr   & \ra{5}{12}{3}{3}& \dec{-15}{41}{52}{2}  & ~~0.0065 &   ~~IIn/IIL/II  & $B$, $H$, $I$, $J$, $K$, $R$, $U$, $V$, $UVW1$, $UVW2$, $UVM2$  &17,21 \\
SN2010bq   & \ra{16}{46}{55}{41}& \dec{34}{9}{35}{4}  & ~~0.031 &   ~~IIn/II  & $B$, $H$, $i'$, $J$, $K$, $R$, $u'$, $V$  &6 \\
SN2012ca   & \ra{18}{41}{07}{25}& \dec{-41}{47}{38}{4}  & ~~0.019 &   ~~Ia-CSM/IIn  & $g$, $H$, $i$, $J$, $K$, $R$, $r$, $V$, $z$  &24 \\
SN2012ec   & \ra{2}{45}{59}{89}& \dec{-7}{34}{25}{0}  & ~~0.004693 &   ~~IIP  & $B$, $H$, $I$, $J$, $Ks$, $R$, $U$, $V$   &25 \\
SN2013dn   & \ra{23}{37}{45}{74}& \dec{14}{42}{37}{1}  & ~~0.056185 &   ~~Ia-CSM/IIn  & $H$, $i$, $J$, $r$, $Y$, $Z$  &26 \\
SN2016bkv   & \ra{10}{18}{19}{31}& \dec{41}{25}{39}{3}  & ~~0.002 &   ~~II  & $B$, $H$, $I$, $J$, $Ks$, $R$, $U$, $V$ &27 \\
PTF12gzk   & \ra{22}{12}{41}{53}& \dec{0}{30}{43}{1}  & ~~0.01377 &   ~~Ic  & $g$, $H$, $i$, $I$, $J$, $K$, $R$, $r$, $V$, $Y$, $Z$, $UVW1$, $UVW2$, $UVM2$  &28 \\
LSQ13zm   & \ra{10}{26}{54}{55}& \dec{19}{52}{55}{2}  & ~~0.029 &   ~~IIn  & $B$, $g$, $H$, $i$, $J$, $K$, $R$, $V$  &29 \\
\hline
\bottomrule
\end{longtable}}
$^a$ {References}. (1) \cite{Benetti2016}; (2) \cite{Di Carlo2002}; (3) \cite{Stritzinger2018}; (4) \cite{Drout2011}; (5) \cite{Bianco2014}; (6) \cite{Hicken2017}; (7) \cite{Faran2014}; (8) \cite{Tominaga2005}; (9) \cite{Pastorello2009};(10) \cite{Drout2013};  (11) \cite{Sako2018}; (12) \cite{Prieto2007}; (13) \cite{Botticella2009}; (14) \cite{Taddia2013}; (15) \cite{Modjaz2006}; (16) \cite{Taddia2012}; (17) \cite{Brown2014}; (18) \cite{Anderson2014}; (19) \cite{Mazzali2008}; (20) \cite{Pastorello2008}; (21) \cite{de Jaeger2019}; (22) \cite{Pignata2011}; (23) \cite{Takats2015}; (24) \cite{Inserra2016}; (25) \cite{Smartt2015}; (26) \cite{Fox2015}; (27) \cite{Nakaoka2018}; (28) \cite{Ben-Ami2012}; (29) \cite{Tartaglia2016}. \\ 
\end{center}

\clearpage

\begin{center}
\setlength{\LTcapwidth}{\textwidth}
\setlength{\tabcolsep}{1pt}{
\begin{longtable}{c c c c c c c c c c c c c}
\caption{The best-fitting parameters of the blackbody plus graphite (silicate) model for the SEDs of the {2} SNe having NIR excesses at the different rest-frame epochs. Here, $T_{\rm ph}$ is the temperature of the SN photosphere, $R_{\rm ph}$ the radius of the SN photosphere, $T_{\rm d}$ the temperature of the dust shell, $M_{\rm d}$ the mass of the dust shell.}
\label{table:SED_PARAM-double}
\\\hline\hline\noalign{\smallskip}
&\multicolumn{5}{c}{{Blackbody plus Graphite}} & \multicolumn{5}{c}{{Blackbody plus Silicate}}\\\noalign{\smallskip}
\cmidrule(lr){2-6} \cmidrule(lr){7-11}
\colhead{Phase$^a$} & \colhead{$T_{\rm ph}$}&\colhead{$R_{\rm ph}$}&\colhead{$T_{\rm d}$}&\colhead{$M_{\rm d}$}&\colhead{$\chi^{\rm 2}$/dof}&\colhead{$T_{\rm ph}$}&\colhead{$R_{\rm ph}$}&\colhead{$T_{\rm d}$} & \colhead{$M_{\rm d}$}  & \colhead{$\chi^{\rm 2}$/dof} \\
& (K) & (10$^{15}$ cm) &  (K) & (10$\rm ^{-3}\ M_{\odot}$) &  & (K) & (10$^{15}$ cm) & (K) & (10$\rm^{-3}\ M_{\odot}$) & \\\noalign{\smallskip} \hline \noalign{\smallskip}
\multicolumn{11}{c}{{SN 2010bq}}\\\noalign{\smallskip} \hline \noalign{\smallskip}
6.9 d & $10901.66^{+463.7}_{-408.3}$ & $1.01^{+0.1}_{-0.1}$ & $1423.55^{+82.8}_{-79.6}$ & $0.31^{+0.1}_{-0.1}$ & 0.5 & $11190.01^{+582.1}_{-475.9}$ & $0.97^{+0.1}_{-0.1}$ & $1683.42^{+109.6}_{-101.5}$ & $1.86^{+0.8}_{-0.6}$ & 0.53\\\noalign{\smallskip} \hline \noalign{\smallskip}
7.8 d & $11313.09^{+1227.3}_{-819.2}$ & $0.97^{+0.1}_{-0.1}$ & $1724.38^{+200.0}_{-184.3}$ & $0.076^{+0.1}_{-0.0}$ & 1.39 & $11589.74^{+1325.0}_{-930.4}$ & $0.93^{+0.1}_{-0.1}$ & $1912.86^{+178.0}_{-211.8}$ & $0.8^{+0.7}_{-0.3}$ & 2.67\\\noalign{\smallskip} \hline \noalign{\smallskip}
9.7 d & $11918.43^{+453.4}_{-389.8}$ & $0.98^{+0.0}_{-0.0}$ & $1478.63^{+94.3}_{-82.4}$ & $0.29^{+0.1}_{-0.1}$ & 11.42 & $12578.89^{+634.2}_{-537.7}$ & $0.9^{+0.1}_{-0.1}$ & $1796.01^{+106.7}_{-103.6}$ & $1.56^{+0.6}_{-0.4}$ & 9.39\\\noalign{\smallskip} \hline \noalign{\smallskip}
15.7 d & $10392.74^{+840.5}_{-672.9}$ & $1.13^{+0.1}_{-0.1}$ & $1536.6^{+134.8}_{-103.0}$ & $0.22^{+0.1}_{-0.1}$ & 2.89 & $11501.15^{+1549.8}_{-1083.8}$ & $0.97^{+0.1}_{-0.2}$ & $1877.39^{+151.9}_{-146.8}$ & $1.2^{+0.7}_{-0.4}$ & 1.73\\\noalign{\smallskip} \hline \noalign{\smallskip}
20.4 d & $10225.32^{+1026.8}_{-706.5}$ & $0.89^{+0.1}_{-0.1}$ & $1728.37^{+168.8}_{-165.5}$ & $0.064^{+0.1}_{-0.0}$ & 2.67 & $10503.13^{+1122.8}_{-800.7}$ & $0.85^{+0.1}_{-0.1}$ & $1916.36^{+162.9}_{-177.5}$ & $0.68^{+0.5}_{-0.2}$ & 4.68\\\noalign{\smallskip} \hline \noalign{\smallskip}
24.3 d & $7848.84^{+416.4}_{-324.2}$ & $1.15^{+0.1}_{-0.1}$ & $1498.22^{+153.5}_{-140.9}$ & $0.14^{+0.1}_{-0.1}$ & 1.93 & $7965.7^{+510.6}_{-380.9}$ & $1.12^{+0.1}_{-0.1}$ & $1683.42^{+183.6}_{-174.2}$ & $1.18^{+1.1}_{-0.5}$ & 3.1\\\noalign{\smallskip} \hline \noalign{\smallskip}
25.3 d & $7708.51^{+299.2}_{-282.1}$ & $1.16^{+0.1}_{-0.1}$ & $1310.28^{+122.4}_{-110.4}$ & $0.34^{+0.3}_{-0.2}$ & 8.56 & $7770.93^{+315.9}_{-281.8}$ & $1.14^{+0.1}_{-0.1}$ & $1469.0^{+130.7}_{-125.8}$ & $2.59^{+2.0}_{-1.1}$ & 8.9\\\noalign{\smallskip} \hline \noalign{\smallskip}
27.3 d & $7429.49^{+350.7}_{-317.7}$ & $1.15^{+0.1}_{-0.1}$ & $1341.98^{+130.9}_{-121.8}$ & $0.26^{+0.3}_{-0.1}$ & 8.33 & $7478.14^{+385.6}_{-328.6}$ & $1.13^{+0.1}_{-0.1}$ & $1511.68^{+152.0}_{-143.4}$ & $1.92^{+1.6}_{-0.9}$ & 8.9\\\noalign{\smallskip} \hline \noalign{\smallskip}
28.2 d & $7311.53^{+350.9}_{-308.7}$ & $1.17^{+0.1}_{-0.1}$ & $1429.46^{+136.5}_{-128.2}$ & $0.15^{+0.1}_{-0.1}$ & 5.91 & $7377.04^{+420.4}_{-328.1}$ & $1.15^{+0.1}_{-0.1}$ & $1612.53^{+165.0}_{-151.8}$ & $1.11^{+0.8}_{-0.5}$ & 6.74\\\noalign{\smallskip} \hline \noalign{\smallskip}
30.1 d & $7085.68^{+371.4}_{-316.6}$ & $1.17^{+0.1}_{-0.1}$ & $1532.11^{+166.4}_{-153.2}$ & $0.091^{+0.1}_{-0.0}$ & 4.65 & $7194.08^{+451.1}_{-350.7}$ & $1.14^{+0.1}_{-0.1}$ & $1717.44^{+190.1}_{-175.0}$ & $0.8^{+0.7}_{-0.4}$ & 5.12\\\noalign{\smallskip} \hline \noalign{\smallskip}
31.1 d & $6860.42^{+377.8}_{-298.3}$ & $1.21^{+0.1}_{-0.1}$ & $1357.33^{+275.6}_{-261.2}$ & $0.18^{+0.8}_{-0.1}$ & 0.49 & $6908.47^{+434.4}_{-332.8}$ & $1.19^{+0.1}_{-0.1}$ & $1513.74^{+314.3}_{-307.3}$ & $1.52^{+6.0}_{-1.0}$ & 0.65\\\noalign{\smallskip} \hline \noalign{\smallskip}
34.9 d & $7149.04^{+542.9}_{-417.3}$ & $0.93^{+0.1}_{-0.1}$ & $1352.65^{+368.7}_{-418.6}$ & $0.055^{+0.3}_{-0.1}$ & 2.68 & $7145.7^{+562.3}_{-420.1}$ & $0.93^{+0.1}_{-0.1}$ & $1481.29^{+374.4}_{-506.6}$ & $0.46^{+2.0}_{-0.5}$ & 3.04\\\noalign{\smallskip} \hline \noalign{\smallskip}
35.9 d & $6660.4^{+1679.7}_{-1016.0}$ & $1.06^{+0.5}_{-0.4}$ & $1537.38^{+454.9}_{-817.4}$ & $0.013^{+0.1}_{-0.0}$ & -- & $6464.79^{+1626.6}_{-884.7}$ & $1.14^{+0.5}_{-0.4}$ & $1556.9^{+456.5}_{-859.4}$ & $0.17^{+1.1}_{-0.2}$ & --\\\noalign{\smallskip} \hline \noalign{\smallskip}
36.9 d & $6707.38^{+538.2}_{-437.4}$ & $0.93^{+0.1}_{-0.1}$ & $1359.67^{+152.0}_{-186.6}$ & $0.12^{+0.1}_{-0.0}$ & 42.22 & $6857.85^{+758.8}_{-530.0}$ & $0.88^{+0.2}_{-0.2}$ & $1616.02^{+218.8}_{-261.2}$ & $0.64^{+0.7}_{-0.2}$ & 106.56\\\noalign{\smallskip} \hline \noalign{\smallskip}
37.8 d & $6975.18^{+842.4}_{-601.1}$ & $0.85^{+0.2}_{-0.2}$ & $1477.11^{+257.0}_{-250.1}$ & $0.079^{+0.2}_{-0.1}$ & 3.14 & $7040.94^{+956.0}_{-652.0}$ & $0.83^{+0.2}_{-0.2}$ & $1629.25^{+280.6}_{-289.3}$ & $0.76^{+1.9}_{-0.5}$ & 3.69\\\noalign{\smallskip} \hline \noalign{\smallskip}
\multicolumn{11}{c}{{SN 2012ca}}\\\noalign{\smallskip} \hline \noalign{\smallskip}
308.9 d & $9075.71^{+461.0}_{-422.2}$ & $1.14^{+0.1}_{-0.1}$ & $659.7^{+163.7}_{-141.2}$ & $7.48^{+107.1}_{-6.5}$ & 5.21 & $9099.73^{+470.8}_{-428.5}$ & $1.14^{+0.1}_{-0.1}$ & $761.16^{+183.9}_{-149.8}$ & $23.14^{+173.8}_{-18.9}$ & 5.24\\\noalign{\smallskip} \hline \noalign{\smallskip}
345.1 d & $7855.93^{+679.4}_{-552.5}$ & $1.05^{+0.1}_{-0.1}$ & $952.31^{+147.7}_{-161.5}$ & $0.6^{+1.9}_{-0.4}$ & 2.74 & $7924.11^{+757.3}_{-592.2}$ & $1.03^{+0.1}_{-0.1}$ & $1123.33^{+201.7}_{-211.4}$ & $2.47^{+7.2}_{-1.5}$ & 2.72\\\noalign{\smallskip} \hline \noalign{\smallskip}
520.5 d & $13692.76^{+2025.8}_{-1509.2}$ & $0.098^{+0.0}_{-0.0}$ & $861.15^{+30.1}_{-29.0}$ & $2.15^{+0.8}_{-0.6}$ & 1.87 & $14188.0^{+2353.1}_{-1648.8}$ & $0.093^{+0.0}_{-0.0}$ & $967.31^{+35.5}_{-34.3}$ & $11.12^{+3.9}_{-2.9}$ & 2.0\\\noalign{\smallskip} \hline \noalign{\smallskip}
554.9 d & $5712.62^{+4312.8}_{-1977.2}$ & $0.061^{+0.1}_{-0.0}$ & $827.43^{+35.4}_{-32.9}$ & $2.51^{+1.2}_{-0.8}$ & -- & $5756.88^{+4273.3}_{-2014.8}$ & $0.057^{+0.1}_{-0.0}$ & $936.28^{+45.2}_{-41.2}$ & $11.78^{+5.7}_{-3.9}$ & --\\\noalign{\smallskip} \hline\hline\noalign{\smallskip}

\end{longtable}}
$^a$ {Phase}. All the phases are relative to the first data which are supposed to be respectively JD 2455296.76 (SN~2010bq) and JD 2456043.08 (SN~2012ca).
\end{center}


\clearpage

\begin{center}
\setlength{\LTcapwidth}{\textwidth}
\setlength{\tabcolsep}{10pt}{
\begin{longtable}{c c c c c c c c c c}
\caption{The derived values of the SN photosphere luminosities ($L_{\rm ph}$), the dust
luminosities ($L_{\rm d}$), the radii ($R_{\rm ph}$) of the SN photosphere, the {blackbody} radii ({$R_{\rm bb}$}) of the dust shells surrounding
SN~2010bq and SN~2012ca, {and their optical depth ($\tau=L_{\rm dust}$/($L_{\rm dust}+L_{\rm ph}$)} at different epochs.
{Note that the values of $R_{\rm bb}$ is the lower limits of that of $R_{\rm d}$.}}
\label{table:SED_L}
\\\hline\hline\noalign{\smallskip}

\colhead{Phase} & \colhead{$L_{\rm ph}$} & \colhead{$L_{\rm d}$} & \colhead{$R_{\rm ph}$} & \colhead{$R_{\rm bb}$} & \colhead{$L_{\rm dust}$/($L_{\rm dust}+L_{\rm ph}$)}\\
(days)& ($\rm 10^{41}$ erg s$^{-1}$)  & ($\rm 10^{41}$ erg s$^{-1}$) & ($\rm 10^{15}$ cm) & ($\rm 10^{15}$ cm) & \\\noalign{\smallskip}\hline\noalign{\smallskip}
\multicolumn{6}{c}{{SN 2010bq}}\\\noalign{\smallskip} \hline \noalign{\smallskip}
6.9 & $102.07^{+5.0}_{-4.3}$ & $25.39^{+2.3}_{-2.3}$ & $1.01^{+0.1}_{-0.1}$ & $29.42^{+4.3}_{-3.7}$ & $0.20^{+0.02}_{-0.02}$\\\noalign{\smallskip}\hline\noalign{\smallskip}
7.8 & $109.16^{+13.4}_{-8.2}$ & $20.34^{+2.9}_{-2.8}$ & $0.97^{+0.1}_{-0.1}$ & $17.65^{+5.3}_{-3.6}$ & $0.16^{+0.02}_{-0.02}$\\\noalign{\smallskip}\hline\noalign{\smallskip}
9.7 & $136.81^{+6.7}_{-5.6}$ & $29.28^{+2.5}_{-2.3}$ & $0.98^{+0.0}_{-0.0}$ & $29.29^{+4.5}_{-4.1}$ & $0.18^{+0.01}_{-0.01}$\\\noalign{\smallskip}\hline\noalign{\smallskip}
15.7 & $106.21^{+11.6}_{-8.3}$ & $28.11^{+2.5}_{-2.4}$ & $1.13^{+0.1}_{-0.1}$ & $26.45^{+4.9}_{-4.7}$ & $0.21^{+0.02}_{-0.02}$\\\noalign{\smallskip}\hline\noalign{\smallskip}
20.4 & $62.07^{+6.0}_{-4.1}$ & $17.06^{+2.3}_{-2.2}$ & $0.89^{+0.1}_{-0.1}$ & $16.16^{+4.2}_{-2.9}$ & $0.22^{+0.03}_{-0.02}$\\\noalign{\smallskip}\hline\noalign{\smallskip}
24.3 & $35.94^{+1.1}_{-1.1}$ & $15.52^{+2.3}_{-1.9}$ & $1.15^{+0.1}_{-0.1}$ & $20.51^{+5.6}_{-4.1}$ & $0.30^{+0.03}_{-0.03}$\\\noalign{\smallskip}\hline\noalign{\smallskip}
25.3 & $33.9^{+1.0}_{-1.0}$ & $17.27^{+3.3}_{-2.8}$ & $1.16^{+0.1}_{-0.1}$ & $28.41^{+7.6}_{-6.1}$ & $0.34^{+0.04}_{-0.04}$\\\noalign{\smallskip}\hline\noalign{\smallskip}
27.3 & $28.81^{+1.0}_{-1.0}$ & $15.07^{+3.0}_{-2.8}$ & $1.15^{+0.1}_{-0.1}$ & $25.36^{+7.2}_{-5.5}$ & $0.34^{+0.05}_{-0.04}$\\\noalign{\smallskip}\hline\noalign{\smallskip}
28.2 & $28.15^{+0.9}_{-0.9}$ & $12.37^{+2.1}_{-2.0}$ & $1.17^{+0.1}_{-0.1}$ & $20.25^{+4.9}_{-3.7}$ & $0.31^{+0.04}_{-0.04}$\\\noalign{\smallskip}\hline\noalign{\smallskip}
30.1 & $24.83^{+0.8}_{-0.8}$ & $11.79^{+2.2}_{-2.1}$ & $1.17^{+0.1}_{-0.1}$ & $17.11^{+5.1}_{-3.8}$ & $0.32^{+0.04}_{-0.04}$\\\noalign{\smallskip}\hline\noalign{\smallskip}
31.1 & $23.13^{+0.7}_{-0.8}$ & $11.93^{+6.7}_{-3.5}$ & $1.21^{+0.1}_{-0.1}$ & $21.4^{+18.8}_{-8.2}$ & $0.34^{+0.13}_{-0.07}$\\\noalign{\smallskip}\hline\noalign{\smallskip}
34.9 & $16.09^{+0.8}_{-0.7}$ & $5.49^{+3.8}_{-5.5}$ & $0.93^{+0.1}_{-0.1}$ & $12.6^{+12.9}_{-11.2}$ & $0.25^{+0.13}_{-0.19}$\\\noalign{\smallskip}\hline\noalign{\smallskip}
35.9 & $16.27^{+2.1}_{-1.4}$ & $4.19^{+4.3}_{-4.2}$ & $1.06^{+0.5}_{-0.4}$ & $7.65^{+9.3}_{-7.6}$ & $0.20^{+0.17}_{-0.16}$\\\noalign{\smallskip}\hline\noalign{\smallskip}
36.9 & $12.51^{+0.7}_{-0.7}$ & $7.45^{+1.3}_{-0.7}$ & $0.93^{+0.1}_{-0.1}$ & $17.34^{+4.7}_{-2.2}$ & $0.37^{+0.04}_{-0.03}$\\\noalign{\smallskip}\hline\noalign{\smallskip}
37.8 & $12.17^{+0.8}_{-0.8}$ & $8.76^{+3.2}_{-2.6}$ & $0.85^{+0.2}_{-0.2}$ & $15.46^{+10.0}_{-5.2}$ & $0.42^{+0.09}_{-0.07}$\\\noalign{\smallskip}\hline\noalign{\smallskip}
\multicolumn{6}{c}{{SN 2012ca}}\\\noalign{\smallskip} \hline \noalign{\smallskip}
308.9 & $63.26^{+6.2}_{-5.4}$ & $6.56^{+19.4}_{-3.4}$ & $1.14^{+0.1}_{-0.1}$ & $69.77^{+154.3}_{-38.5}$ & $0.09^{+0.25}_{-0.04}$\\\noalign{\smallskip}\hline\noalign{\smallskip}
345.1 & $29.66^{+3.2}_{-2.8}$ & $4.62^{+1.7}_{-0.7}$ & $1.05^{+0.1}_{-0.1}$ & $27.65^{+19.6}_{-7.9}$ & $0.13^{+0.04}_{-0.02}$\\\noalign{\smallskip}\hline\noalign{\smallskip}
520.5 & $2.38^{+0.7}_{-0.4}$ & $8.9^{+1.2}_{-1.0}$ & $0.098^{+0.0}_{-0.0}$ & $47.64^{+6.6}_{-5.7}$ & $0.79^{+0.05}_{-0.04}$\\\noalign{\smallskip}\hline\noalign{\smallskip}
554.9 & $0.052^{+0.2}_{-0.0}$ & $8.21^{+1.5}_{-1.3}$ & $0.061^{+0.1}_{-0.0}$ & $49.58^{+8.8}_{-7.6}$ & $0.99^{+0.03}_{-0.01}$\\\noalign{\smallskip}\hline\hline\noalign{\smallskip}
\end{longtable}}
\end{center}


\clearpage

\begin{figure}[tbph]
\begin{center}
\includegraphics[width=0.245\textwidth,angle=0]{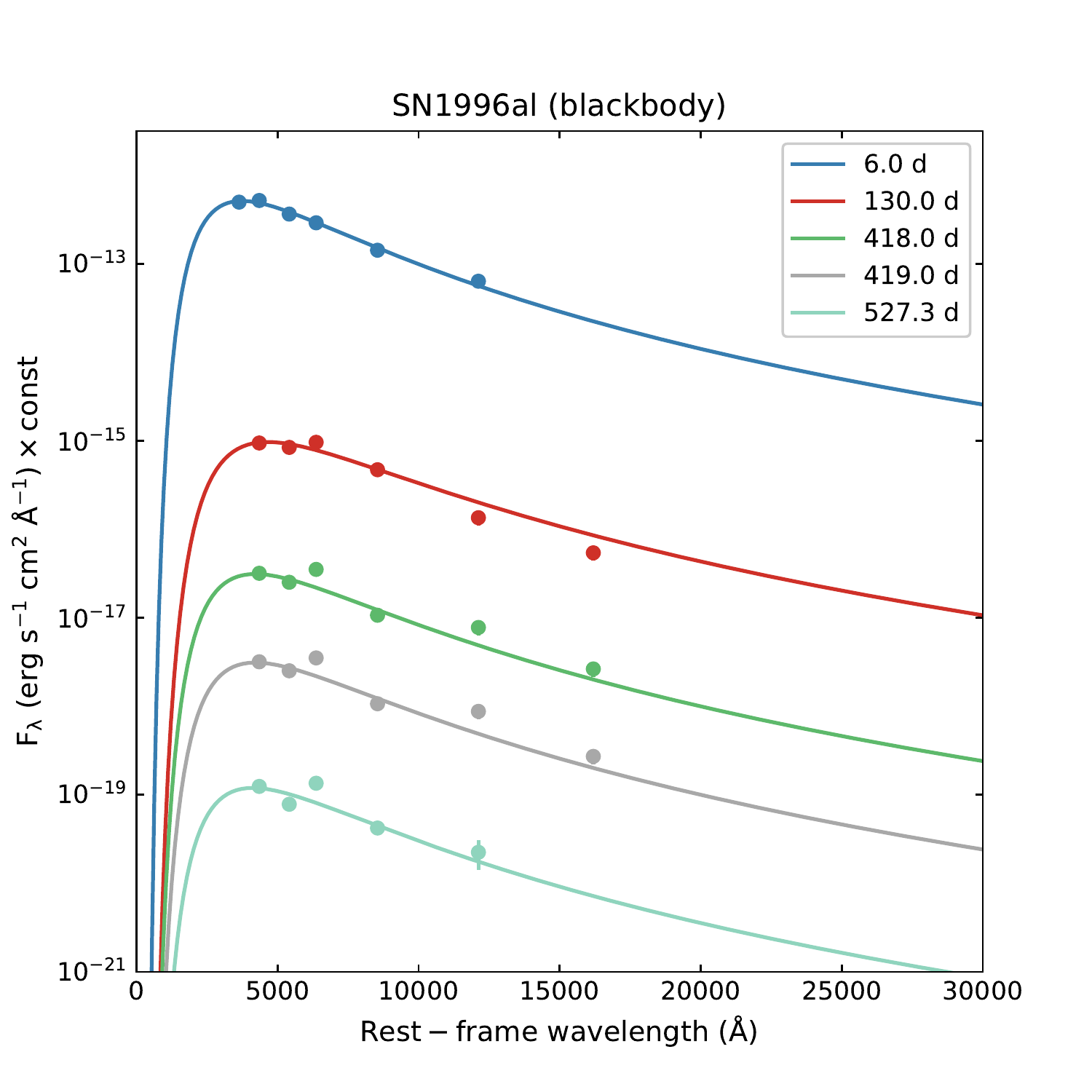}
\includegraphics[width=0.245\textwidth,angle=0]{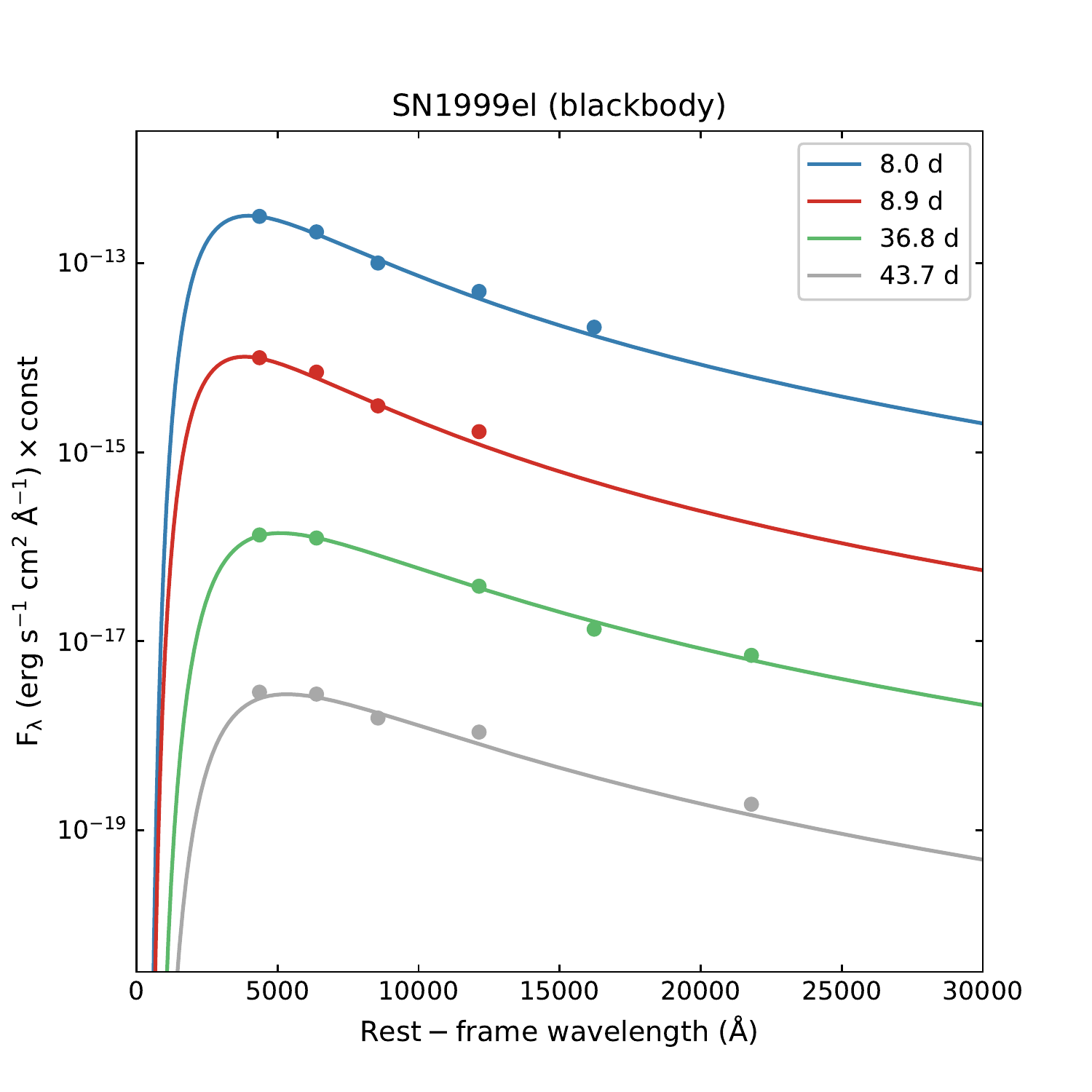}
\includegraphics[width=0.245\textwidth,angle=0]{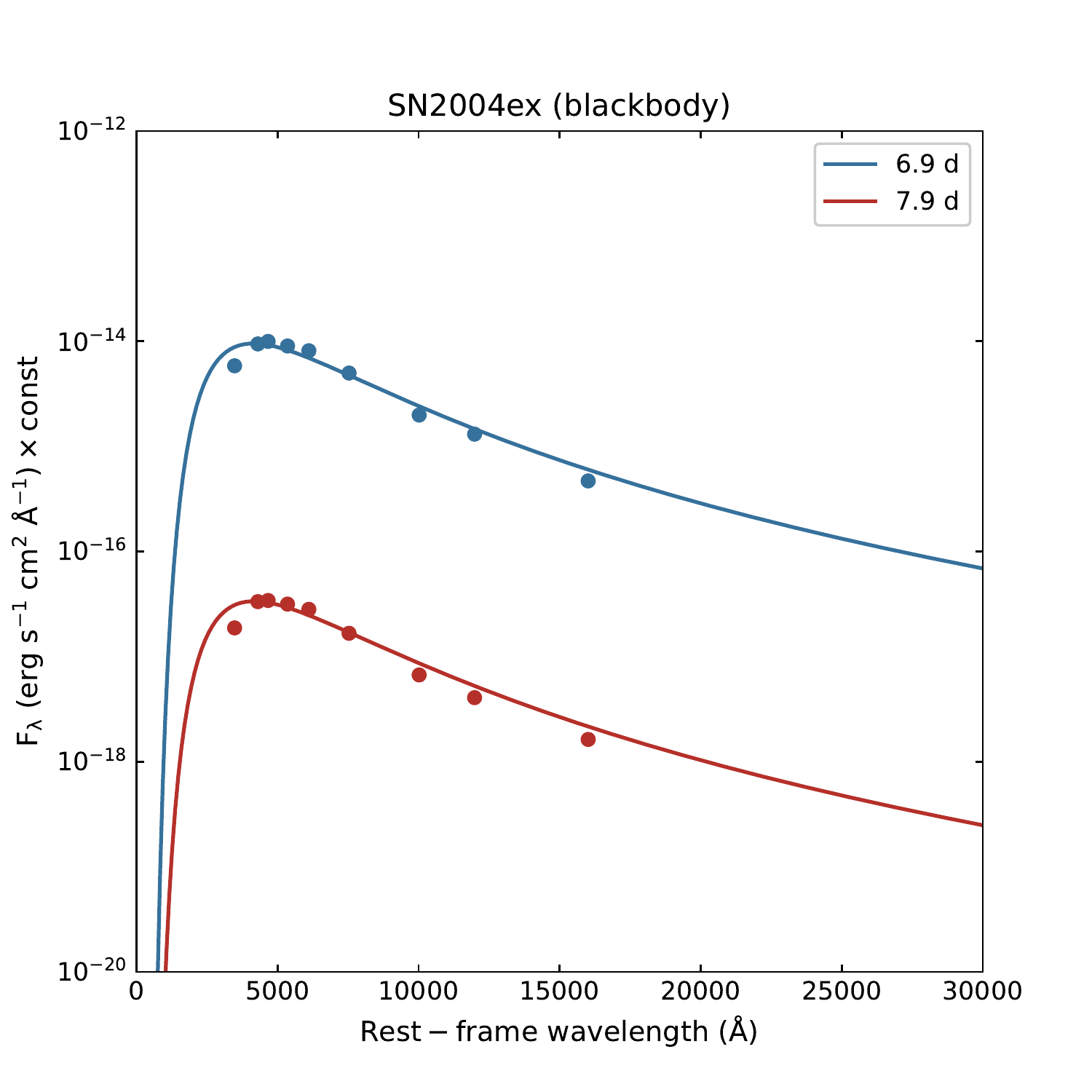}
\includegraphics[width=0.245\textwidth,angle=0]{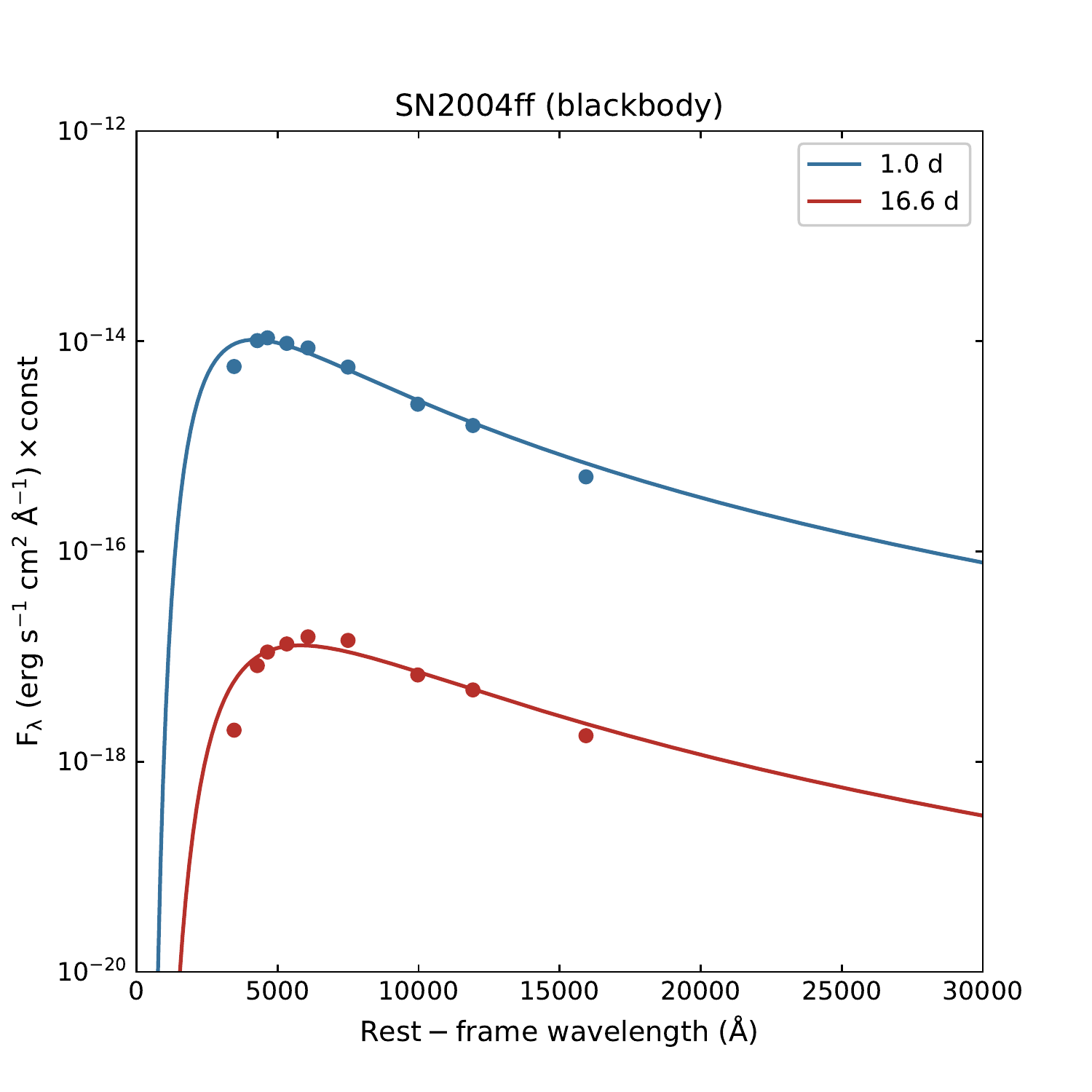}
\includegraphics[width=0.245\textwidth,angle=0]{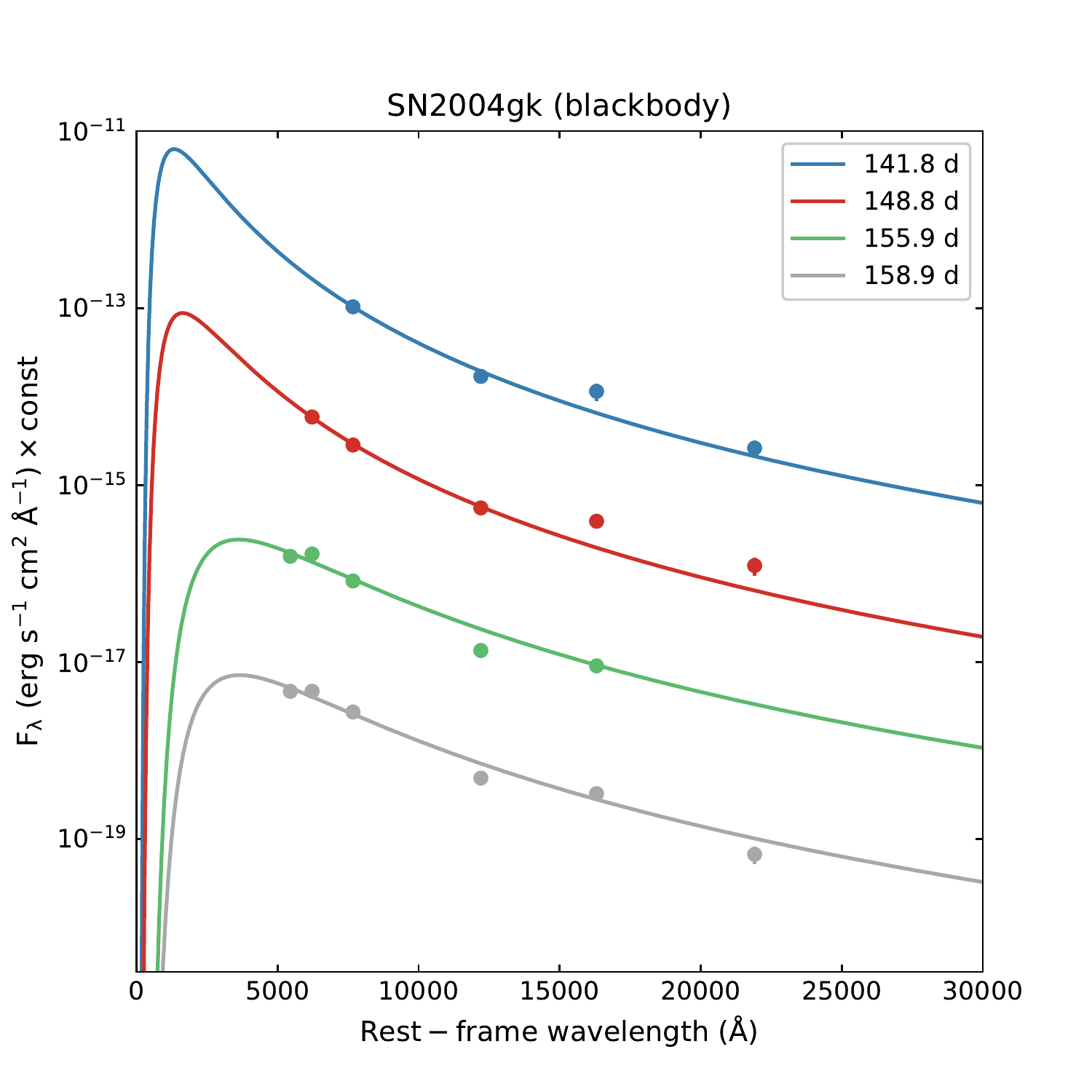}
\includegraphics[width=0.245\textwidth,angle=0]{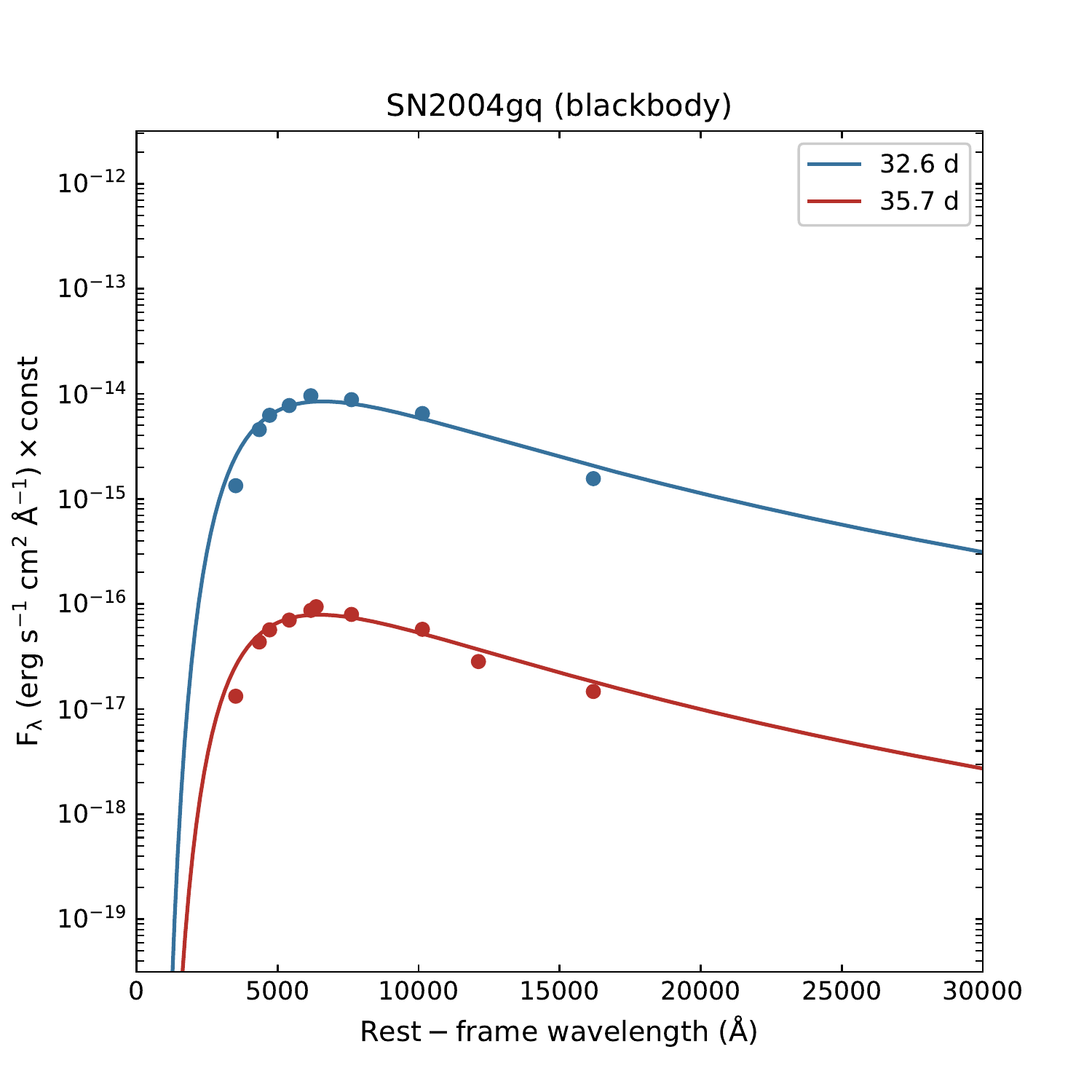}
\includegraphics[width=0.245\textwidth,angle=0]{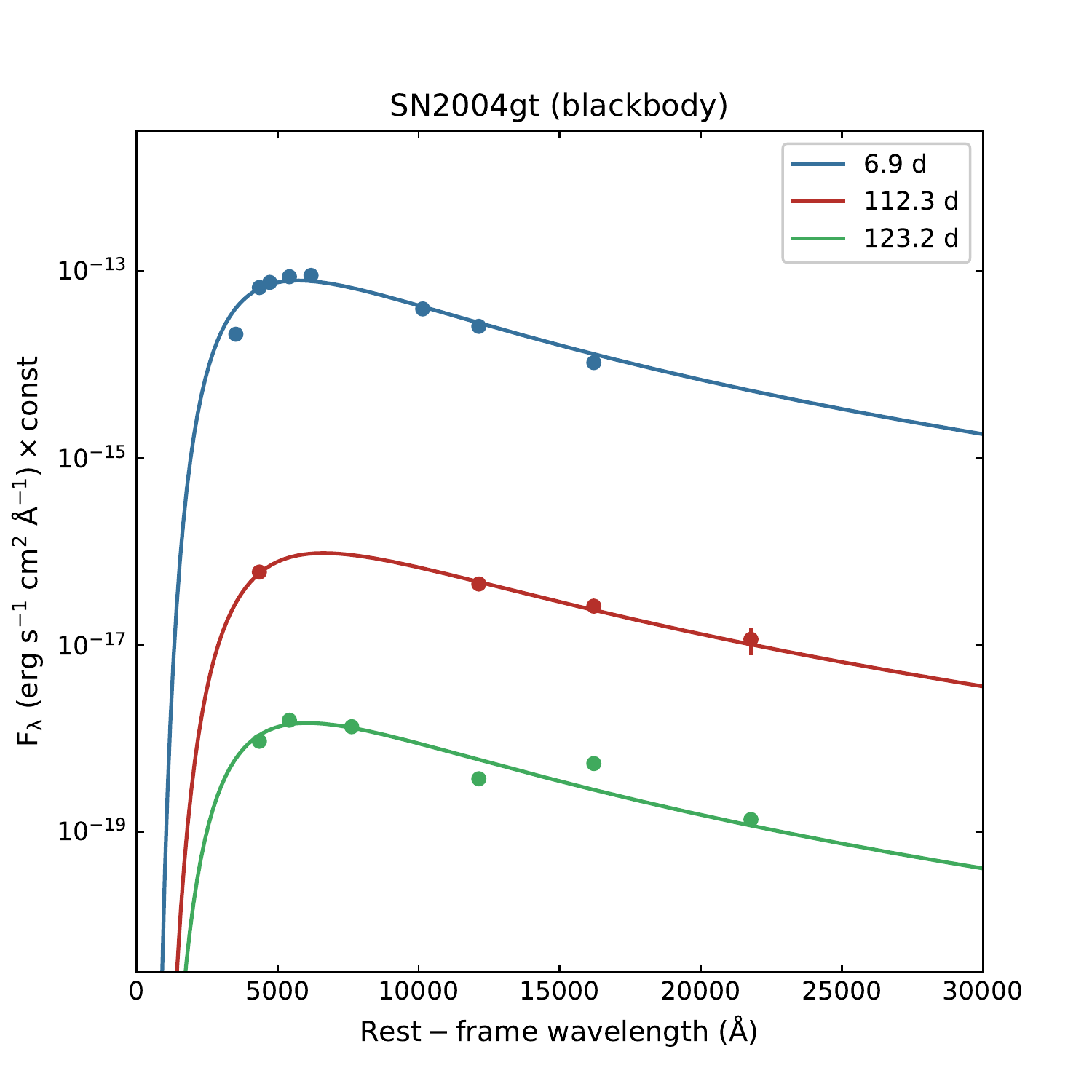}
\includegraphics[width=0.245\textwidth,angle=0]{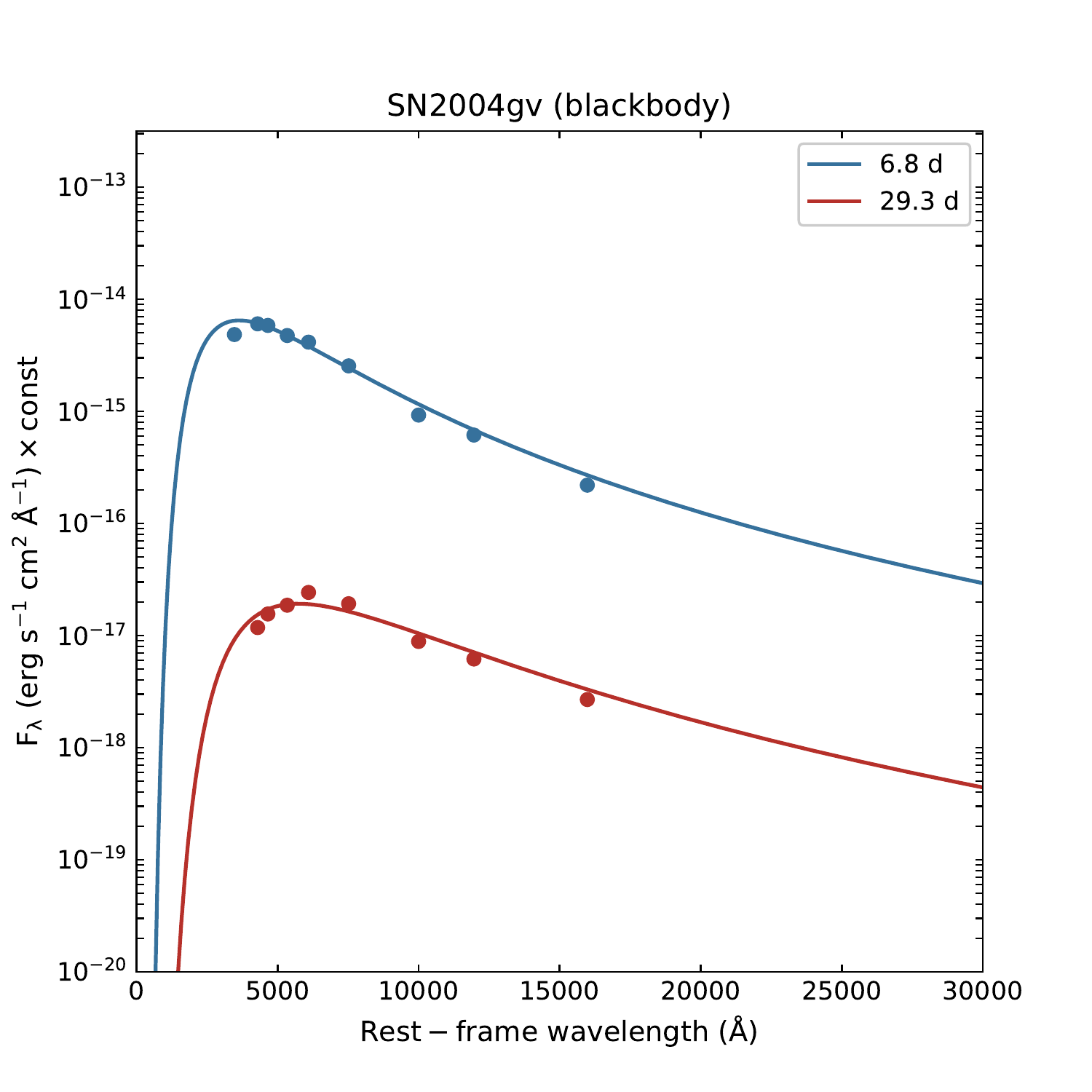}
\includegraphics[width=0.245\textwidth,angle=0]{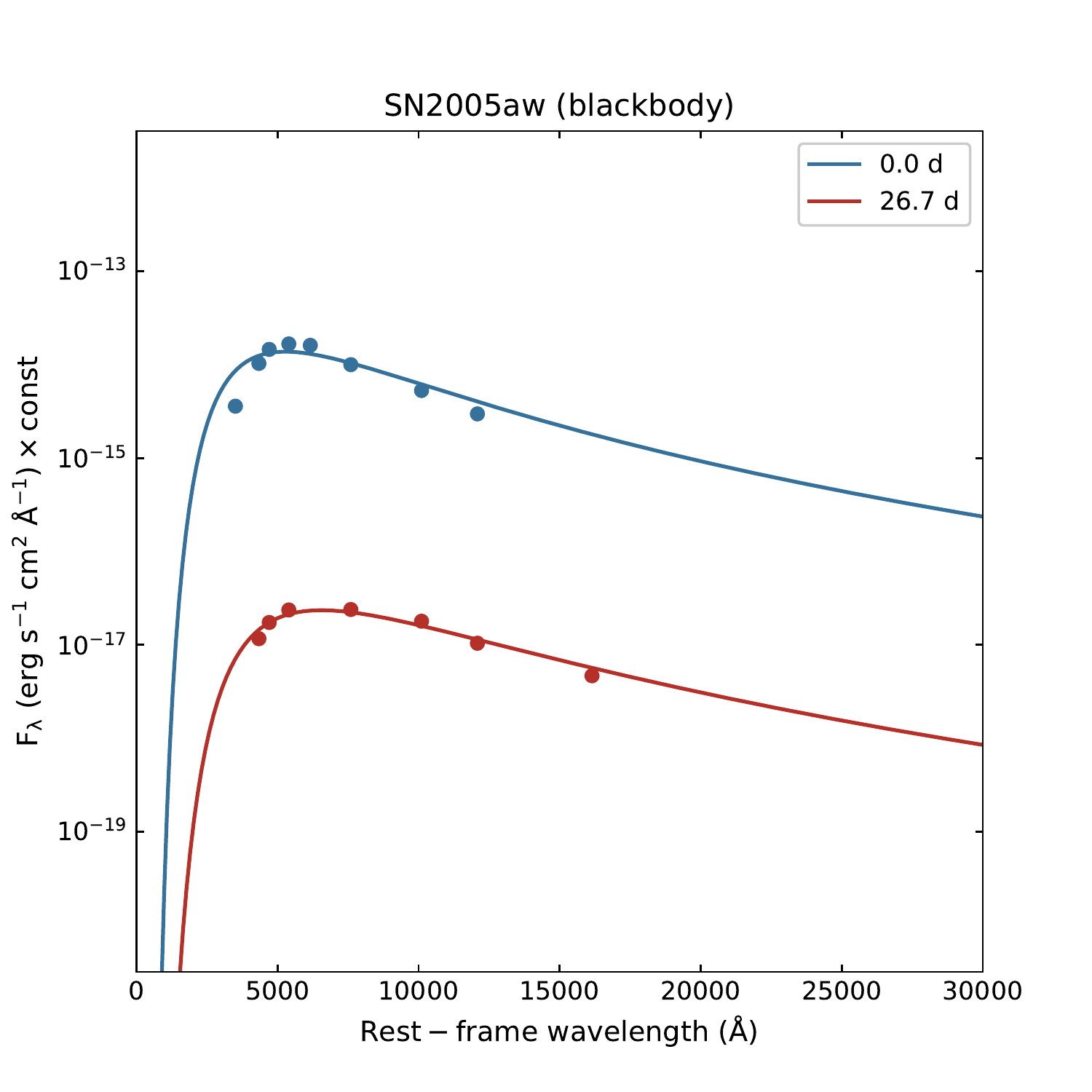}
\includegraphics[width=0.245\textwidth,angle=0]{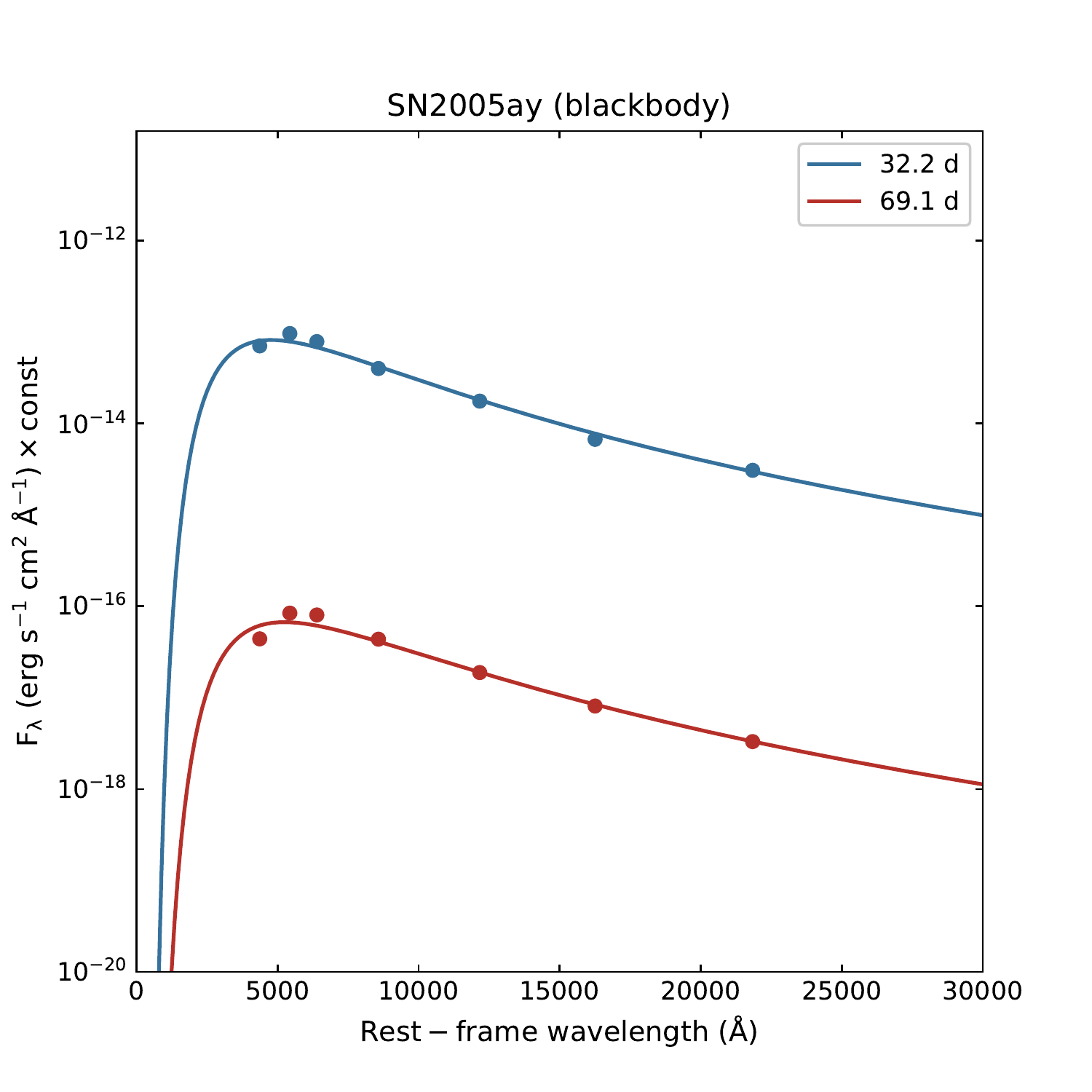}
\includegraphics[width=0.245\textwidth,angle=0]{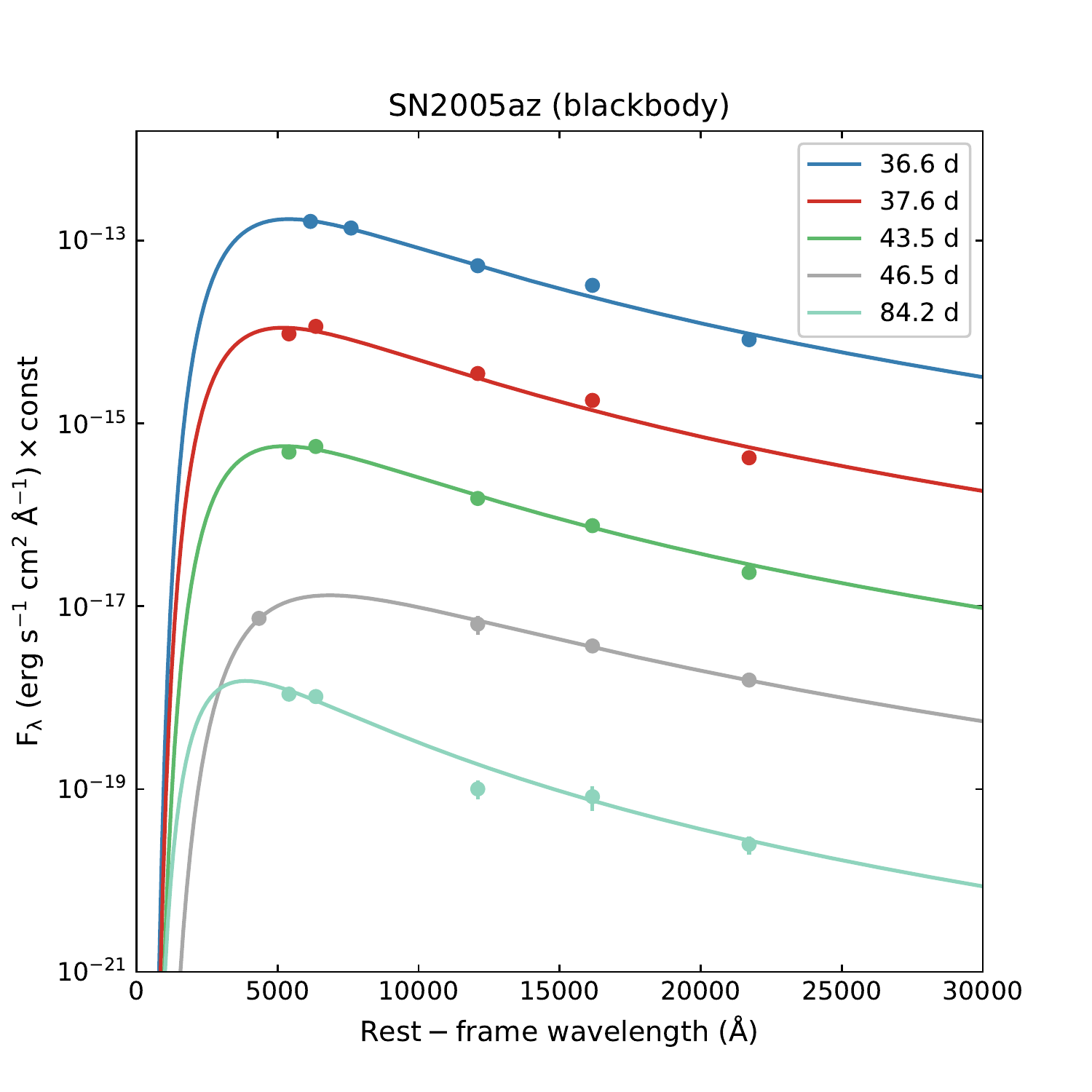}
\includegraphics[width=0.245\textwidth,angle=0]{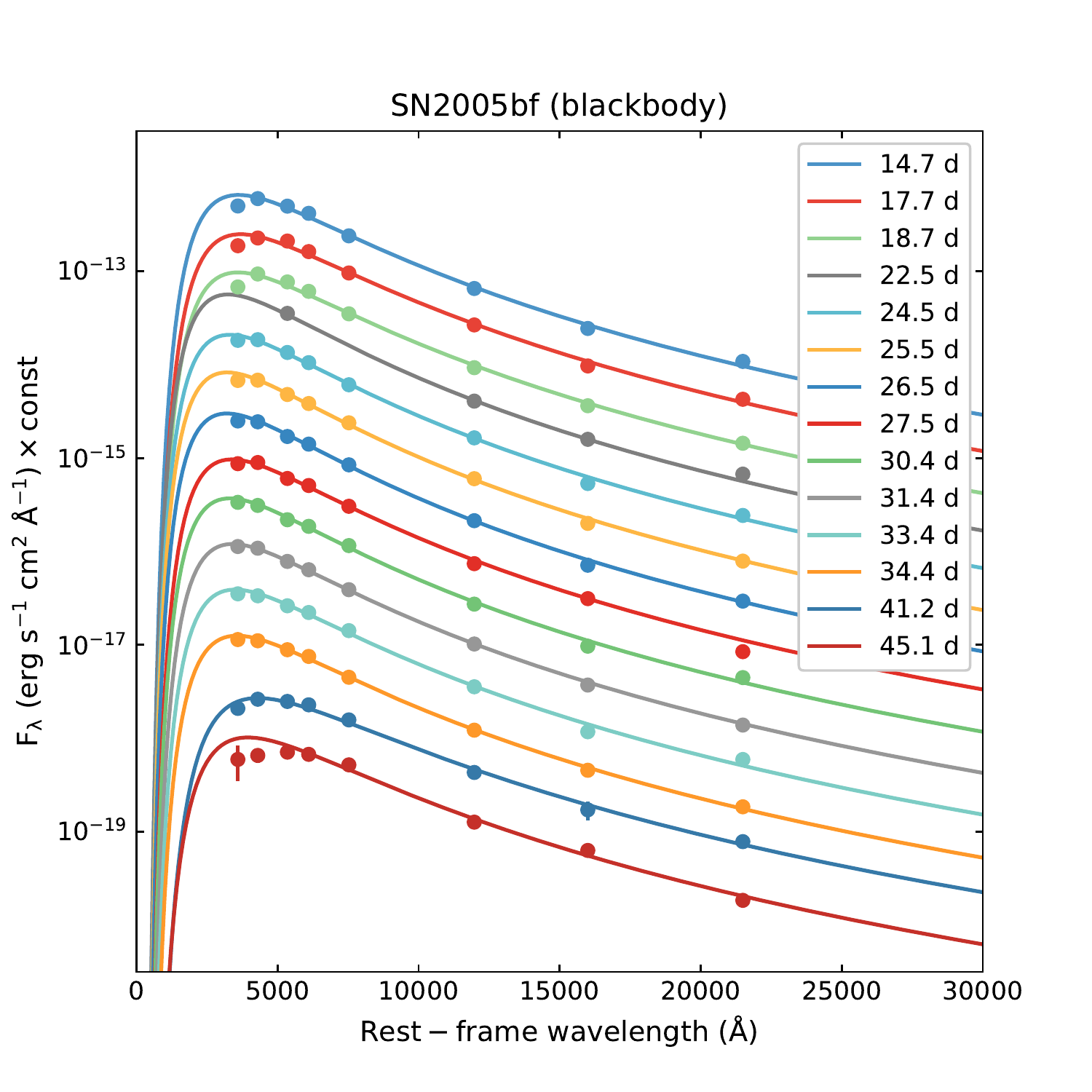}
\includegraphics[width=0.245\textwidth,angle=0]{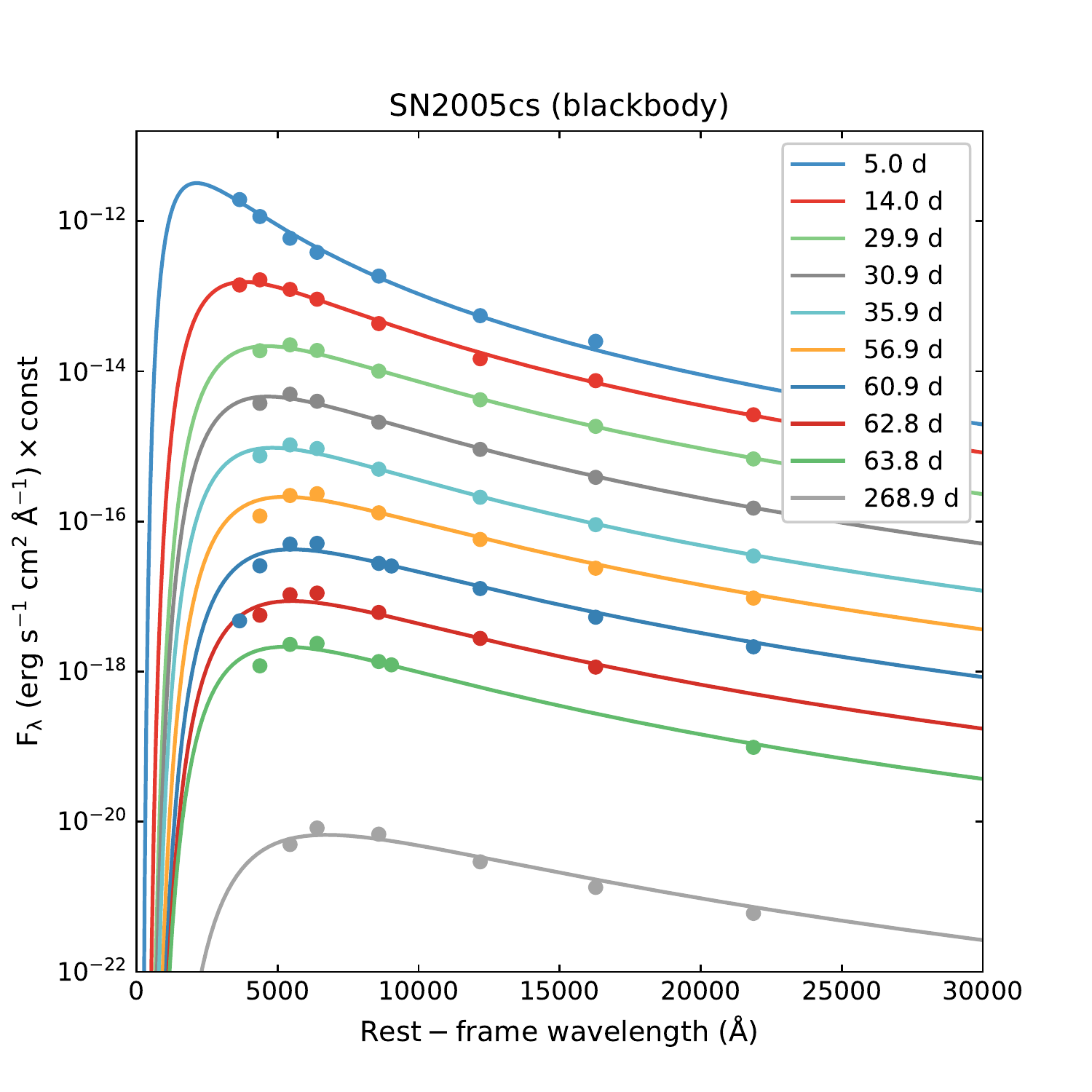}
\includegraphics[width=0.245\textwidth,angle=0]{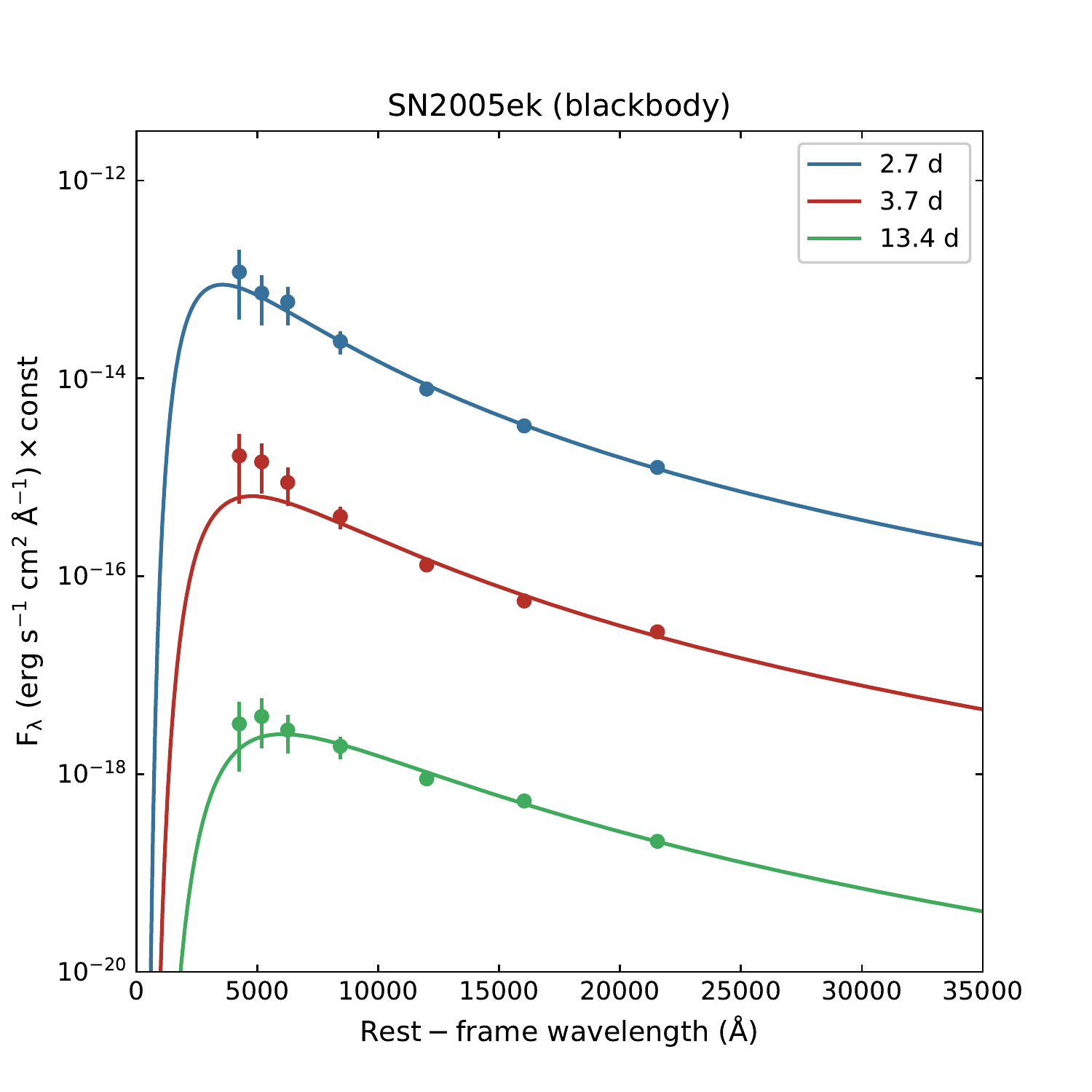}
\includegraphics[width=0.245\textwidth,angle=0]{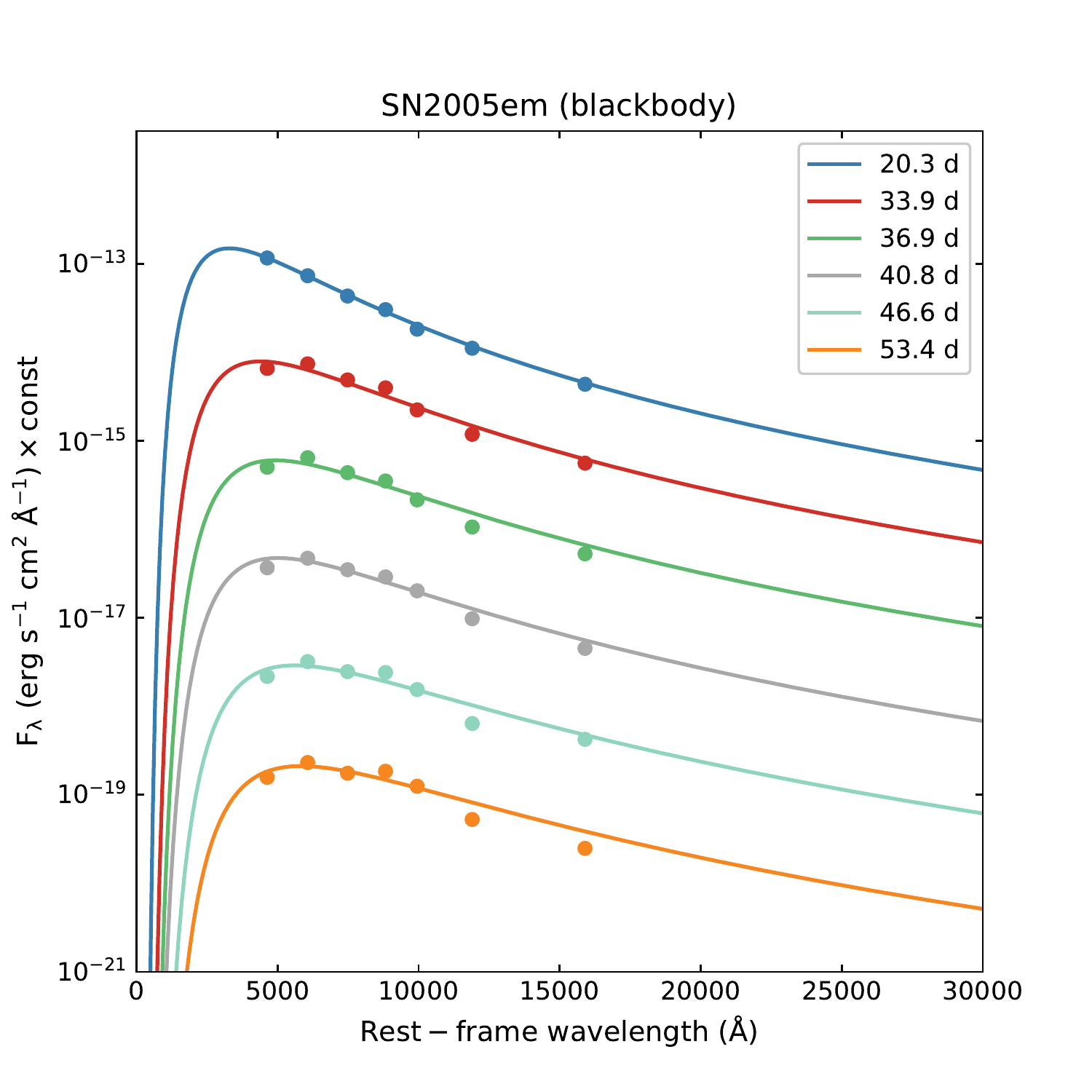}
\includegraphics[width=0.245\textwidth,angle=0]{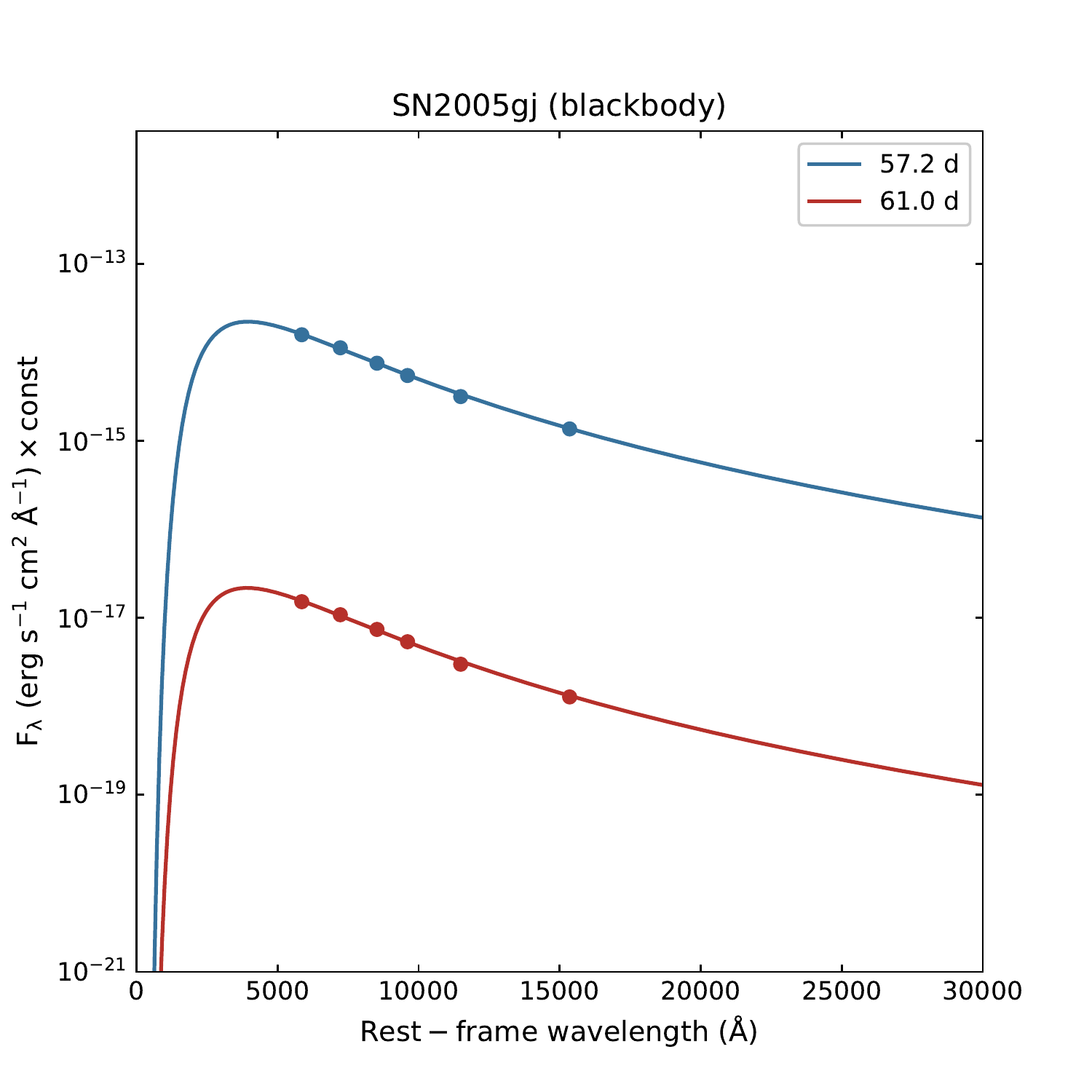}
\includegraphics[width=0.245\textwidth,angle=0]{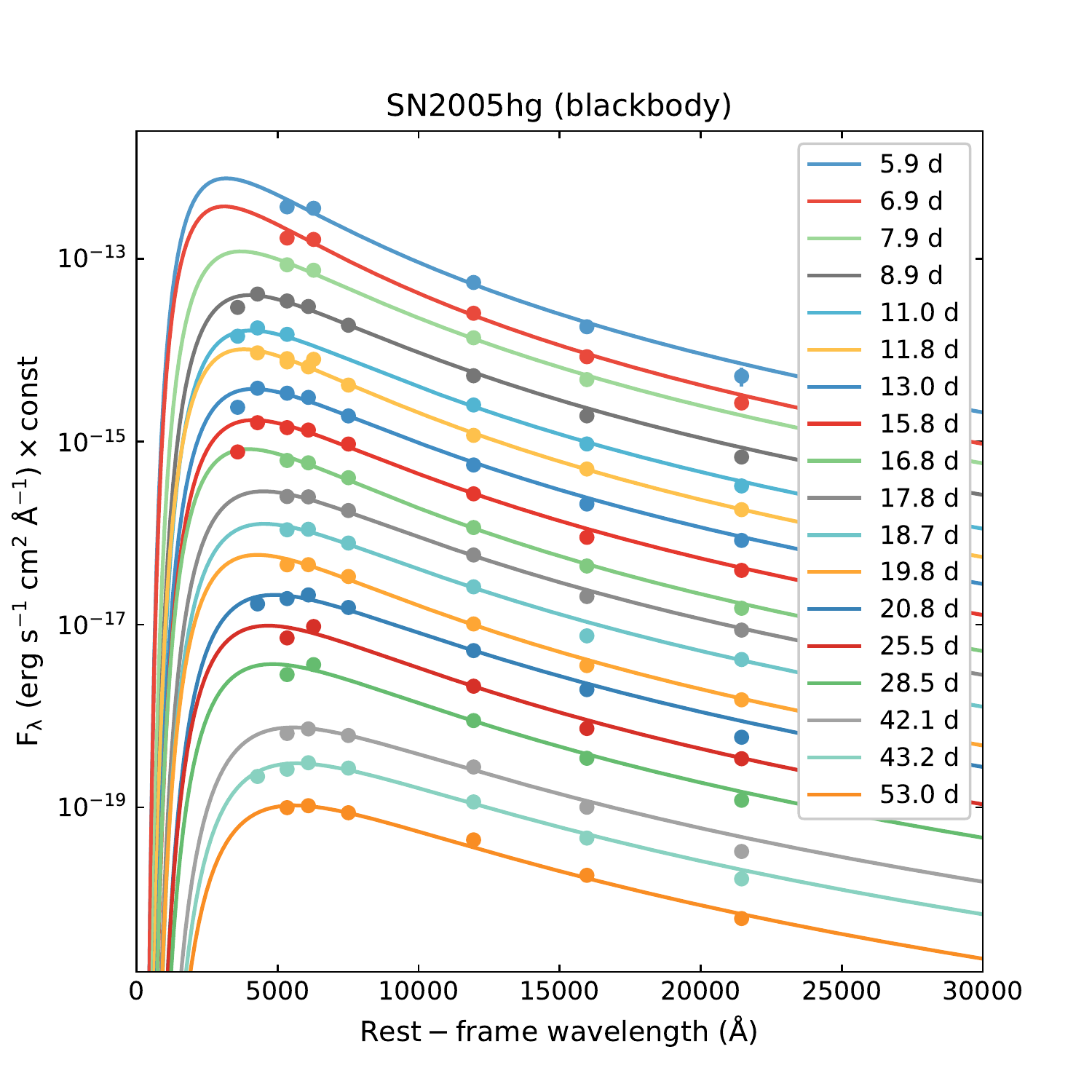}
\includegraphics[width=0.245\textwidth,angle=0]{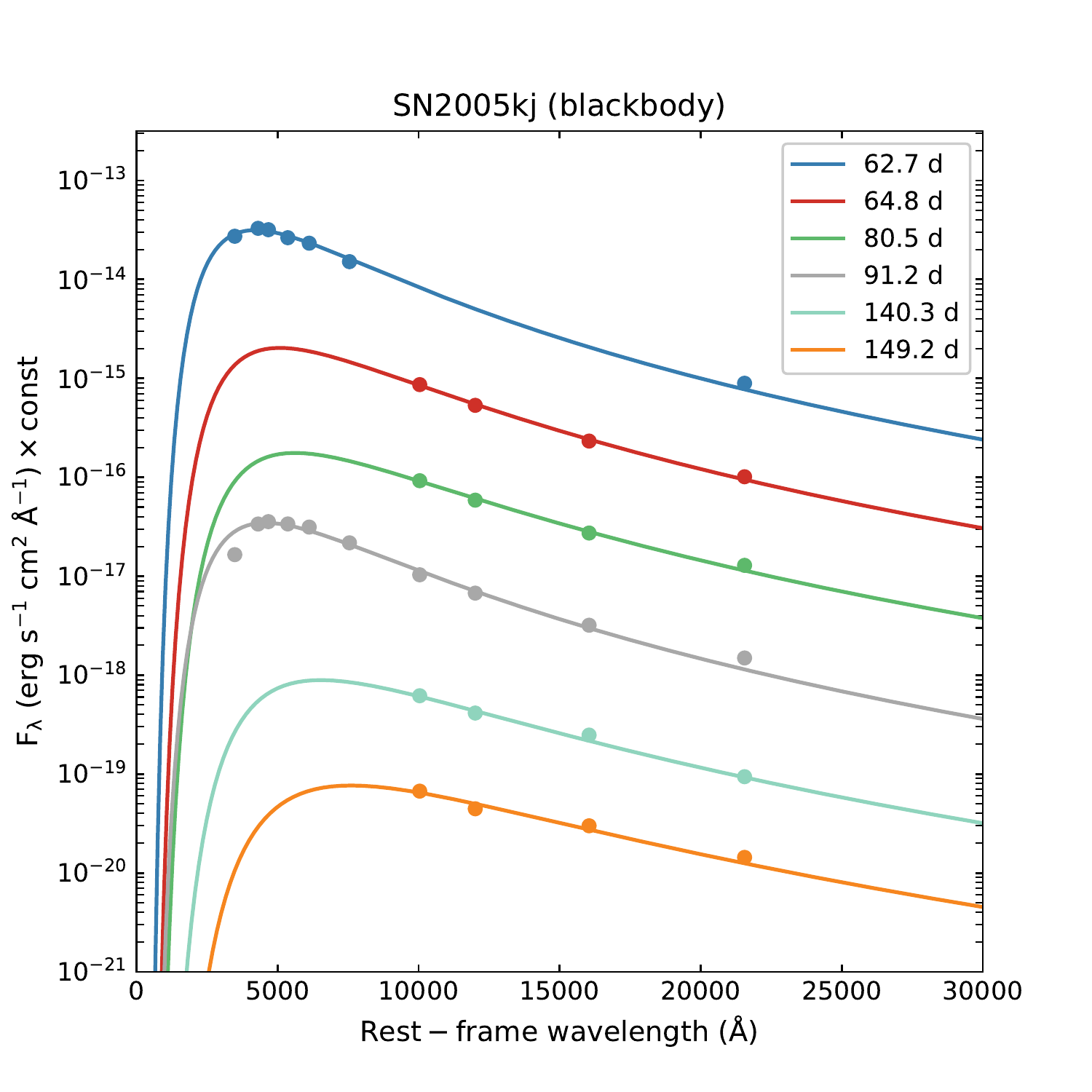}
\includegraphics[width=0.245\textwidth,angle=0]{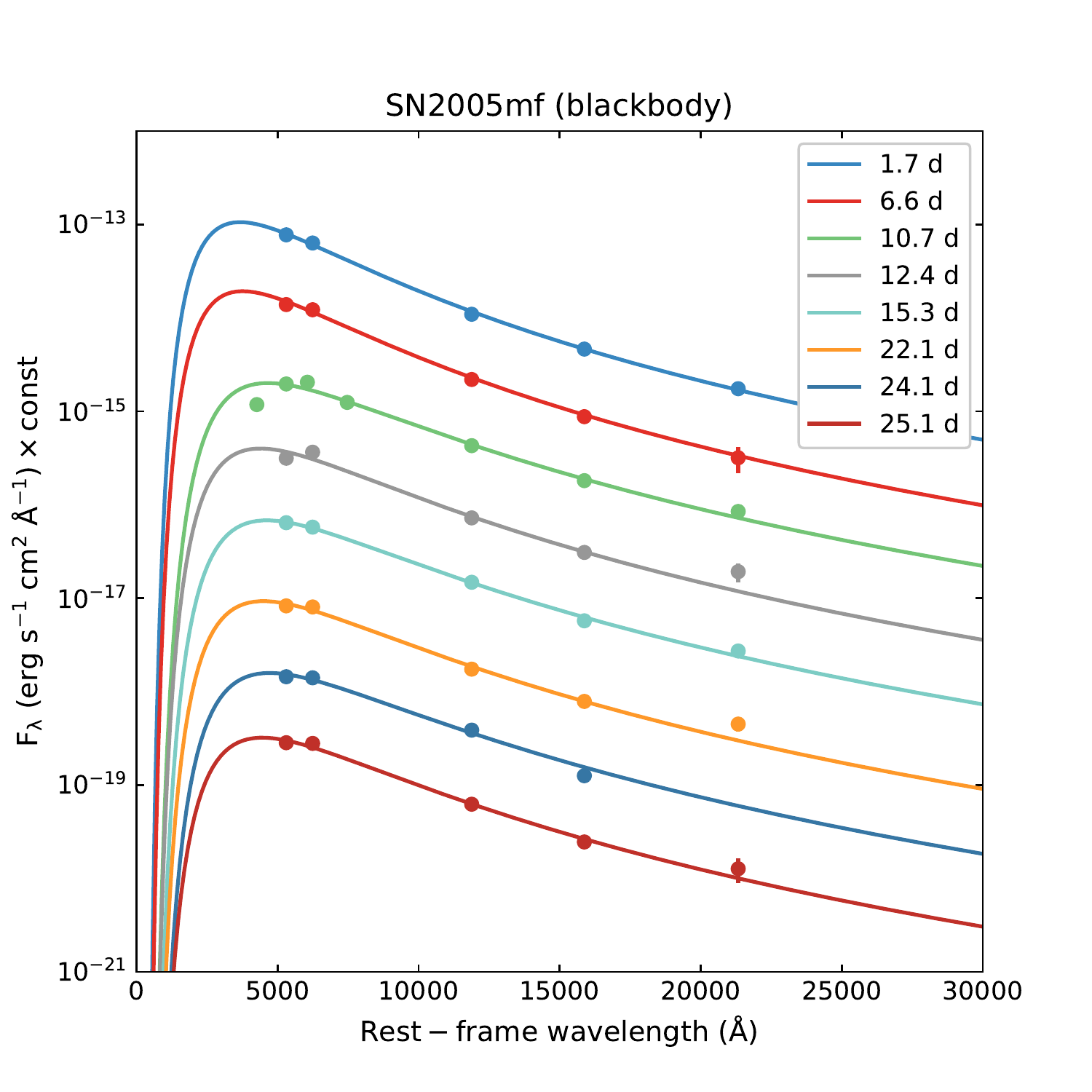}
\includegraphics[width=0.245\textwidth,angle=0]{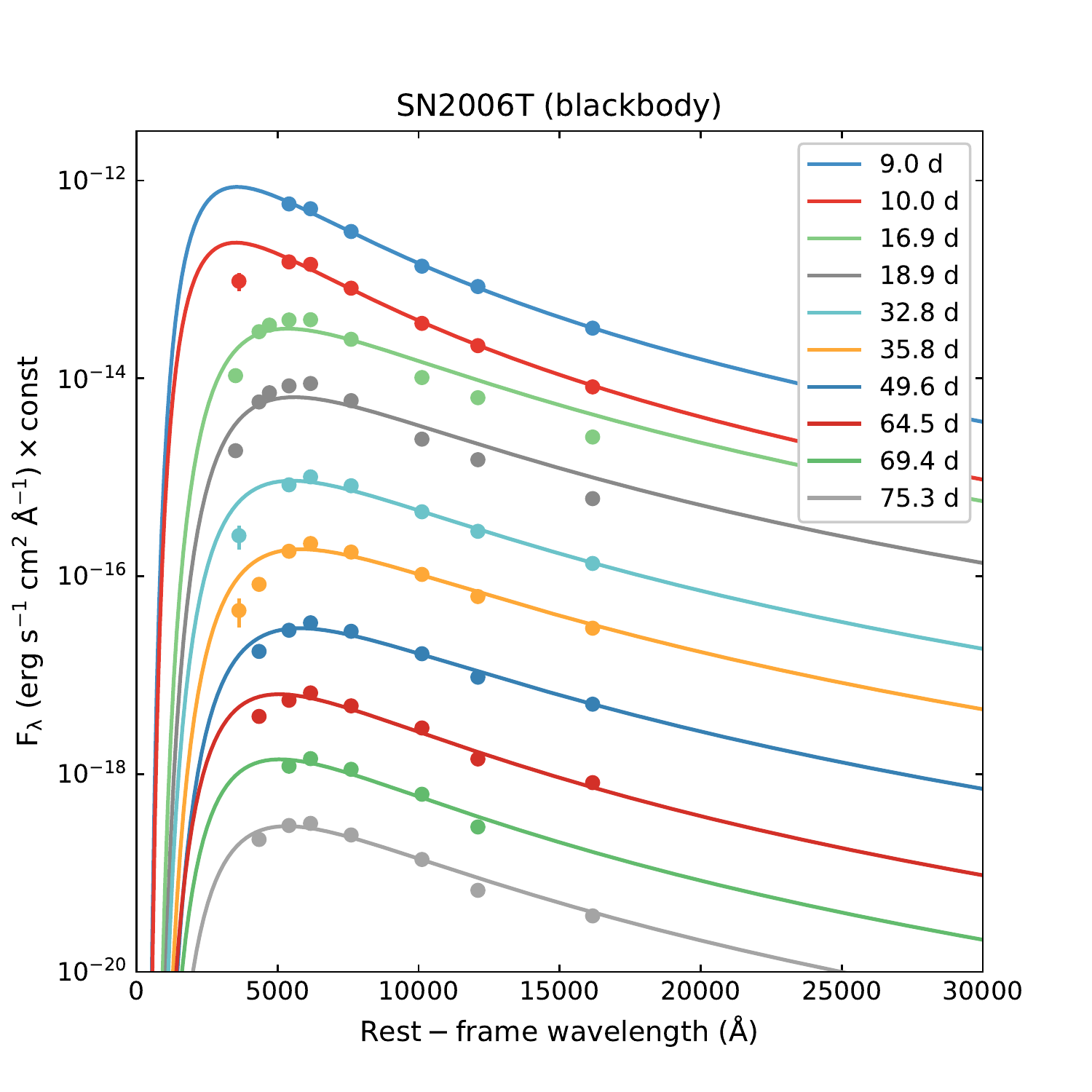}
\end{center}
\caption{The blackbody fits for the optical and NIR SEDs of SNe in our sample
(don't show evident IR excesses).
The data are from the references listed in Table \ref{table:details}. For clarity,
the flux at all epochs are shifted
by adding different constants.}
\label{fig:SED-BB-1}
\end{figure}

\clearpage

\begin{figure}[tbph]
\begin{center}
\ContinuedFloat
\includegraphics[width=0.245\textwidth,angle=0]{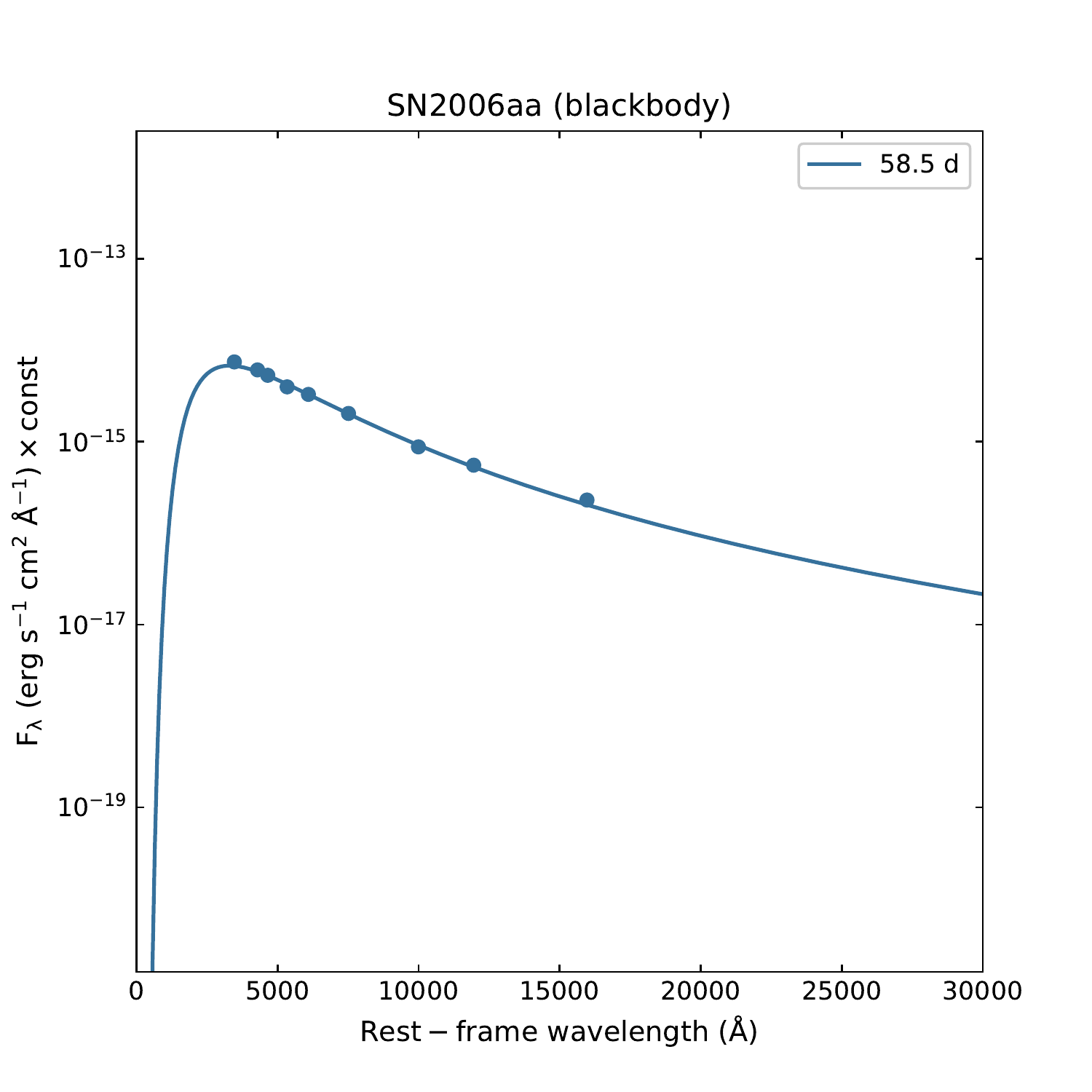}
\includegraphics[width=0.245\textwidth,angle=0]{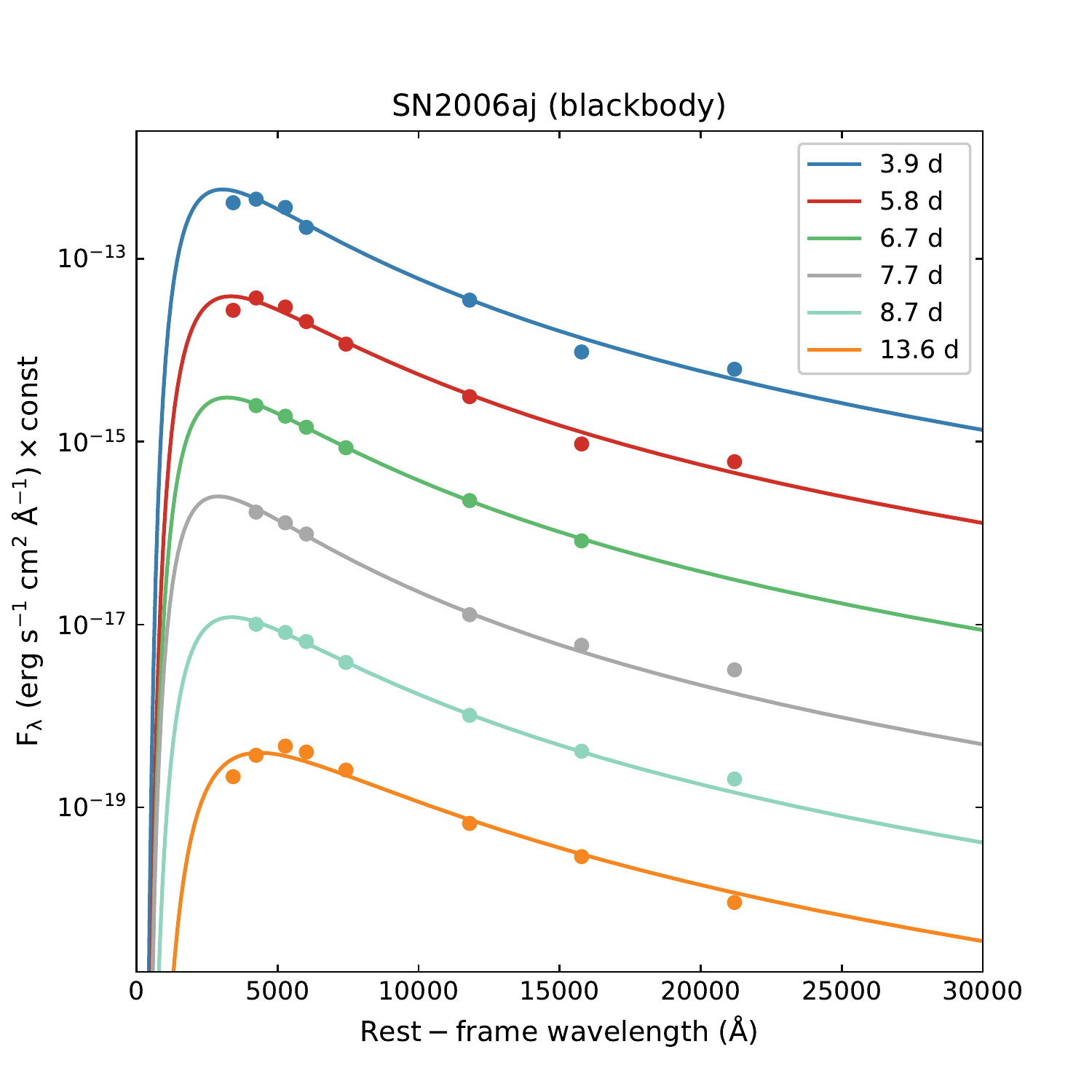}
\includegraphics[width=0.245\textwidth,angle=0]{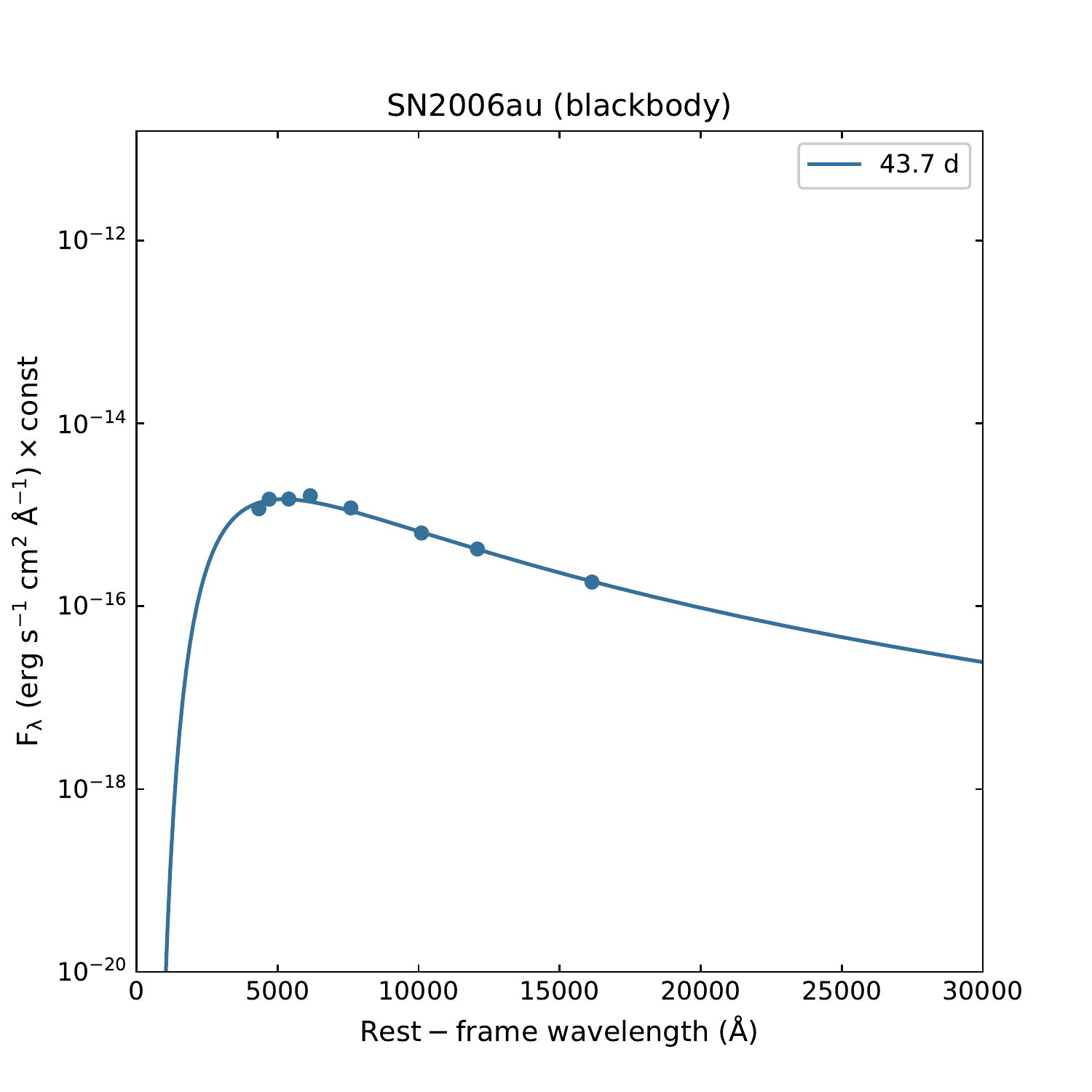}
\includegraphics[width=0.245\textwidth,angle=0]{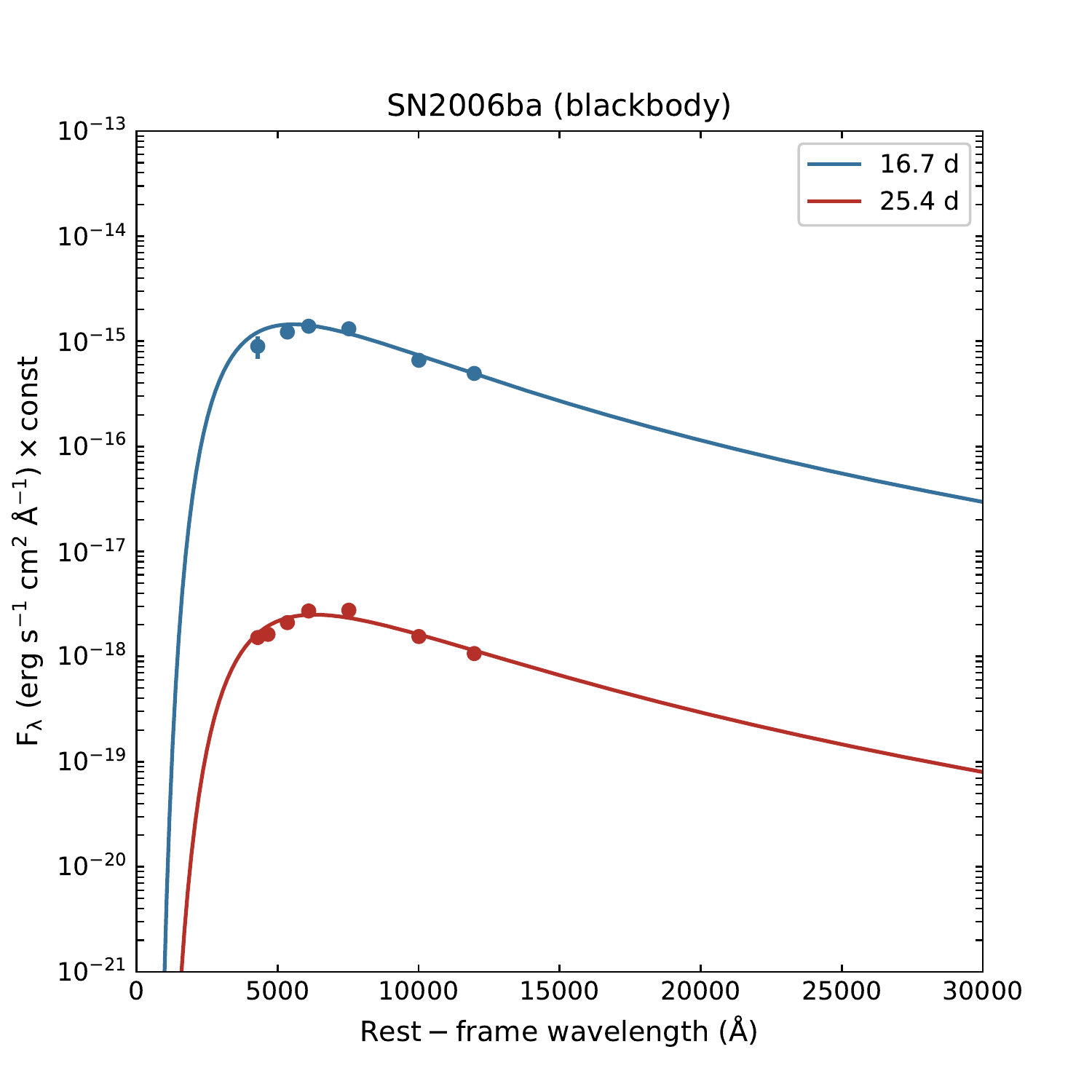}
\includegraphics[width=0.245\textwidth,angle=0]{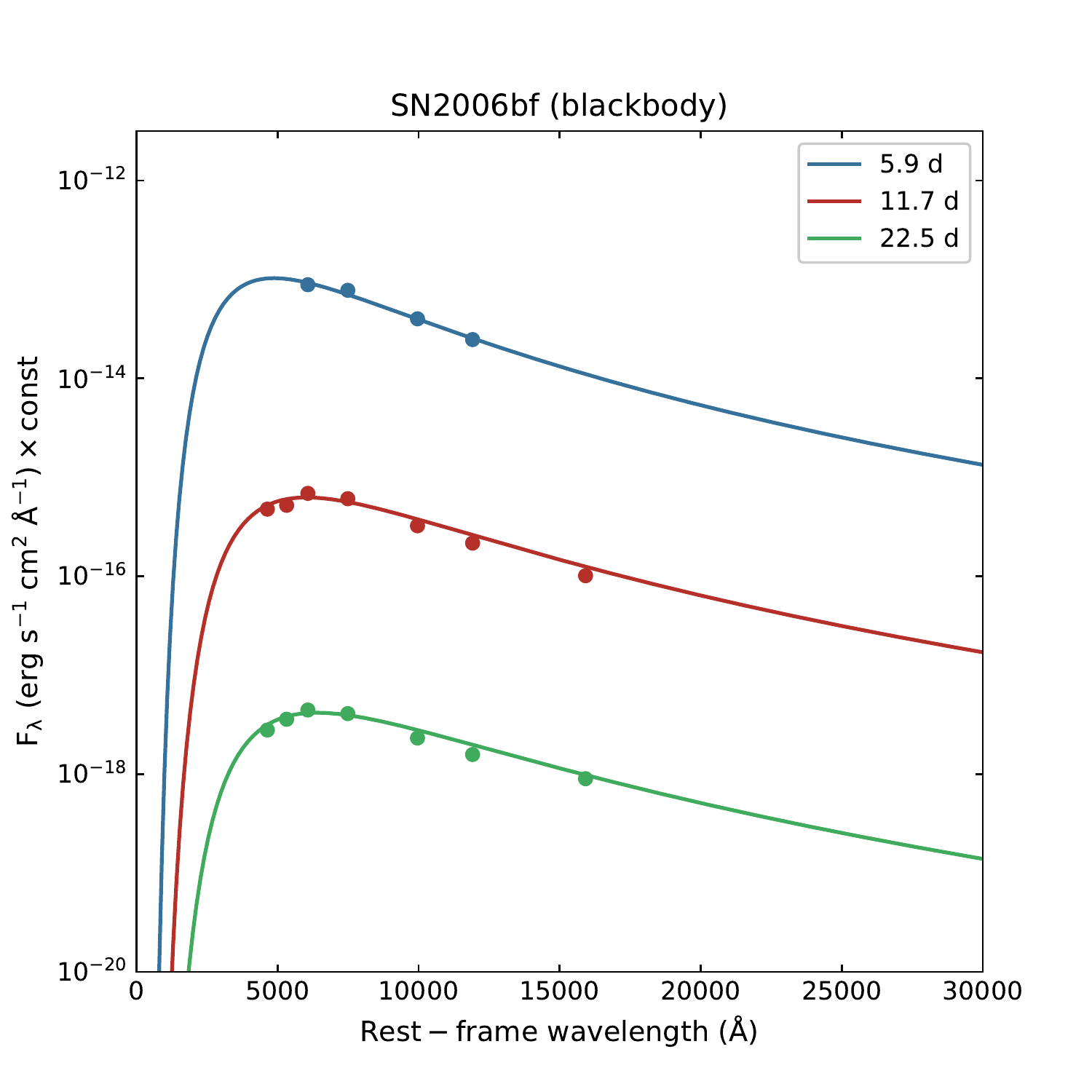}
\includegraphics[width=0.245\textwidth,angle=0]{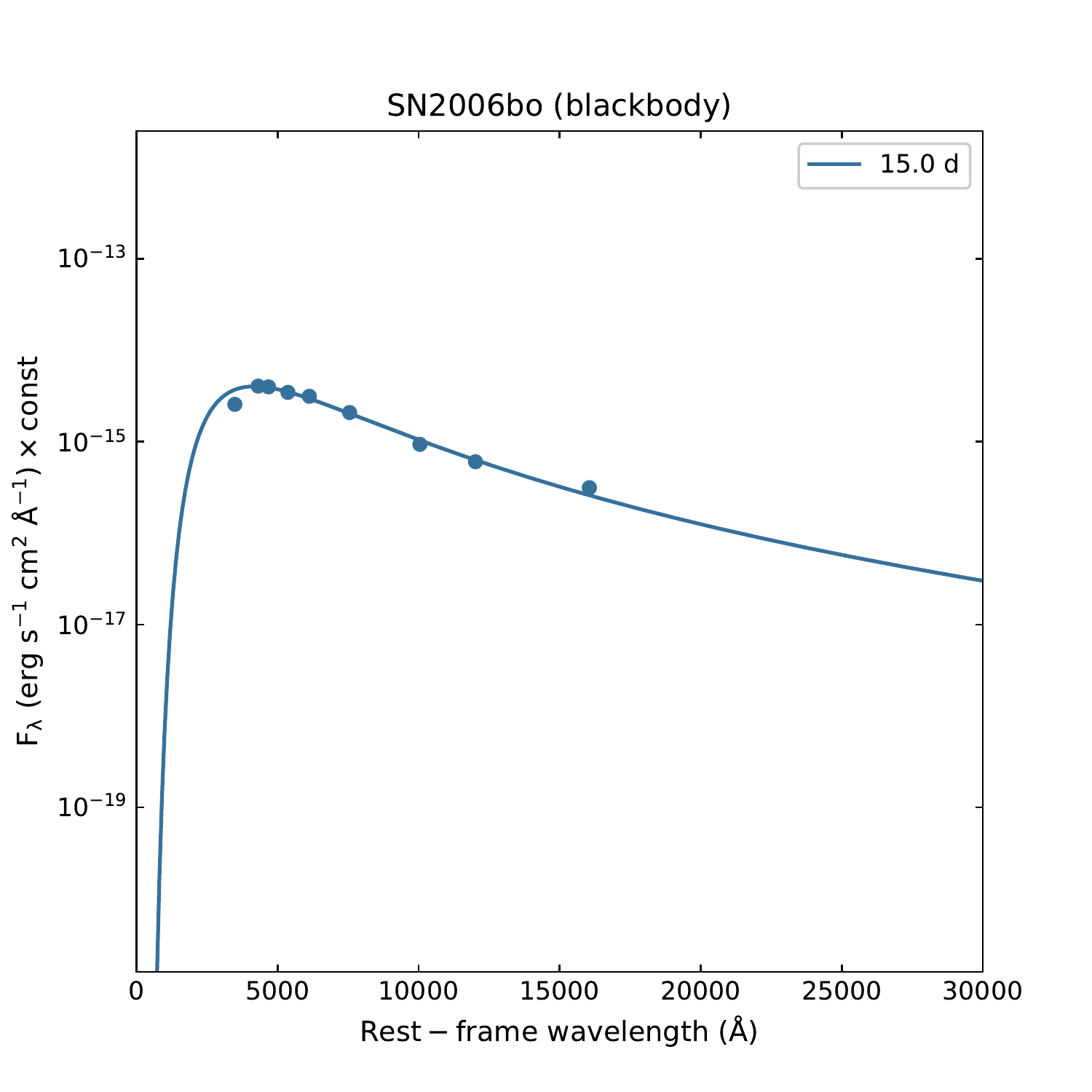}
\includegraphics[width=0.245\textwidth,angle=0]{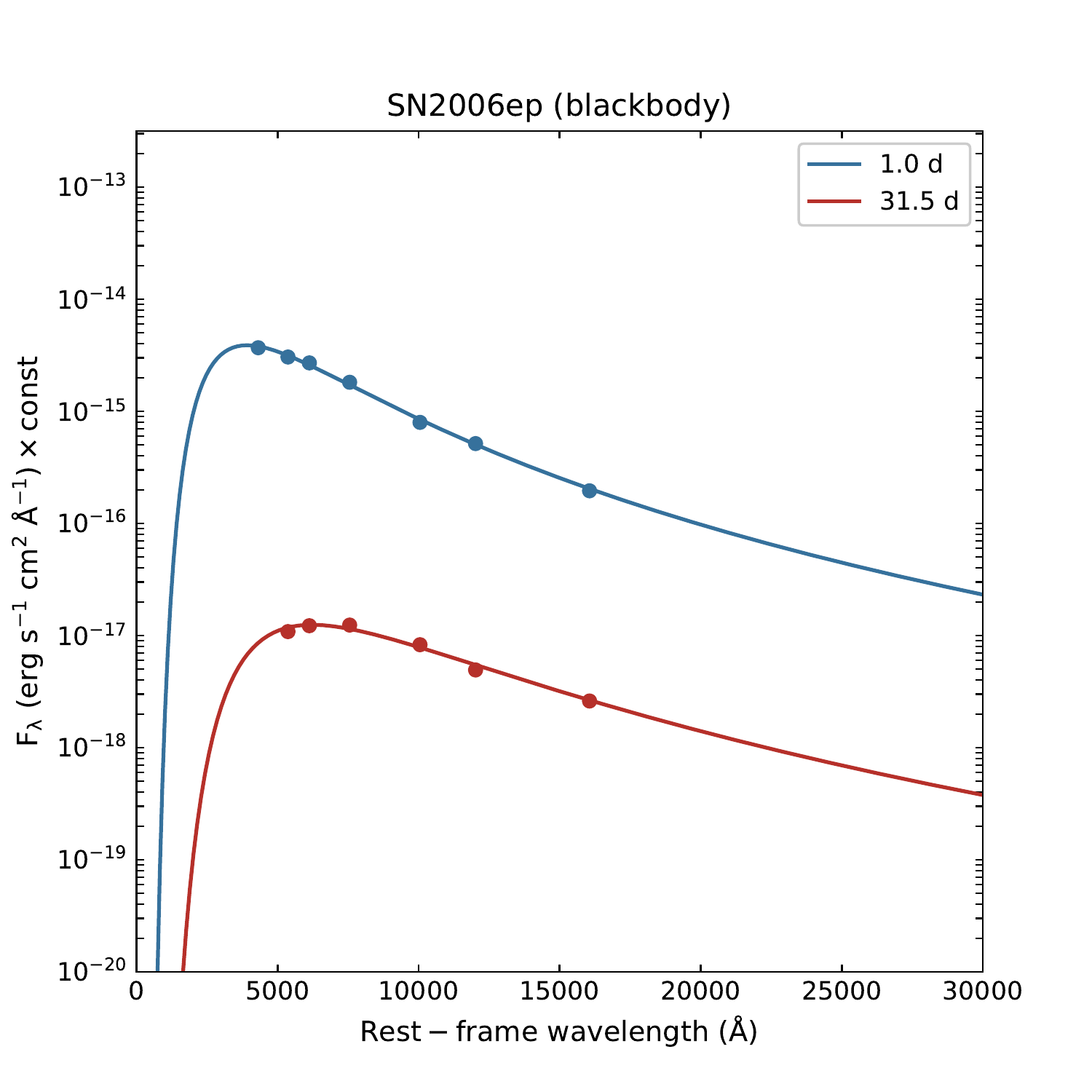}
\includegraphics[width=0.245\textwidth,angle=0]{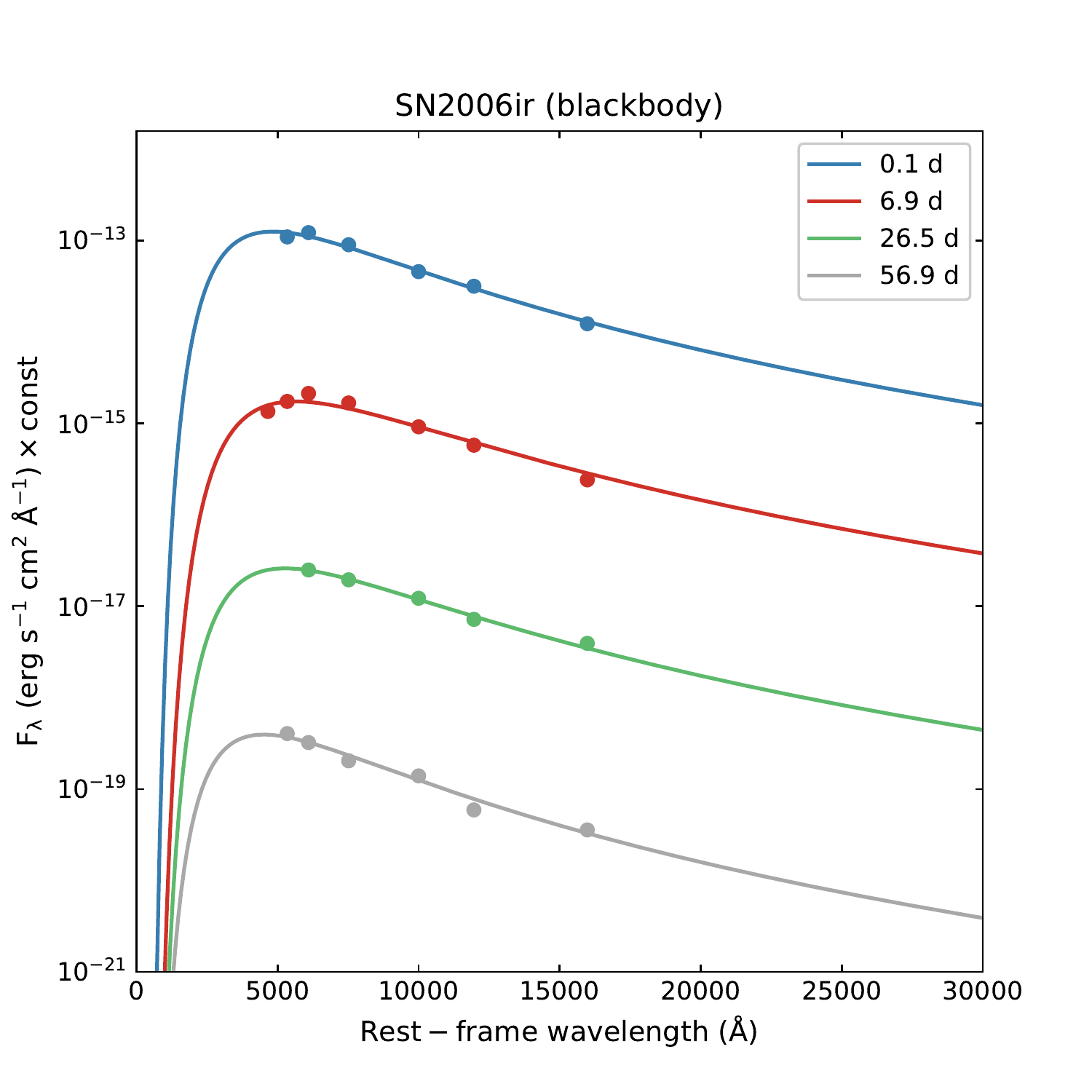}
\includegraphics[width=0.245\textwidth,angle=0]{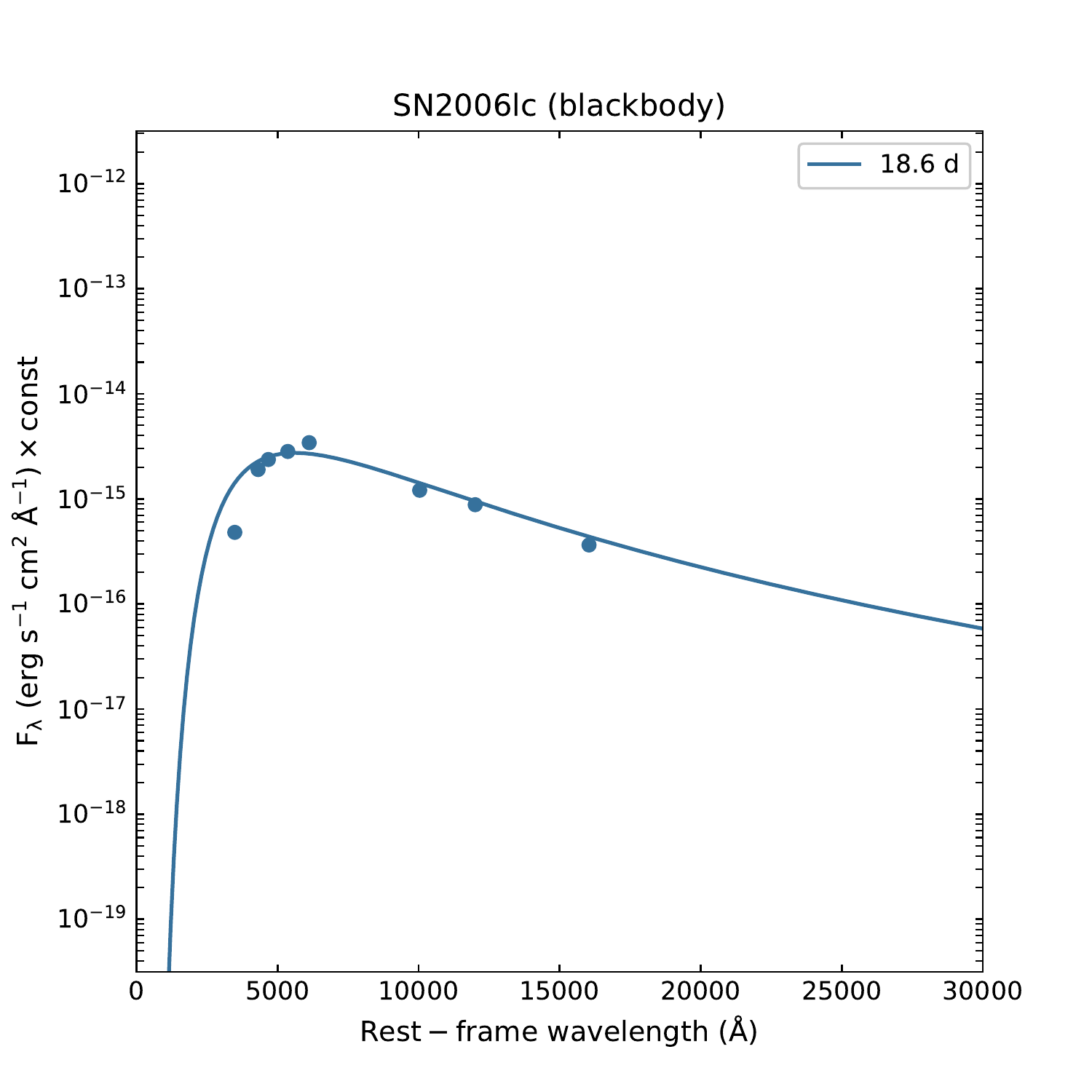}
\includegraphics[width=0.245\textwidth,angle=0]{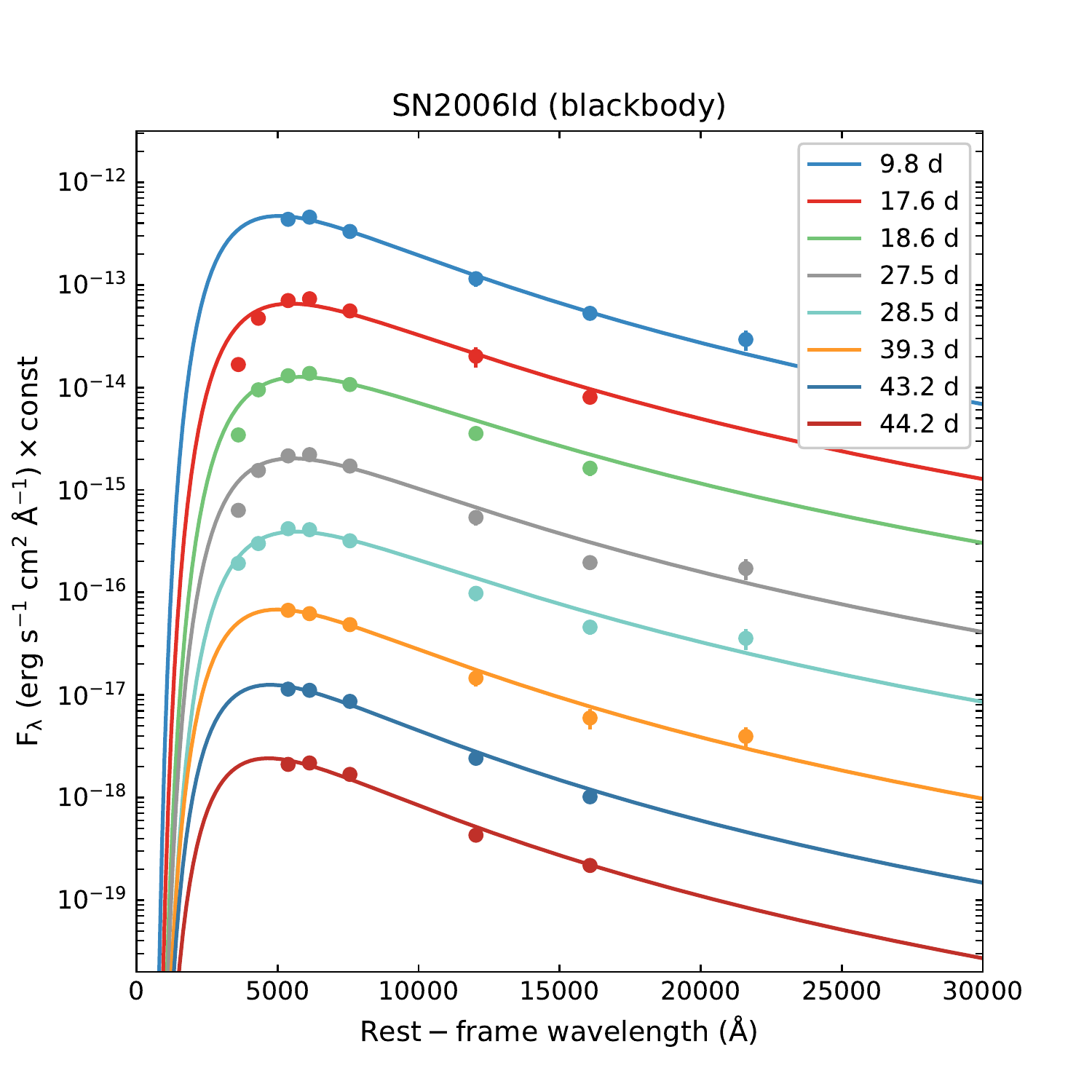}
\includegraphics[width=0.245\textwidth,angle=0]{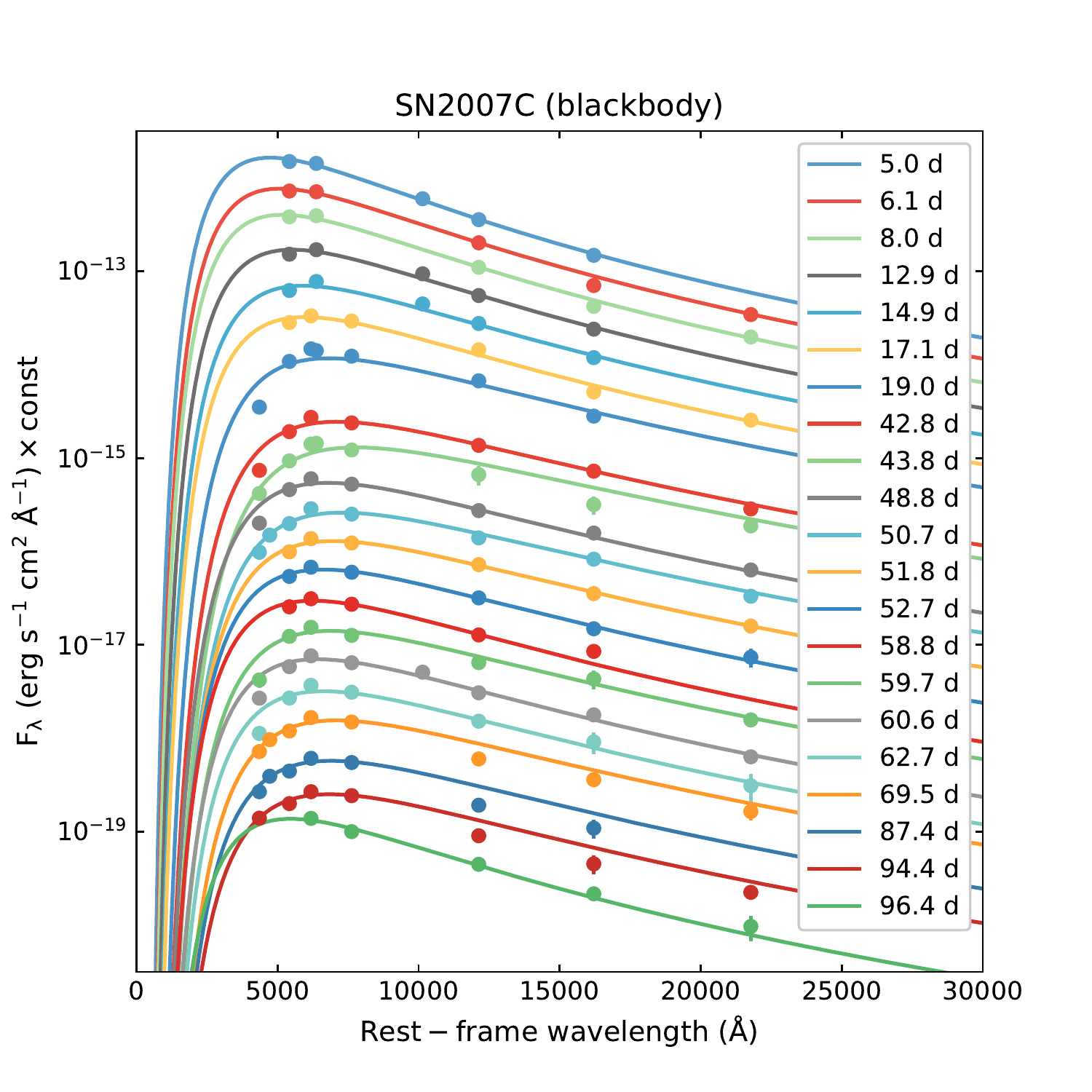}
\includegraphics[width=0.245\textwidth,angle=0]{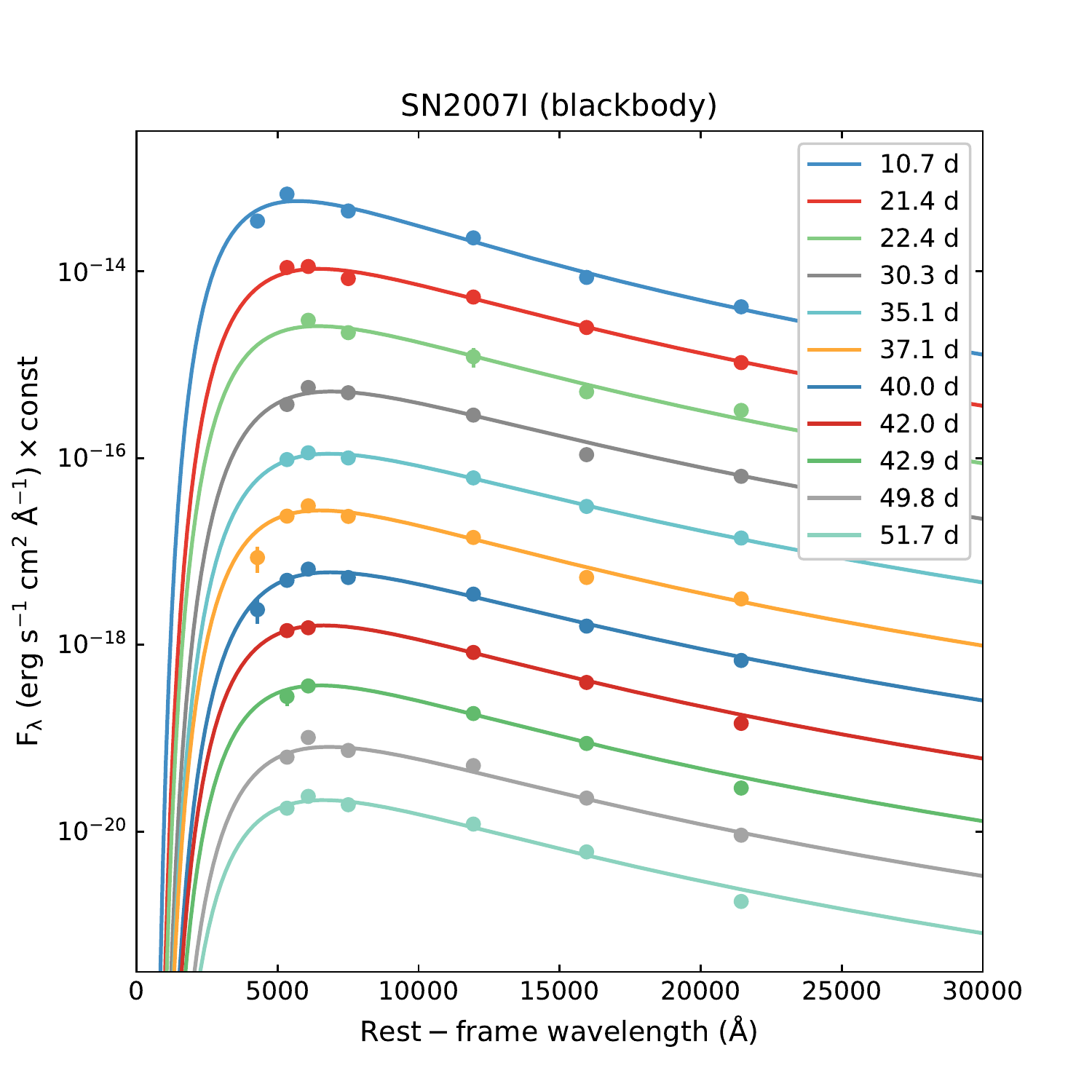}
\includegraphics[width=0.245\textwidth,angle=0]{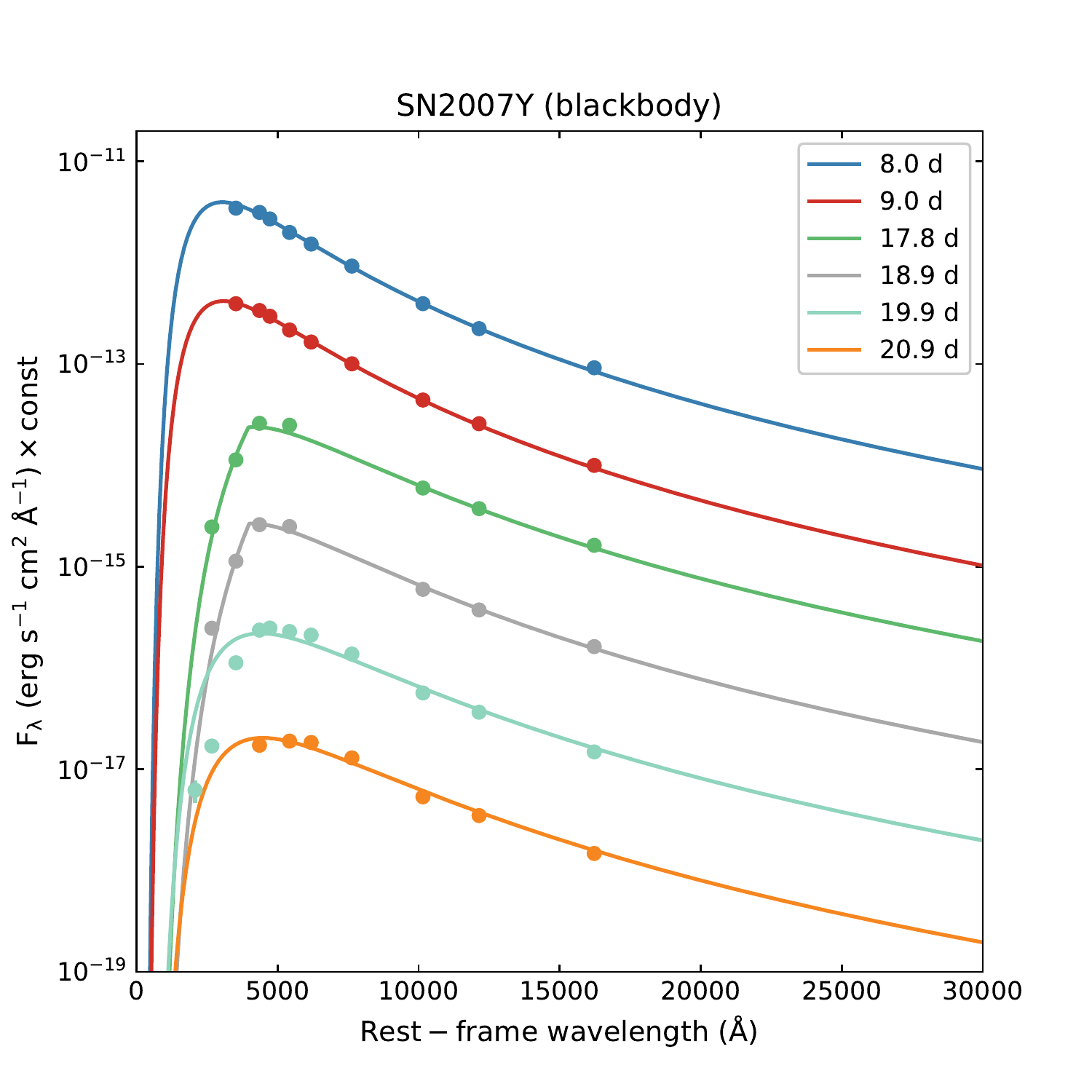}
\includegraphics[width=0.245\textwidth,angle=0]{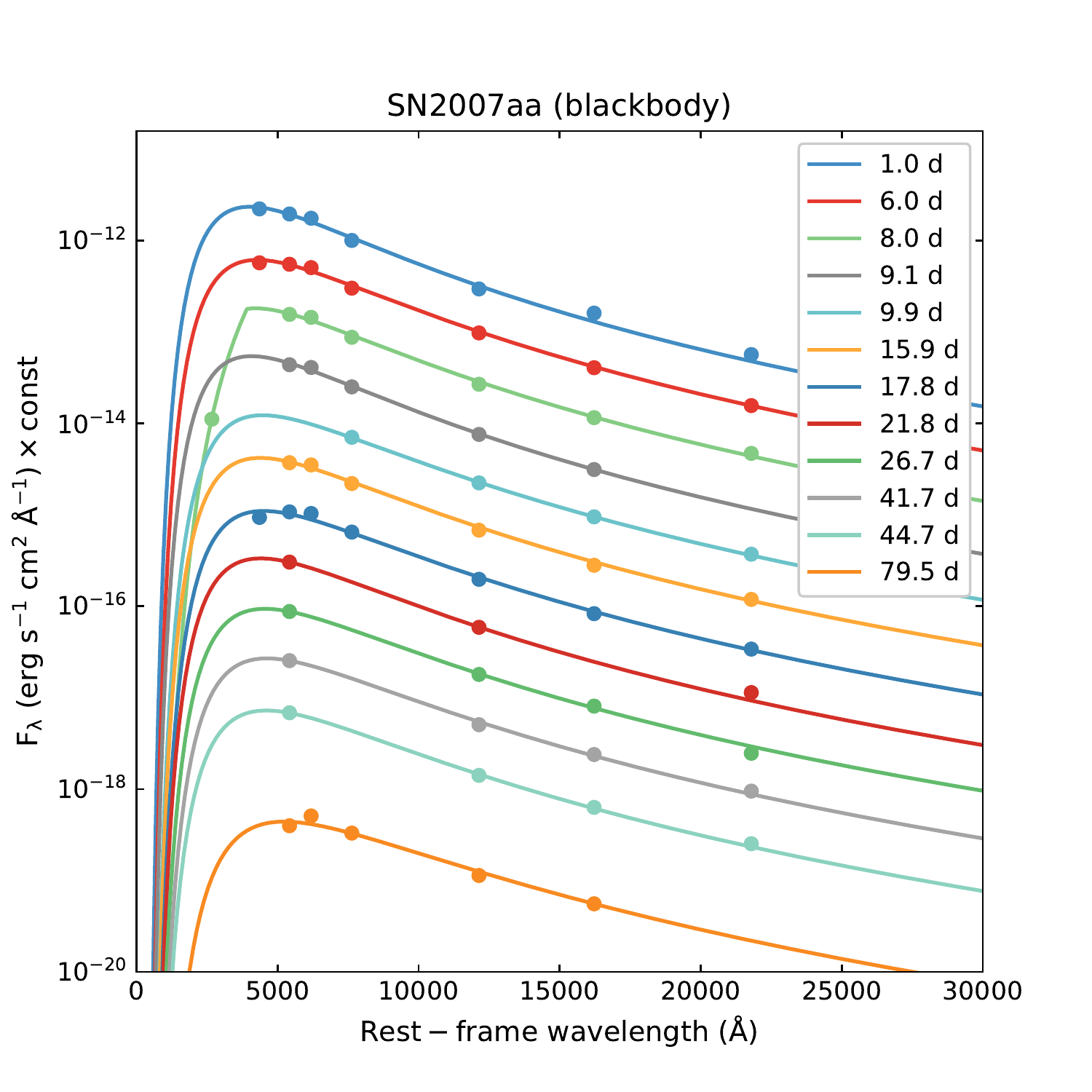}
\includegraphics[width=0.245\textwidth,angle=0]{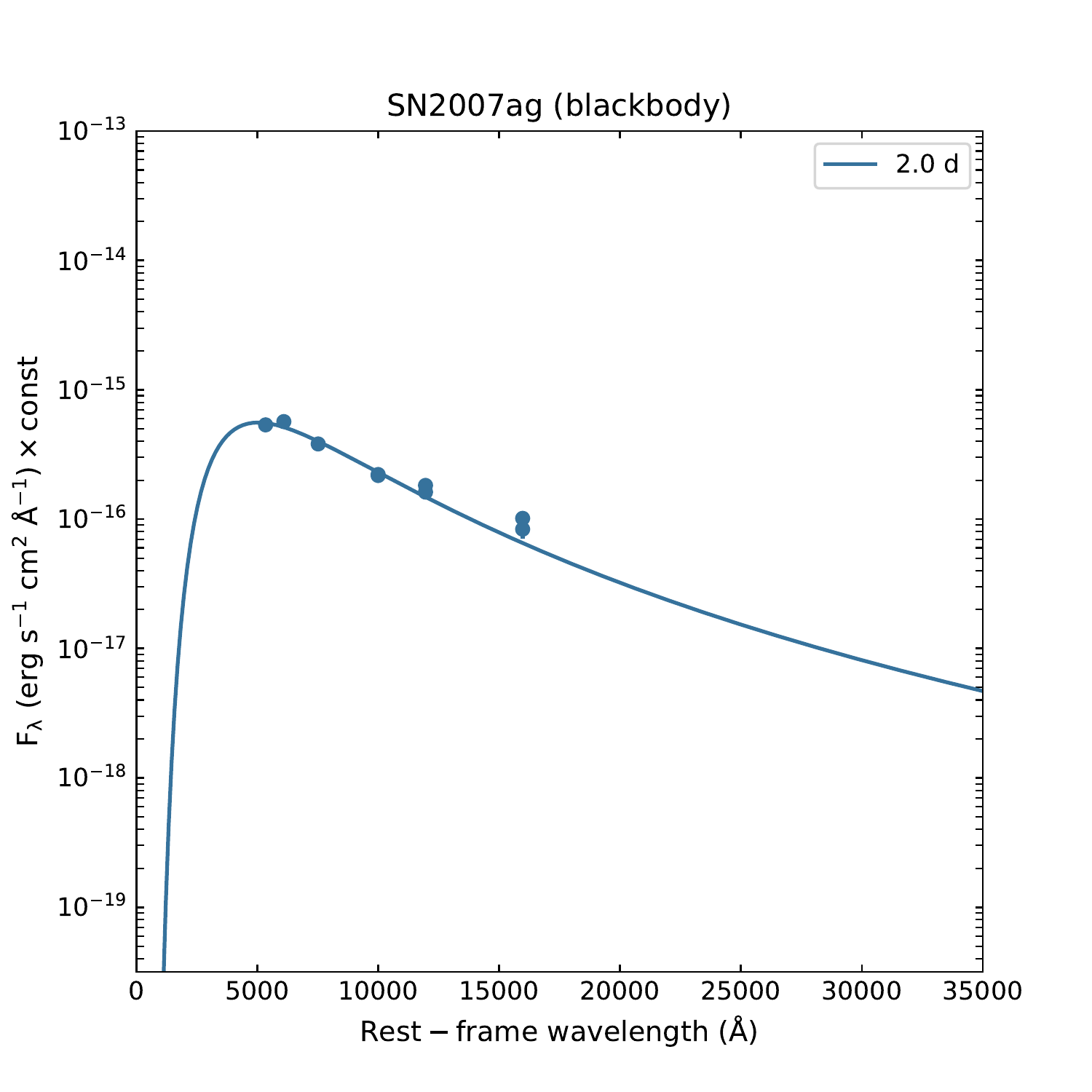}
\includegraphics[width=0.245\textwidth,angle=0]{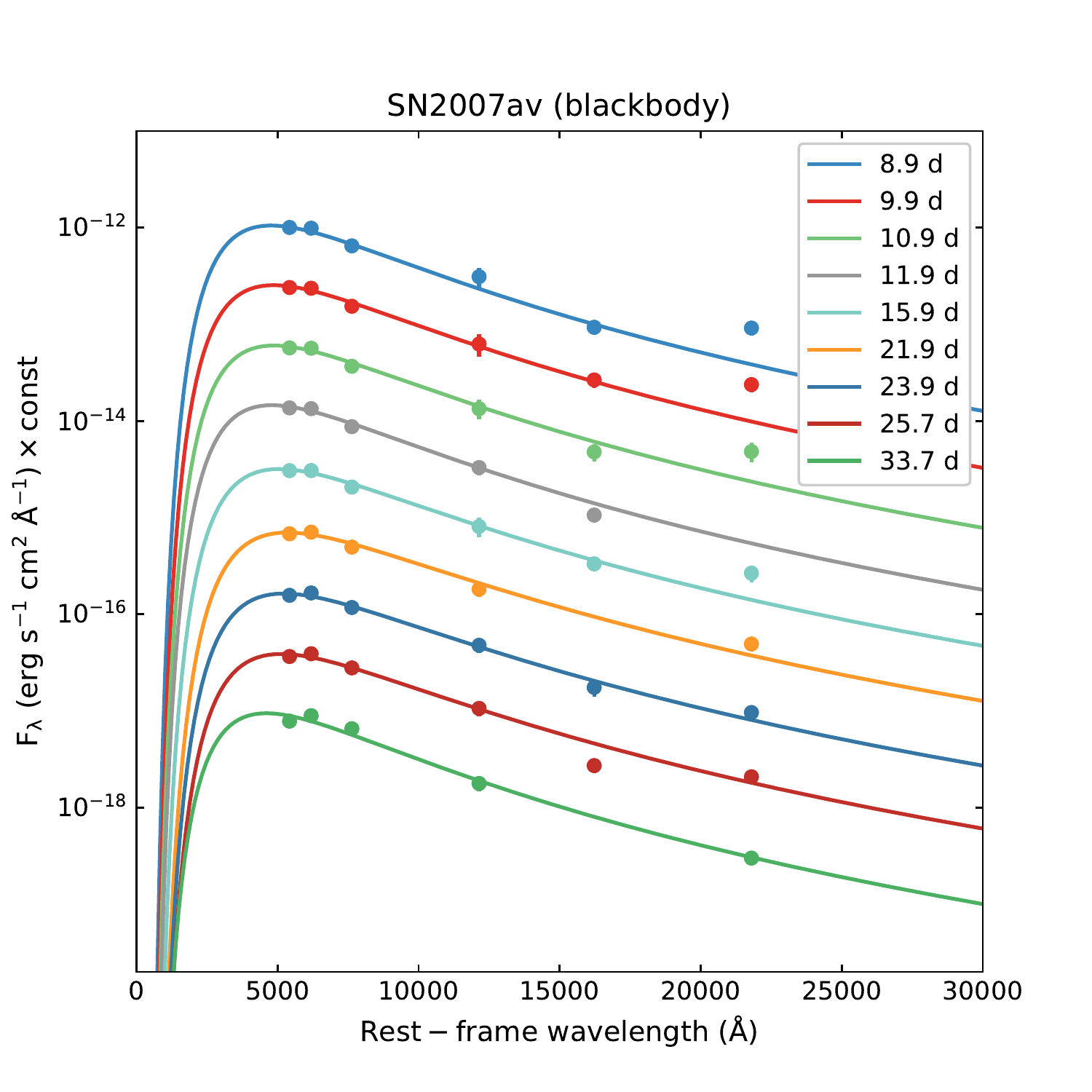}
\includegraphics[width=0.245\textwidth,angle=0]{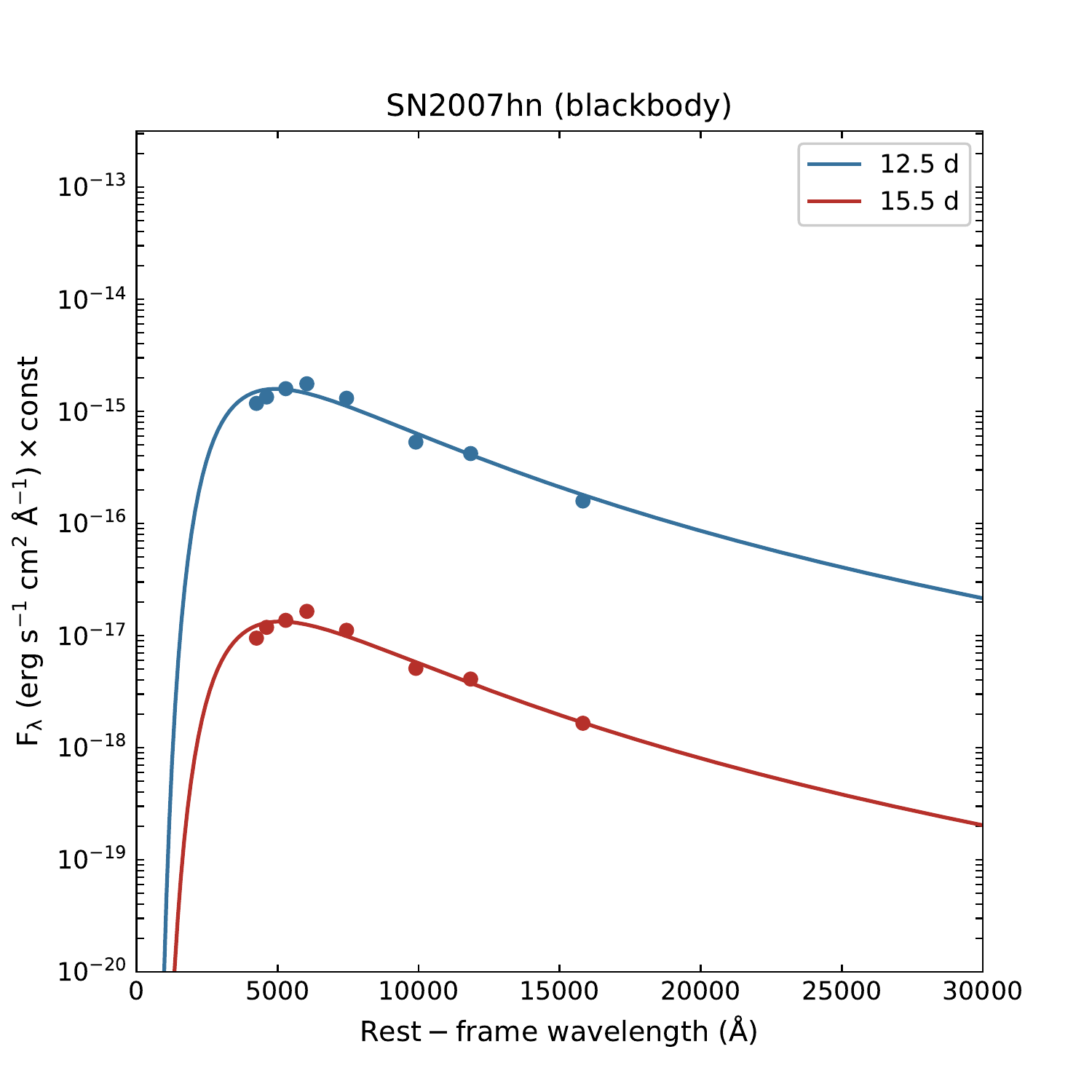}
\includegraphics[width=0.245\textwidth,angle=0]{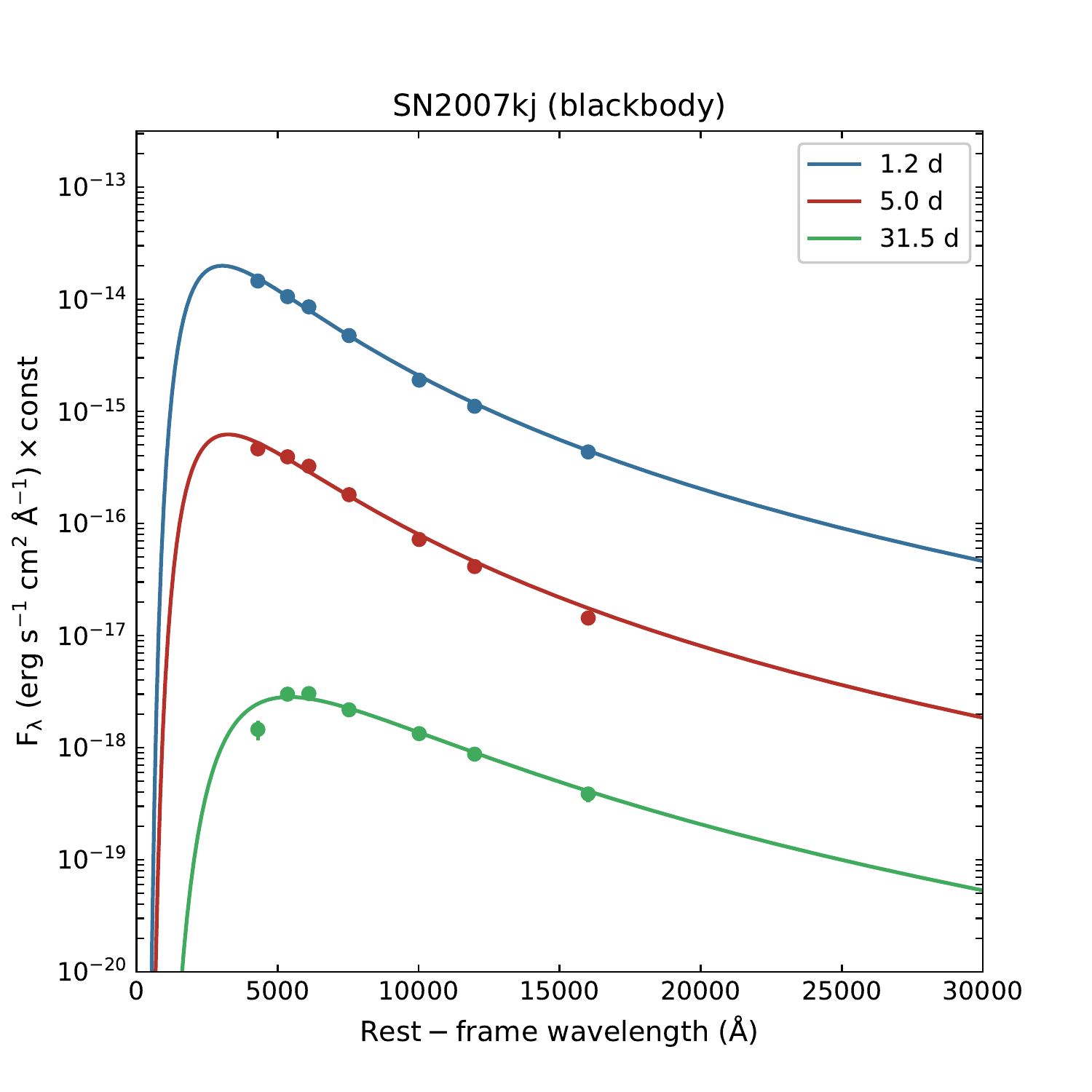}
\includegraphics[width=0.245\textwidth,angle=0]{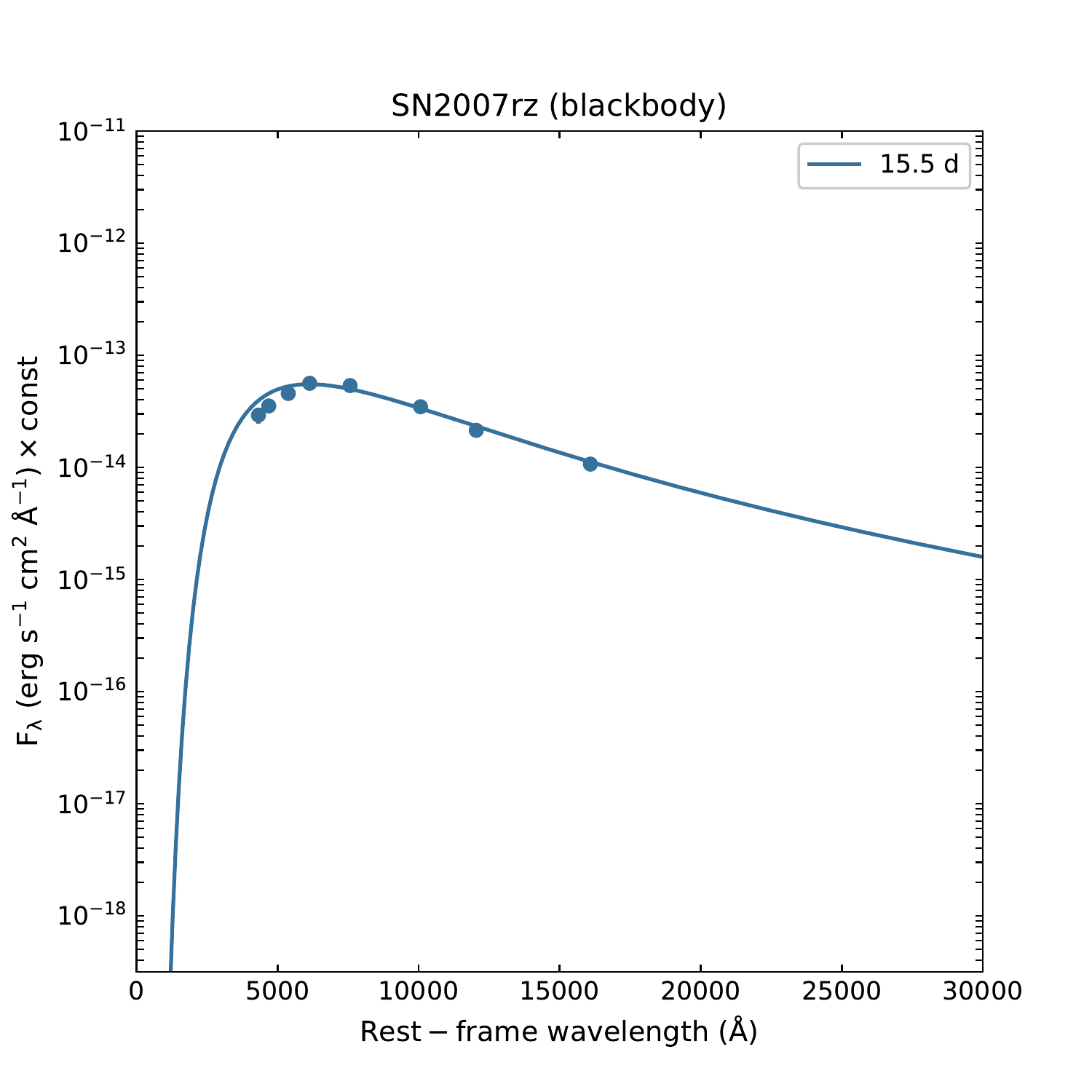}
\includegraphics[width=0.245\textwidth,angle=0]{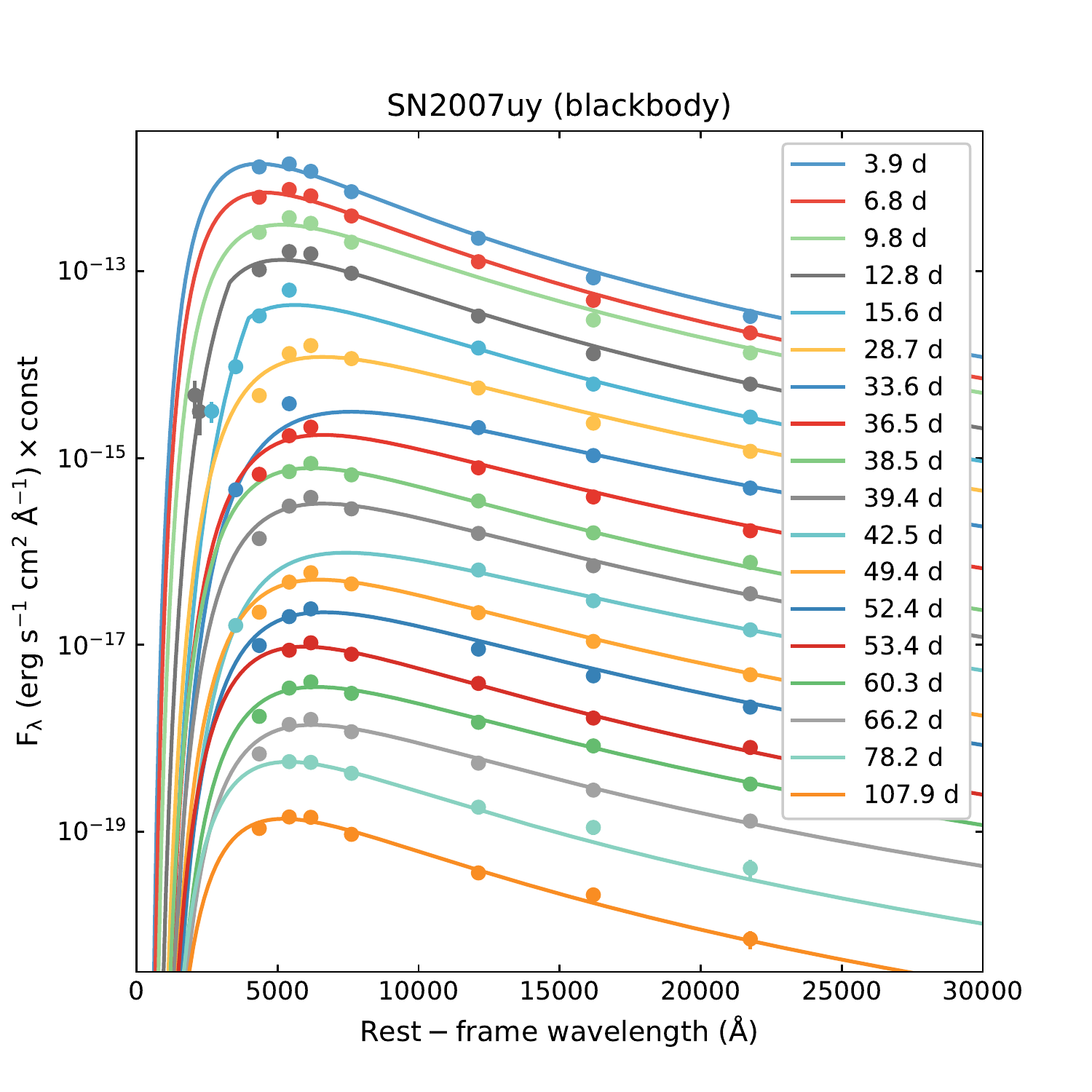}
\end{center}
\caption{(Continued).}
\label{fig:SED}
\end{figure}

\clearpage

\begin{figure}[tbph]
\begin{center}
\ContinuedFloat
\includegraphics[width=0.245\textwidth,angle=0]{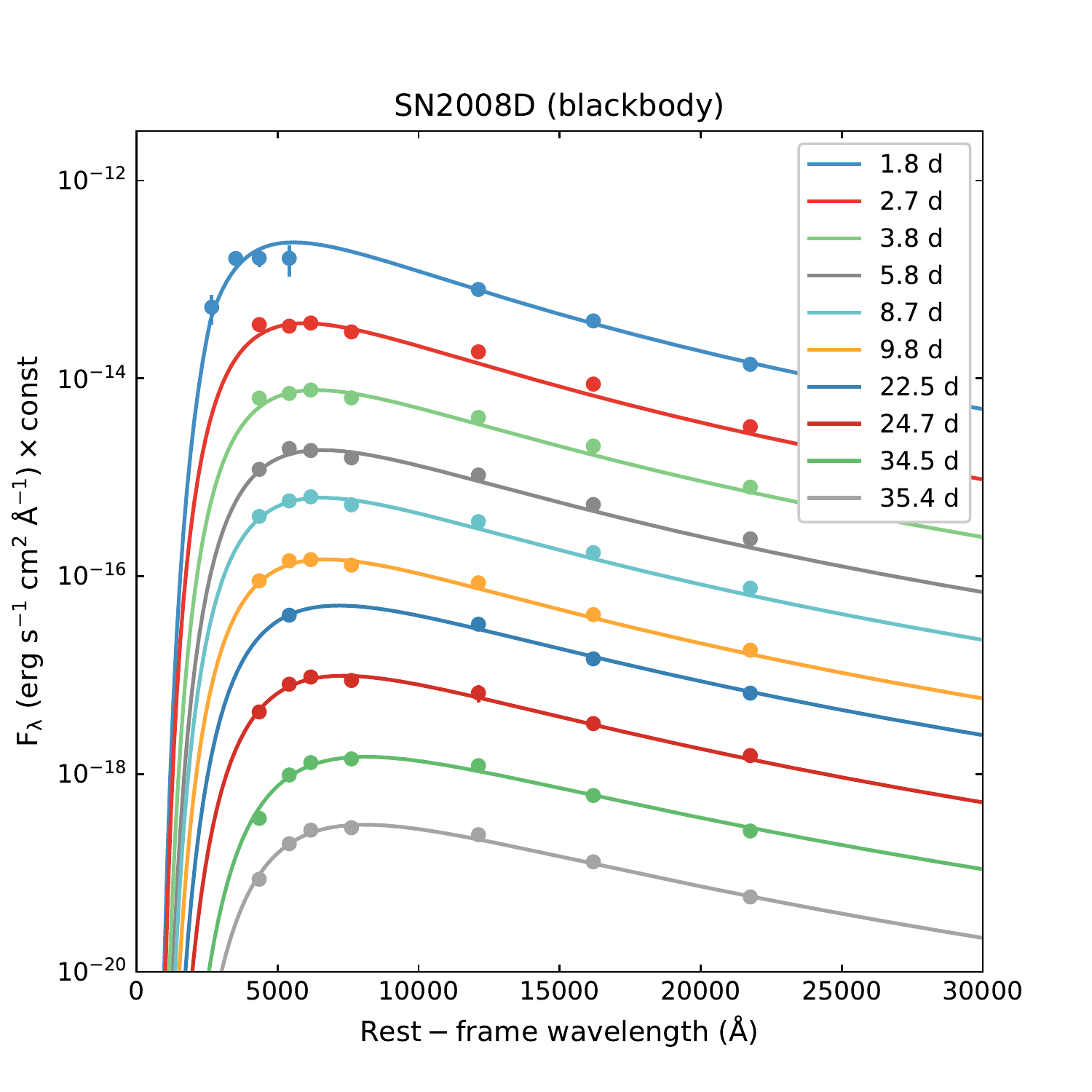}
\includegraphics[width=0.245\textwidth,angle=0]{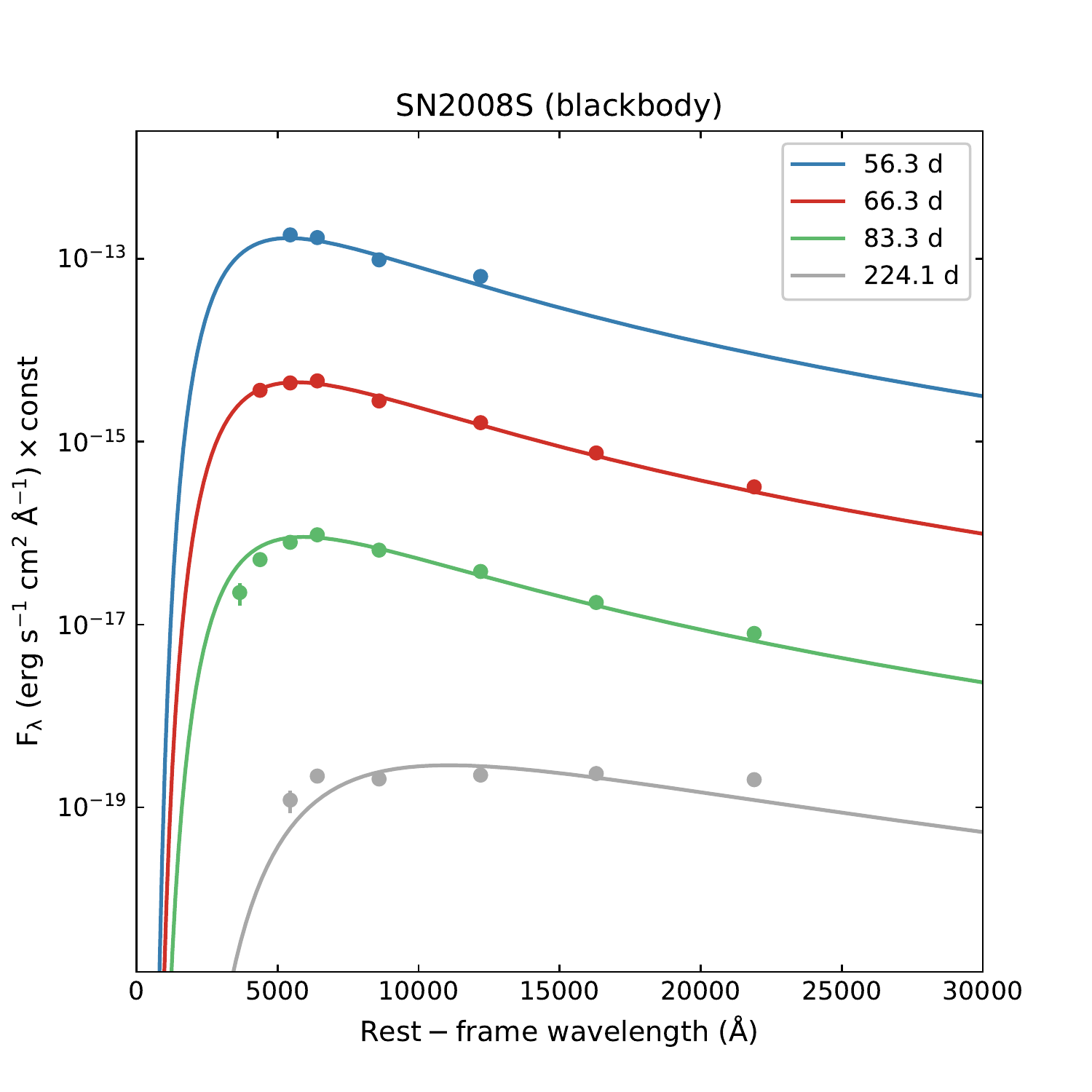}
\includegraphics[width=0.245\textwidth,angle=0]{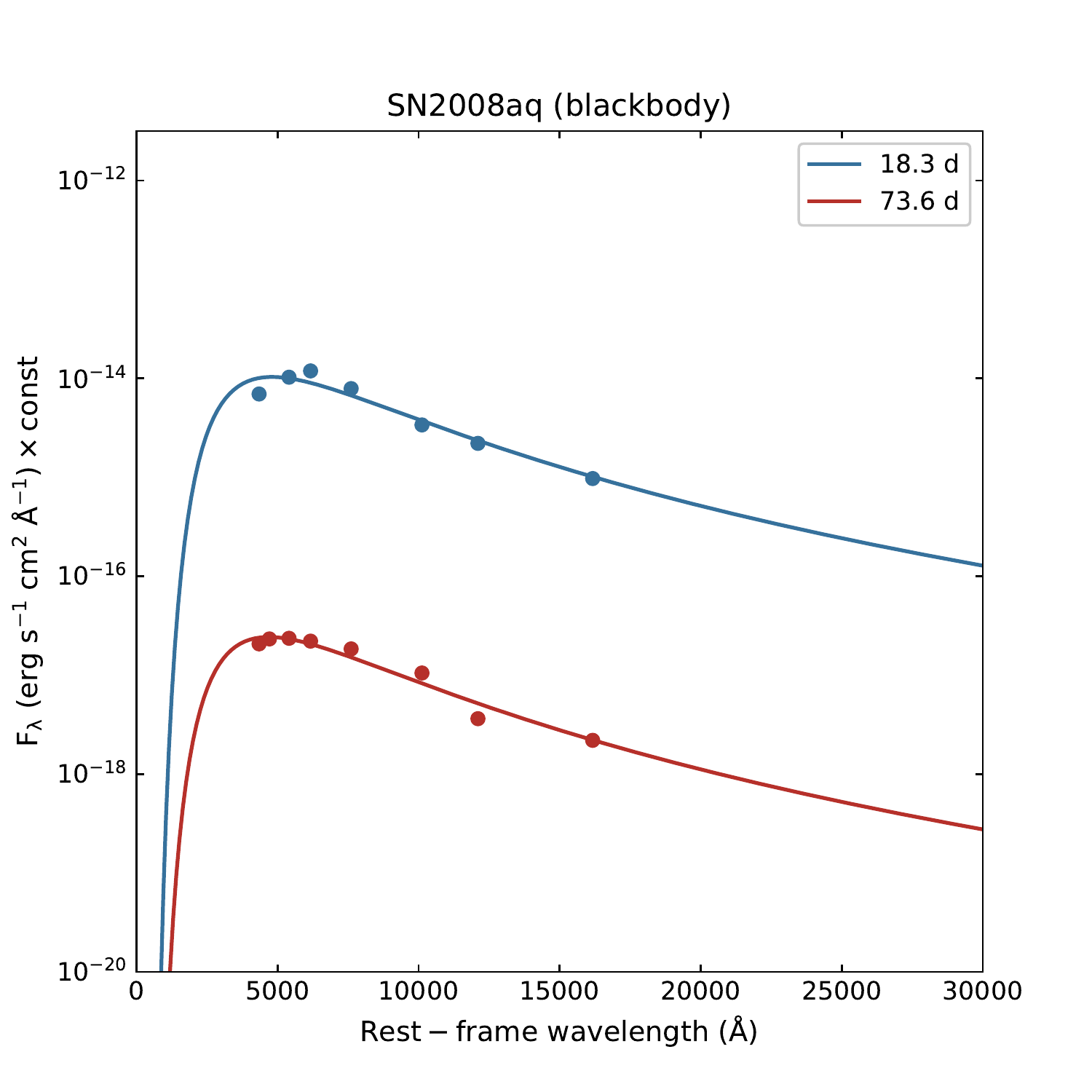}
\includegraphics[width=0.245\textwidth,angle=0]{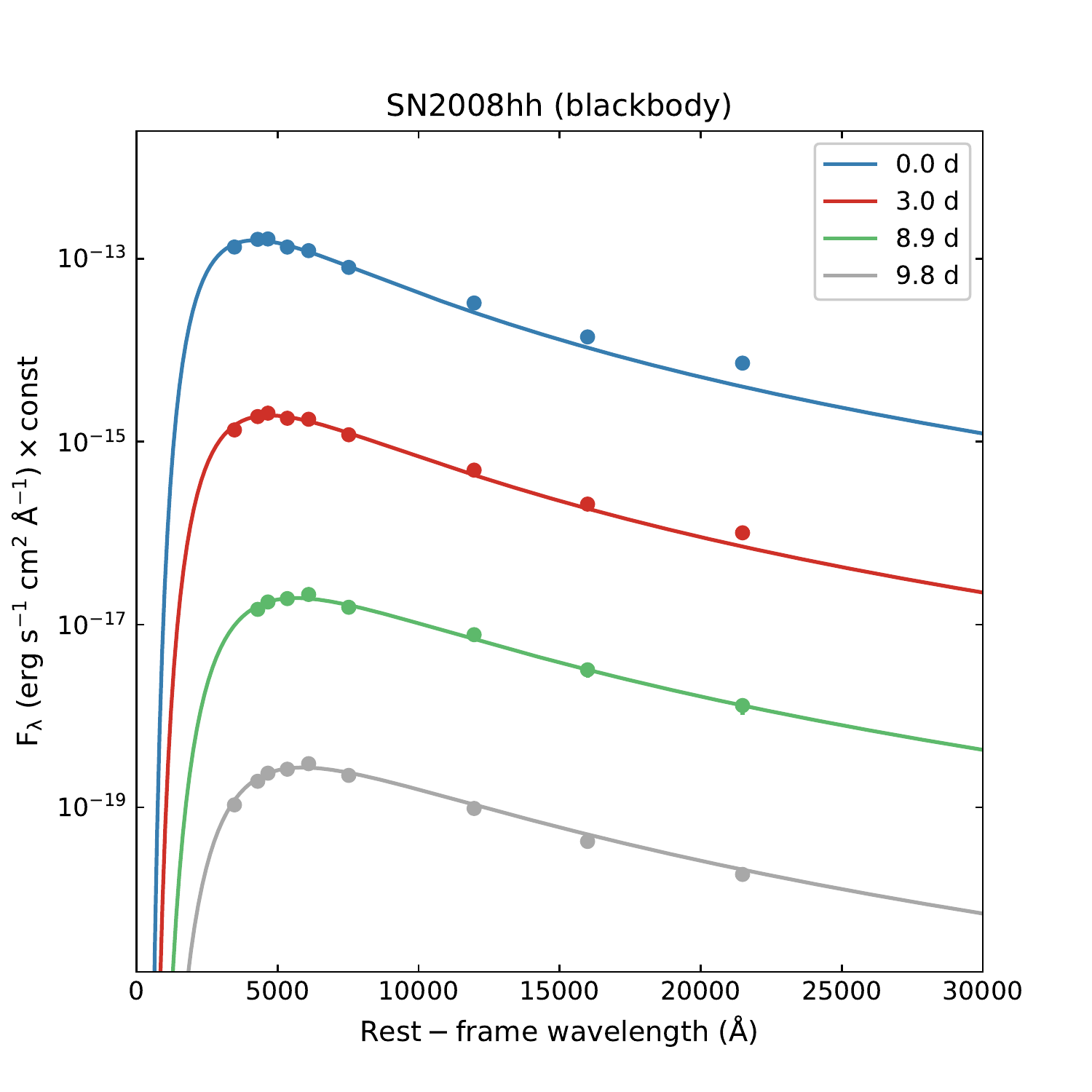}
\includegraphics[width=0.245\textwidth,angle=0]{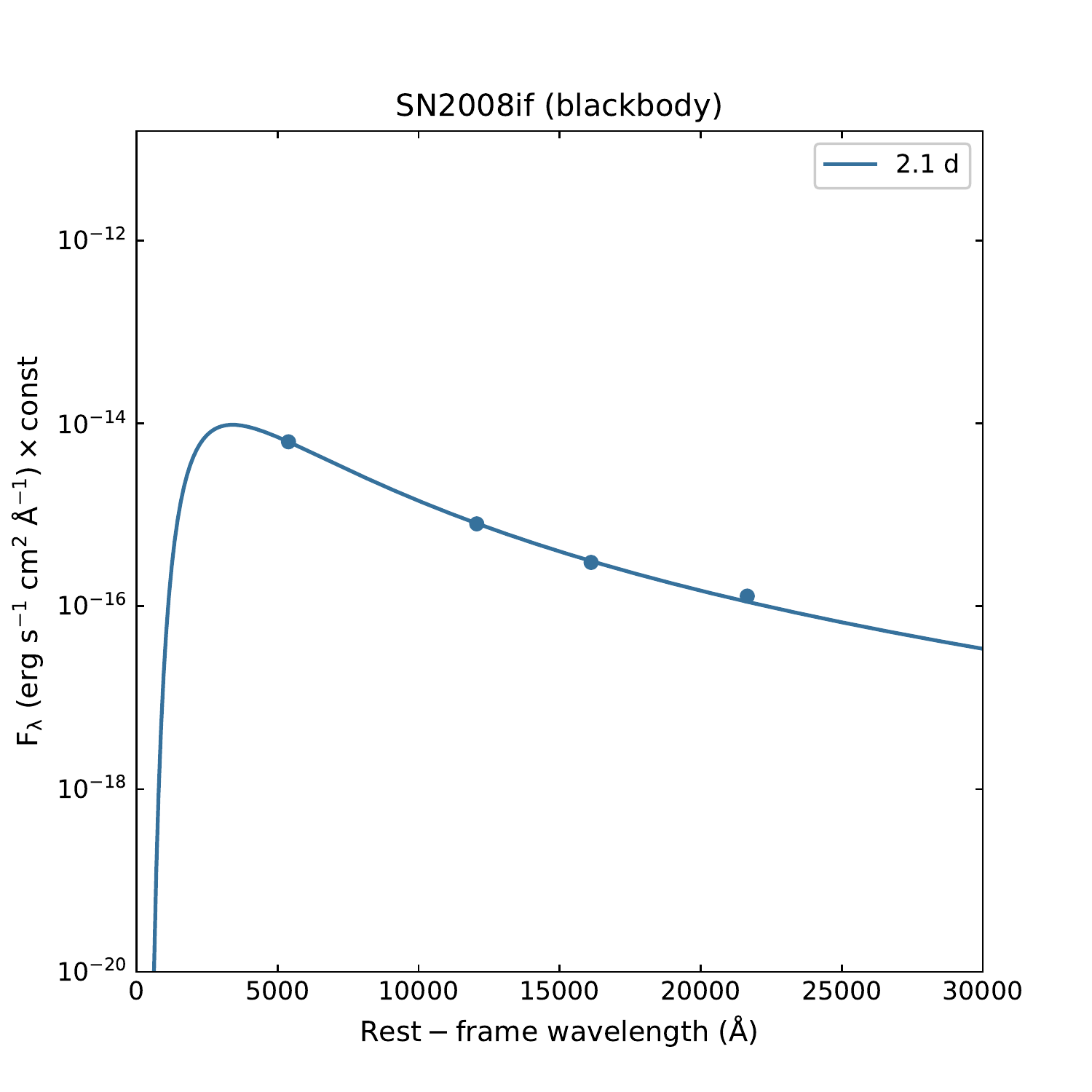}
\includegraphics[width=0.245\textwidth,angle=0]{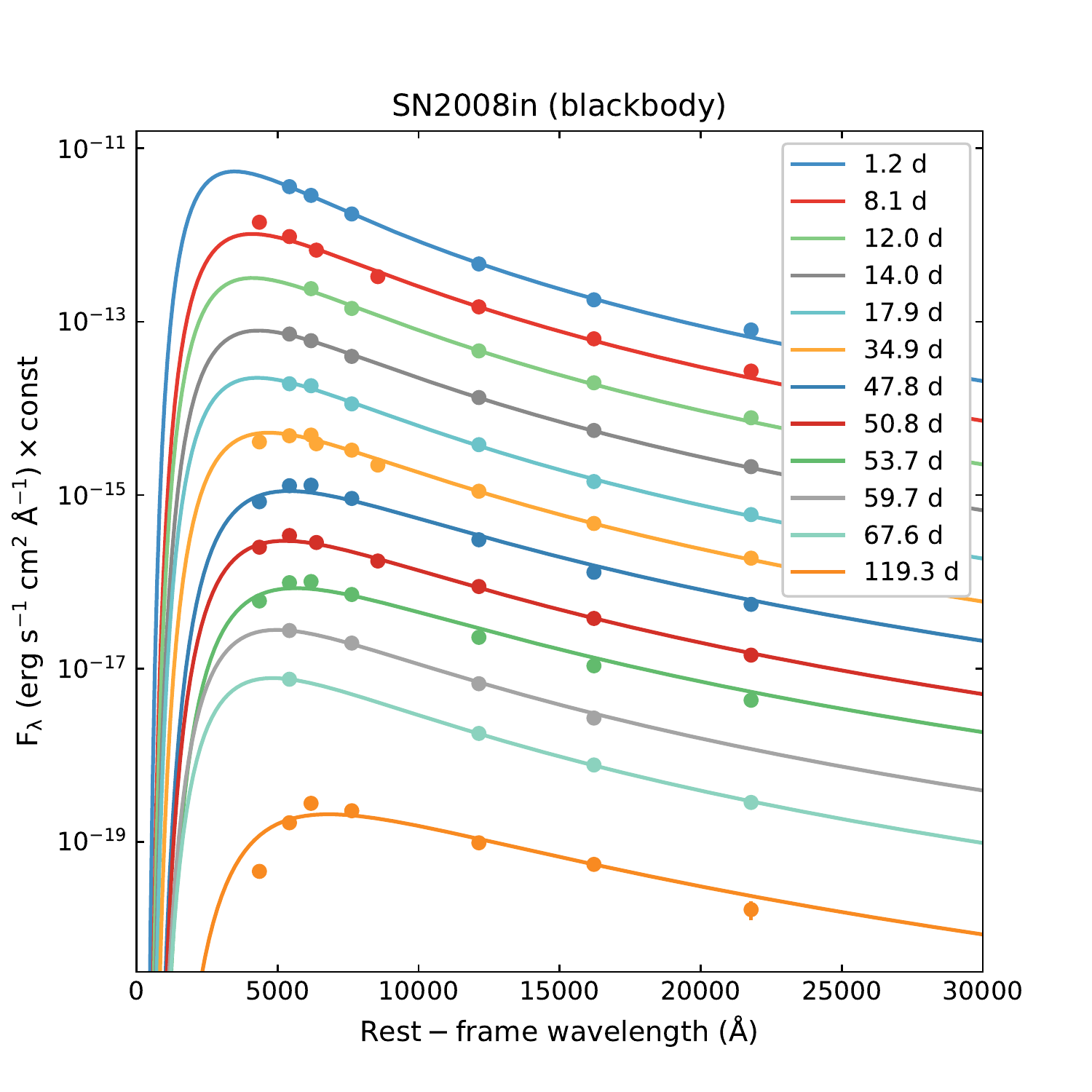}
\includegraphics[width=0.245\textwidth,angle=0]{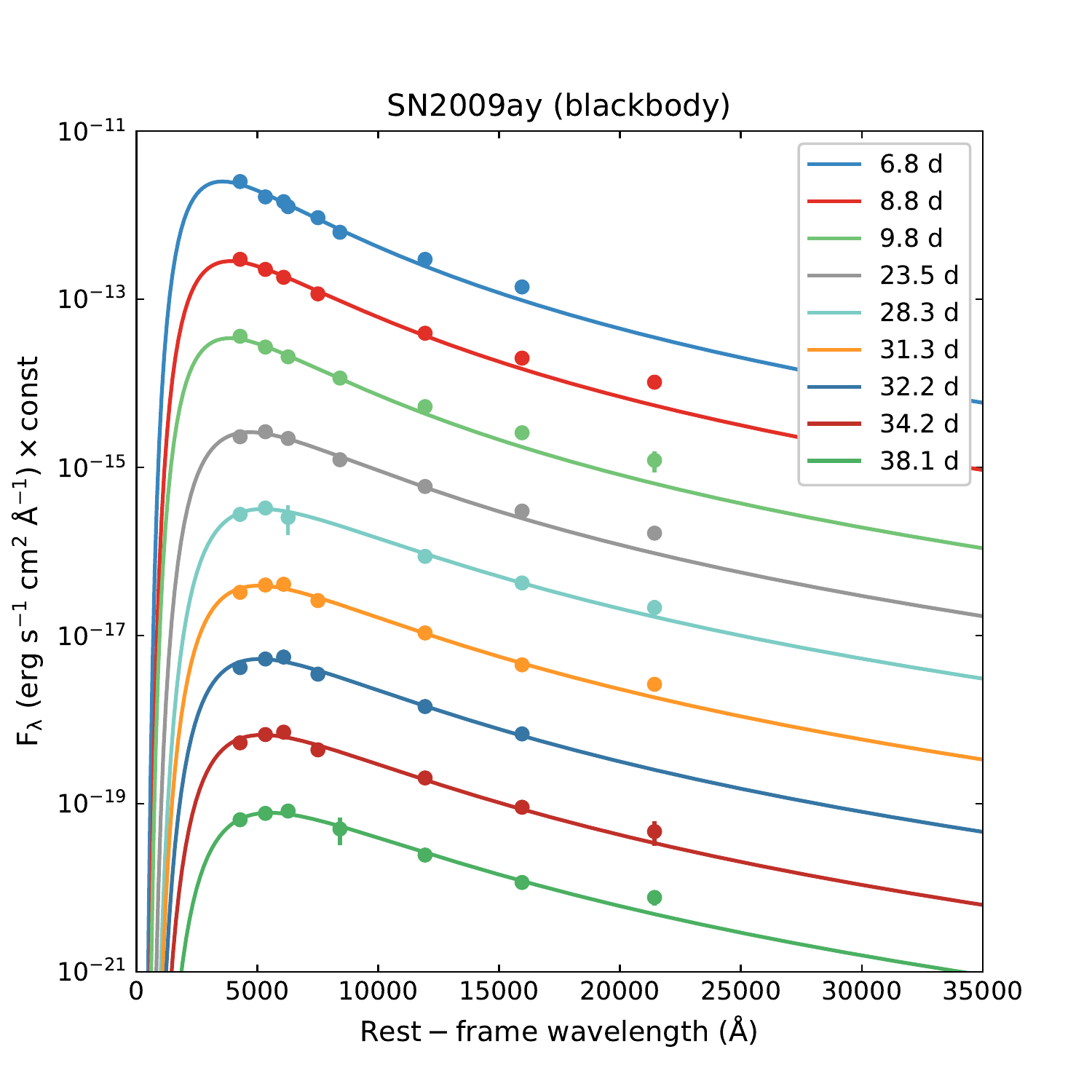}
\includegraphics[width=0.245\textwidth,angle=0]{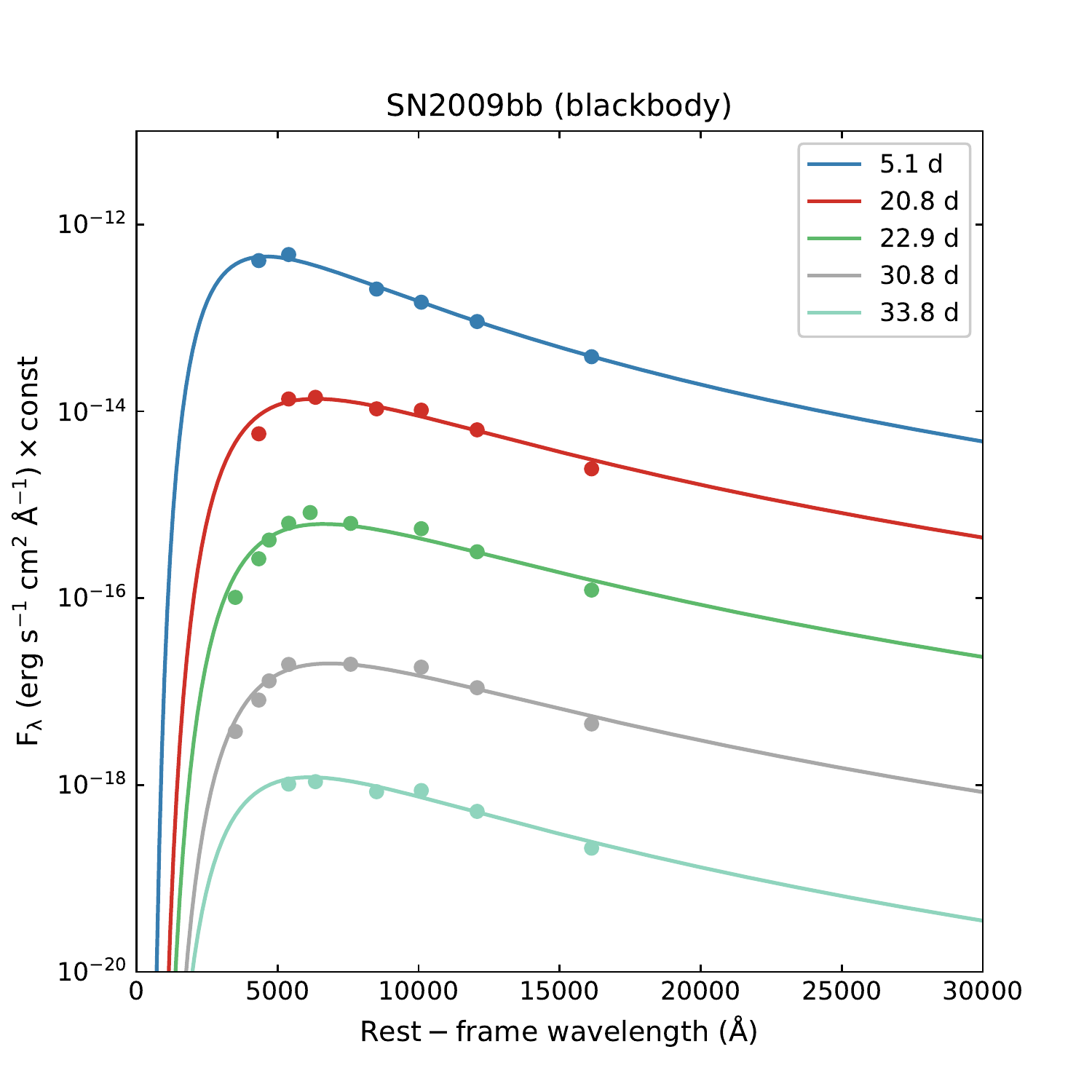}
\includegraphics[width=0.245\textwidth,angle=0]{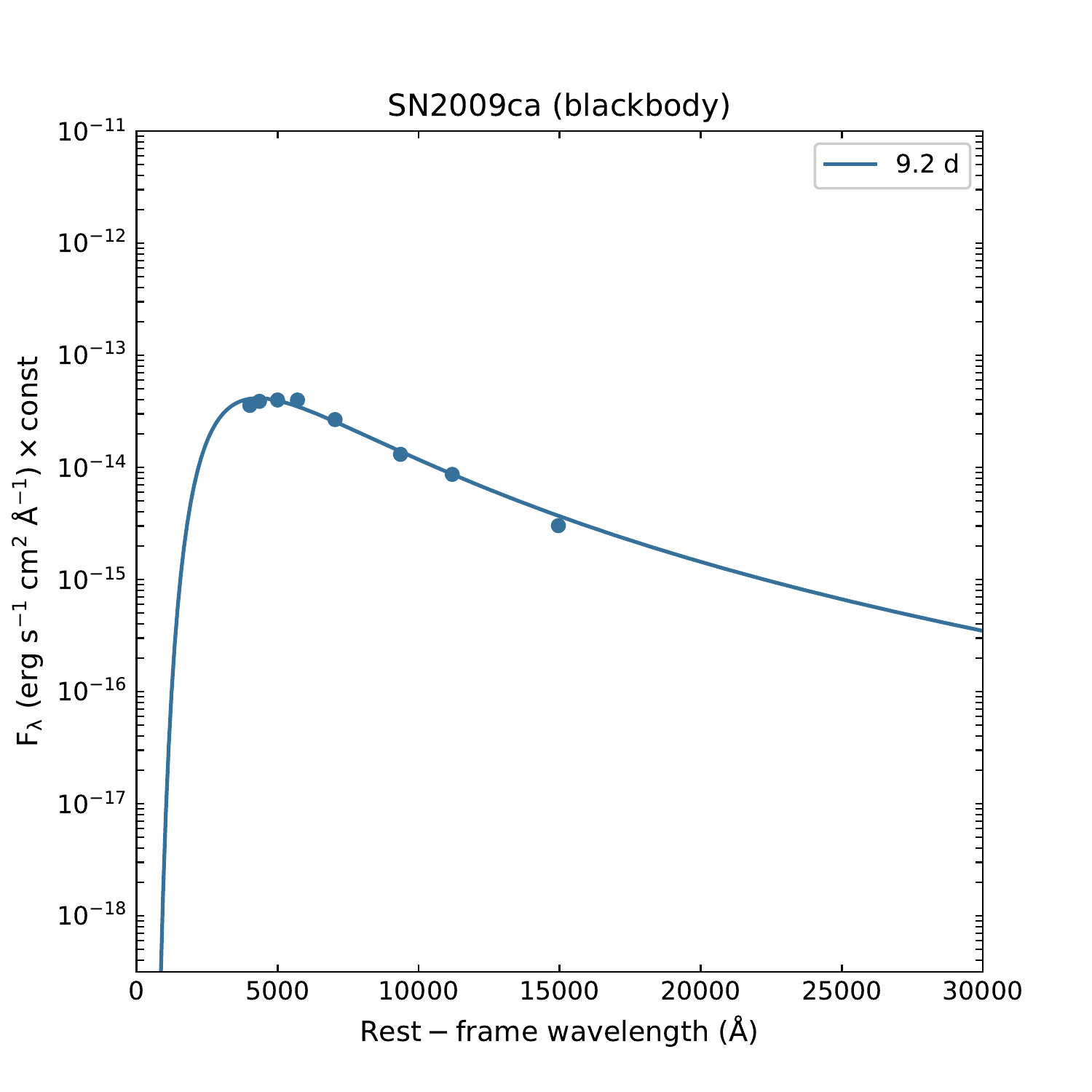}
\includegraphics[width=0.245\textwidth,angle=0]{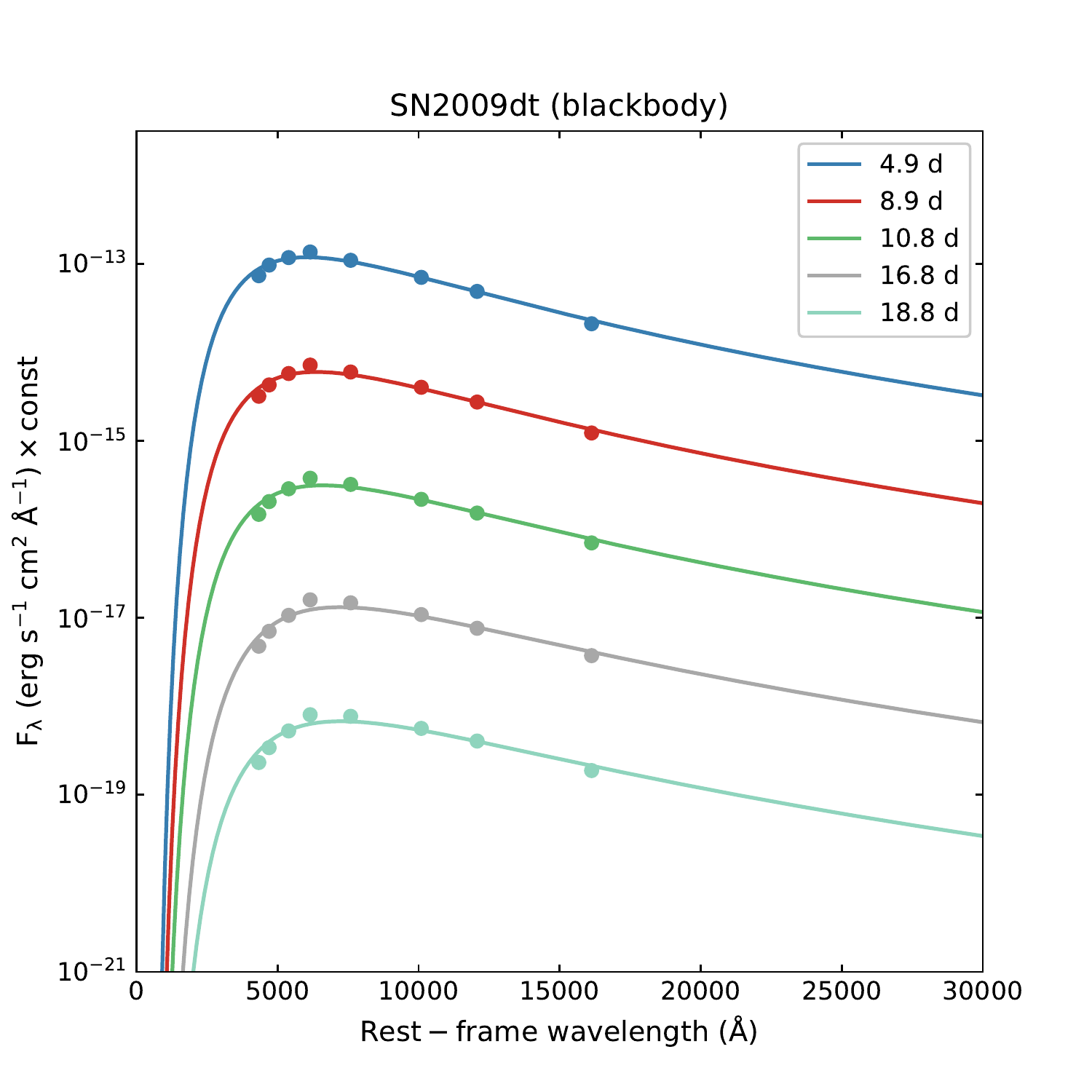}
\includegraphics[width=0.245\textwidth,angle=0]{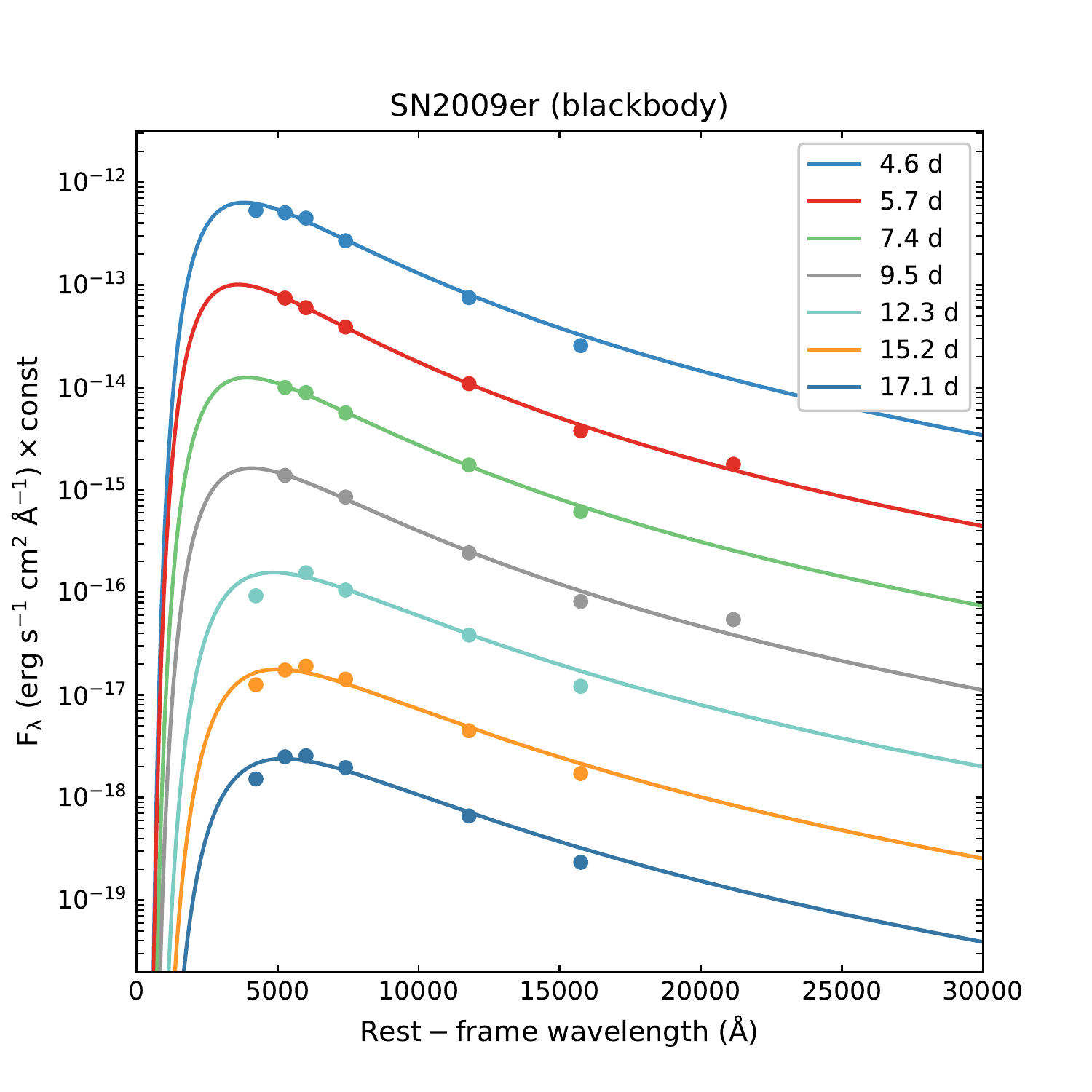}
\includegraphics[width=0.245\textwidth,angle=0]{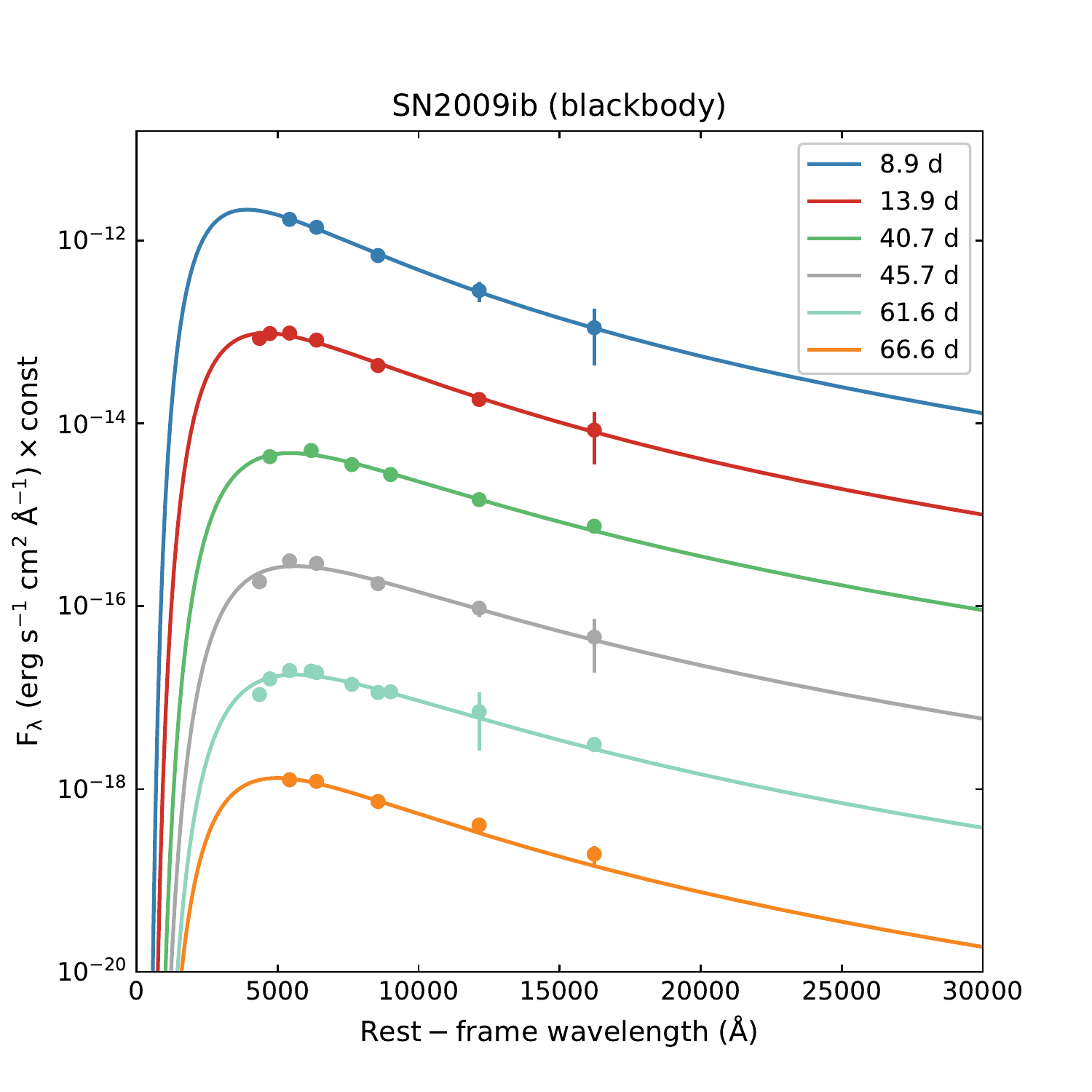}
\includegraphics[width=0.245\textwidth,angle=0]{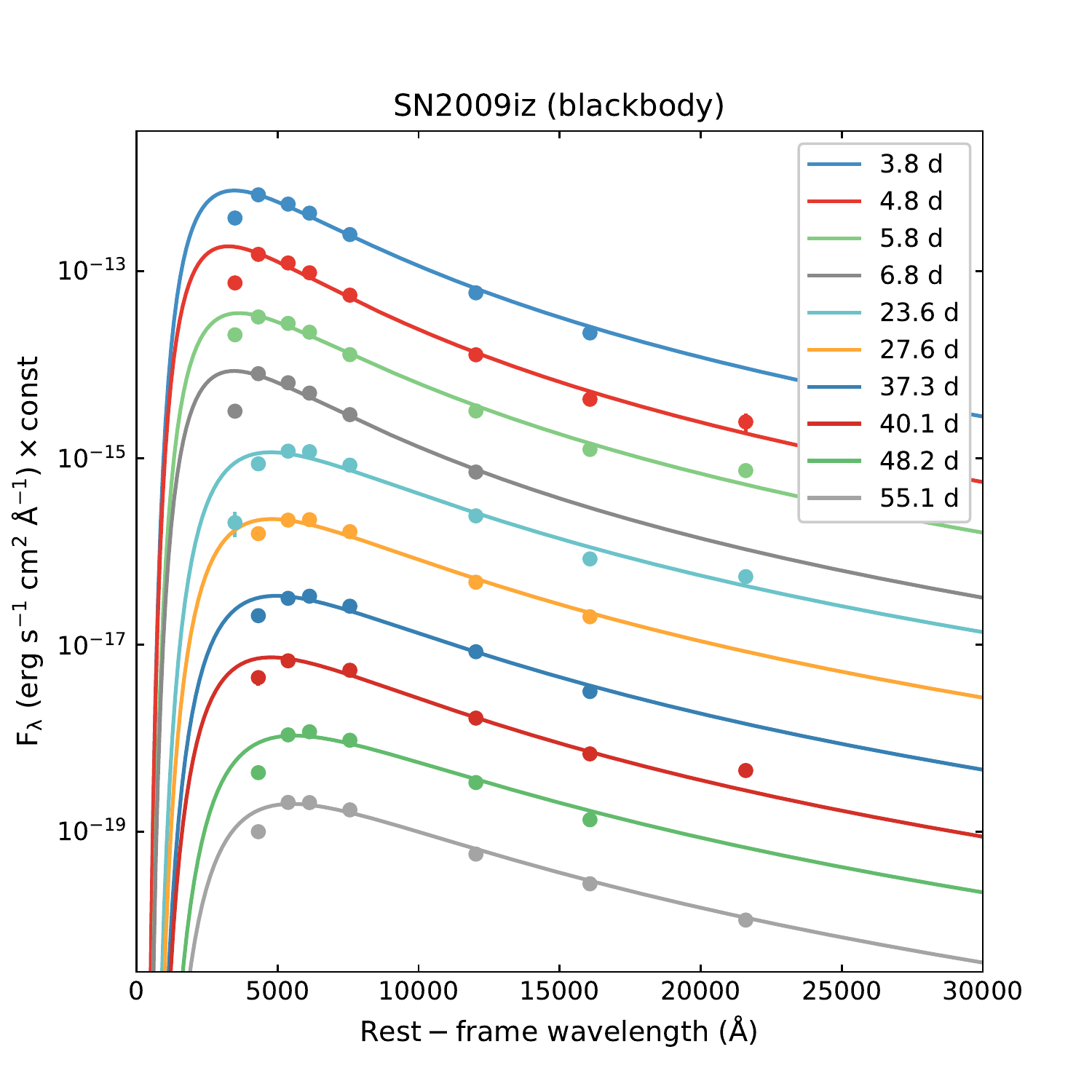}
\includegraphics[width=0.245\textwidth,angle=0]{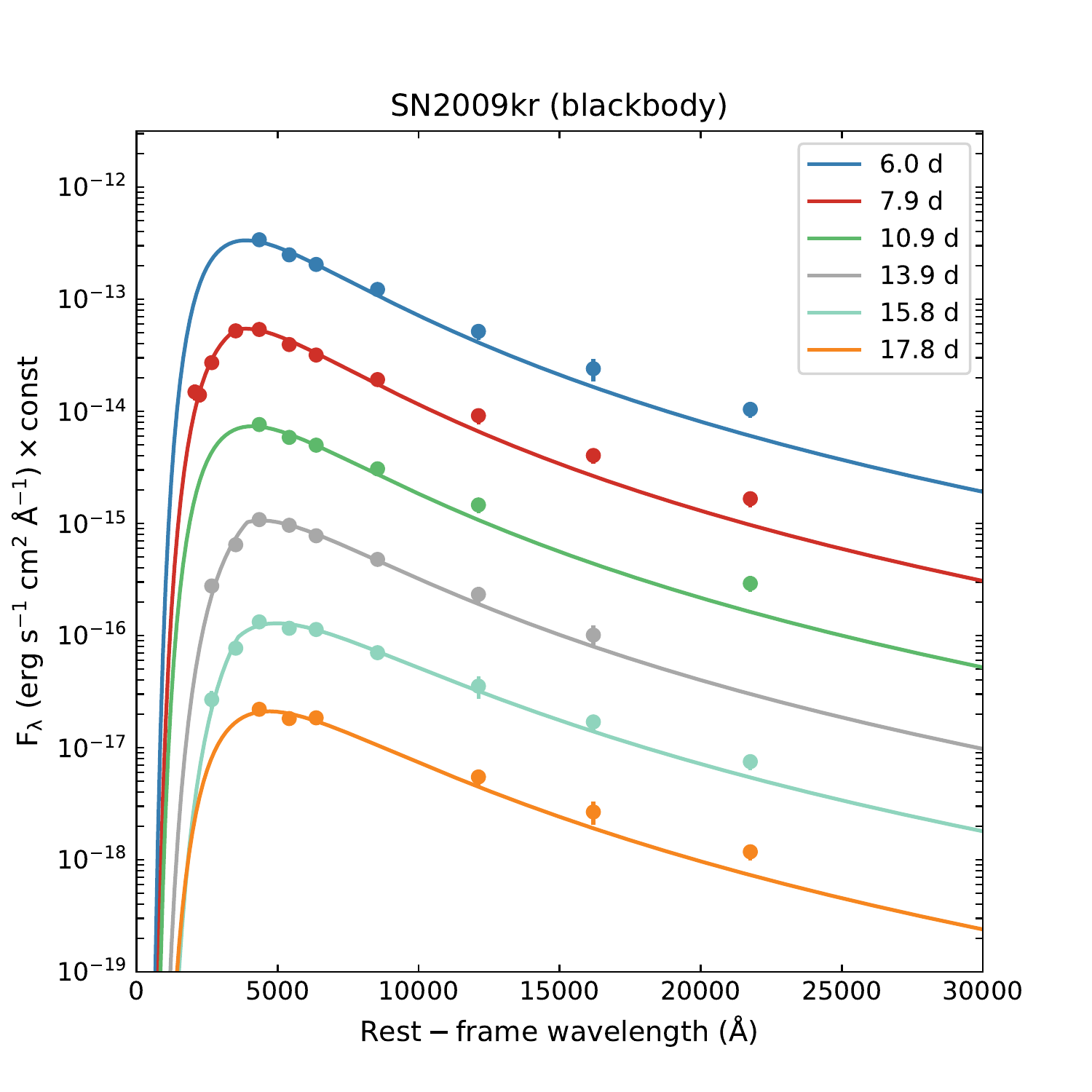}
\includegraphics[width=0.245\textwidth,angle=0]{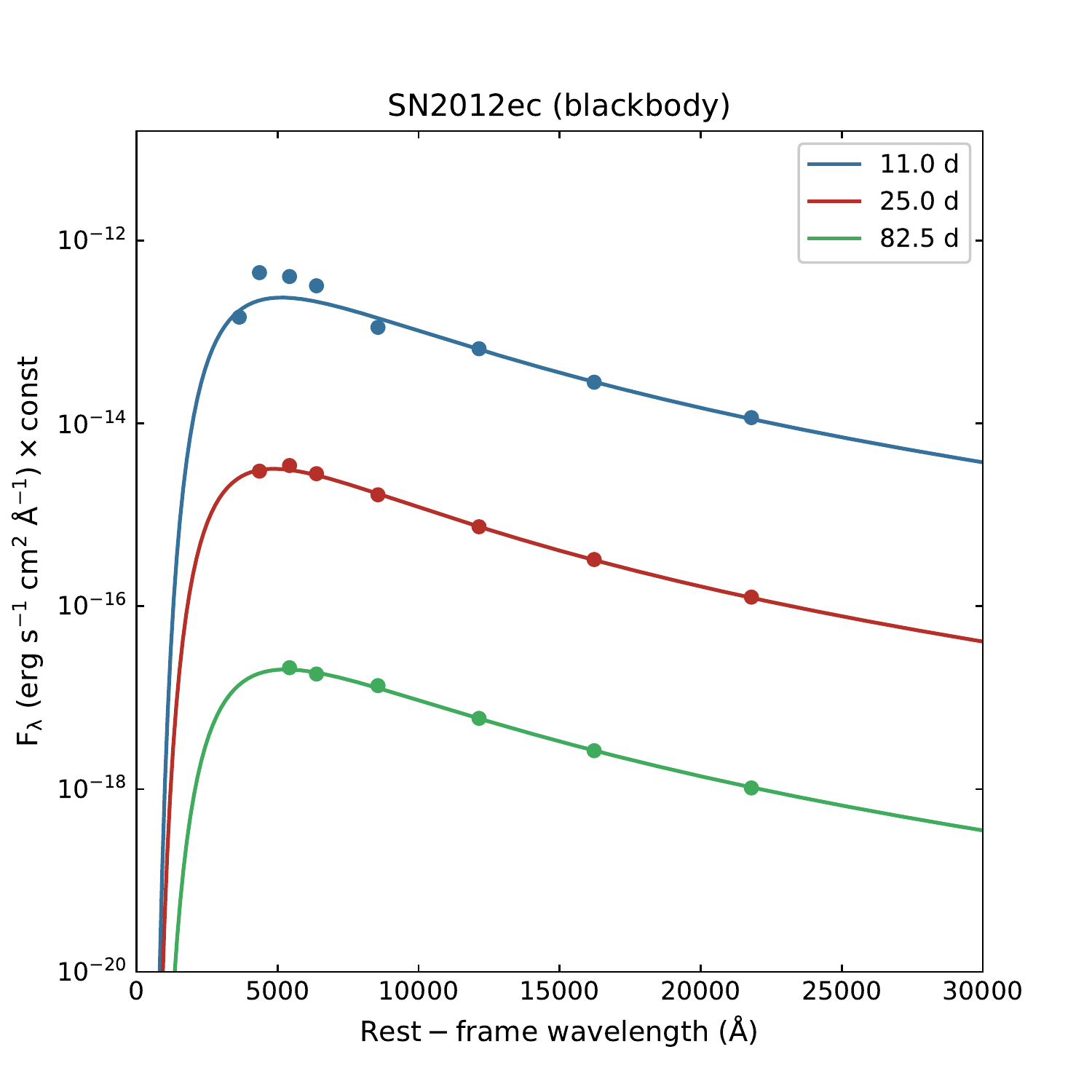}
\includegraphics[width=0.245\textwidth,angle=0]{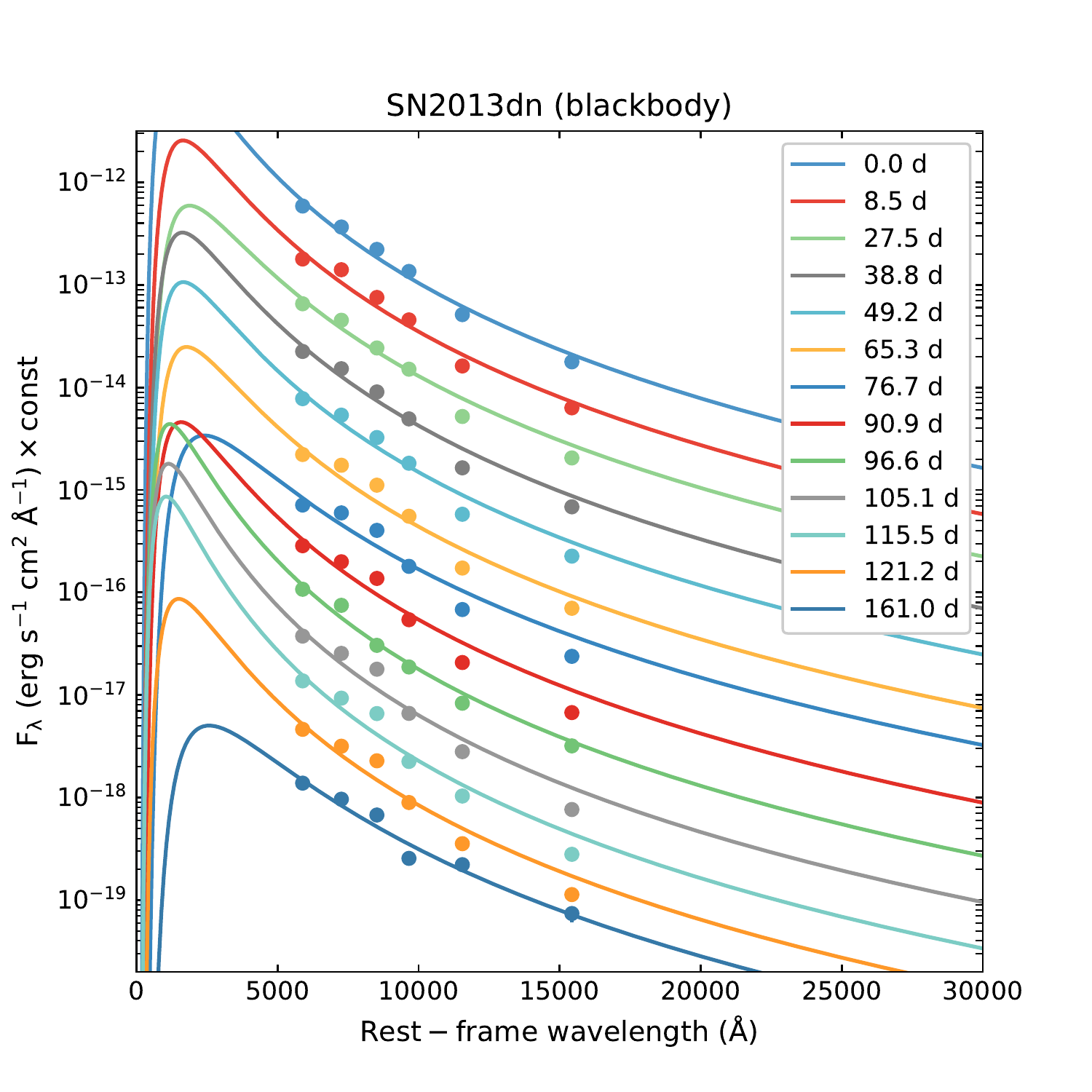}
\includegraphics[width=0.245\textwidth,angle=0]{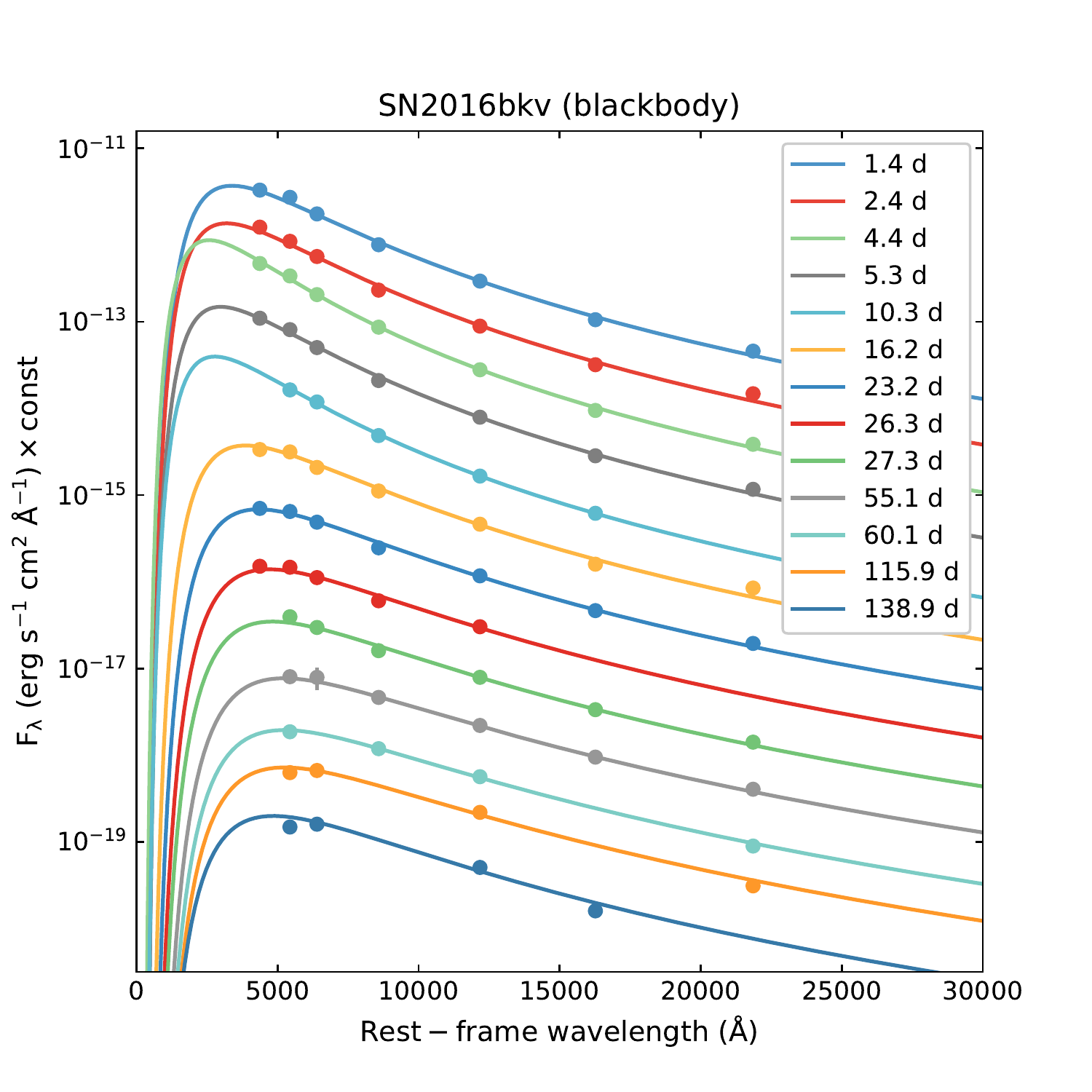}
\includegraphics[width=0.245\textwidth,angle=0]{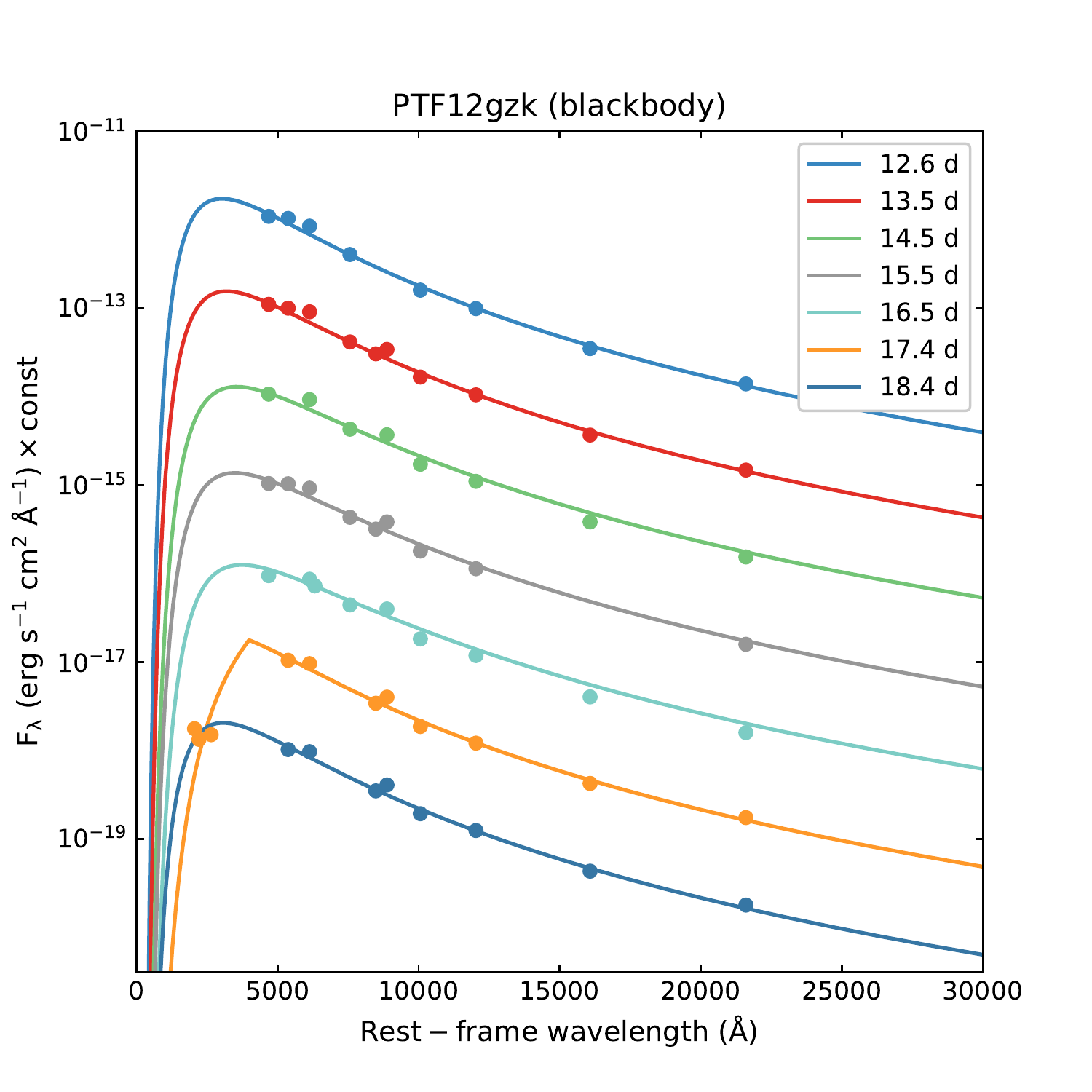}
\includegraphics[width=0.245\textwidth,angle=0]{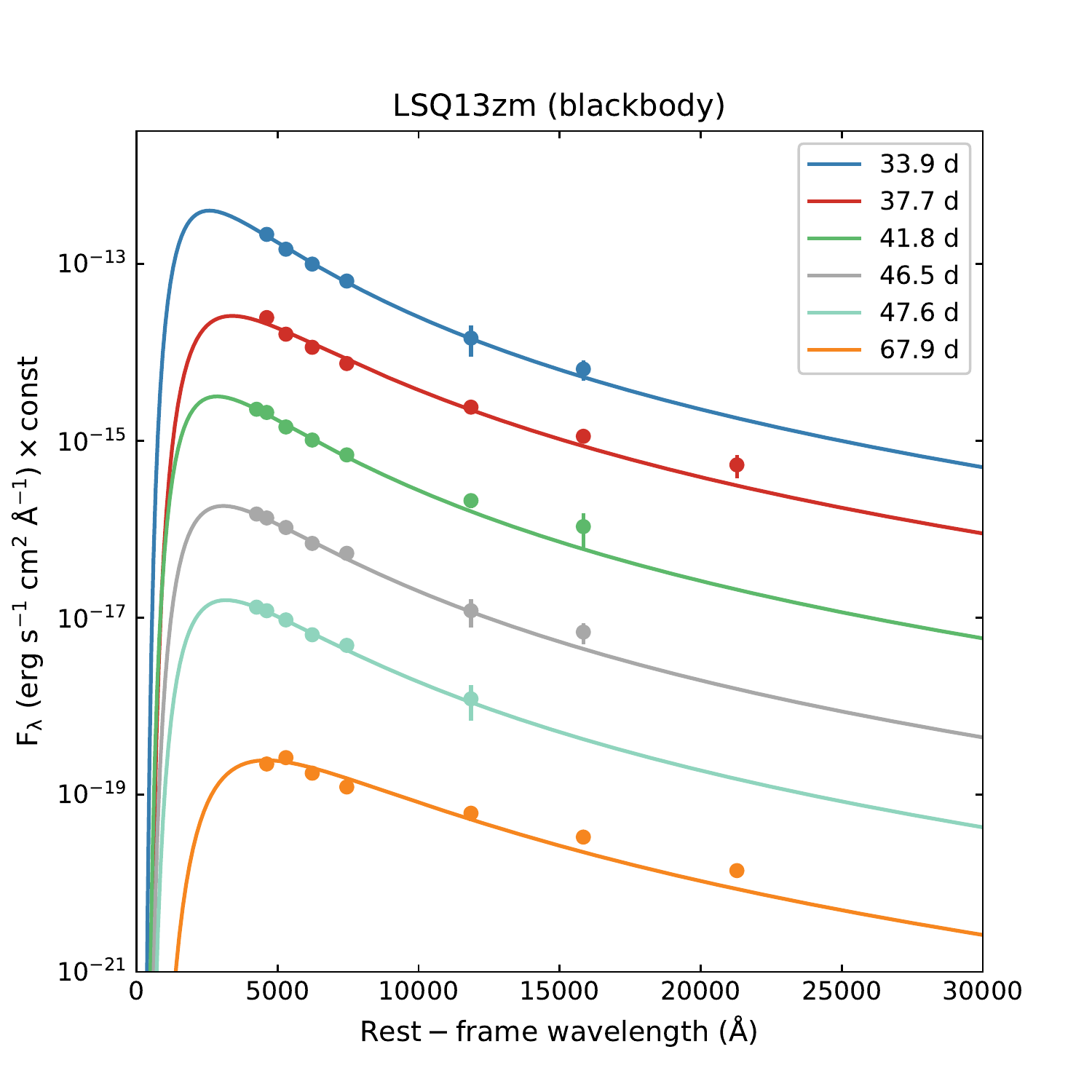}
\end{center}
\caption{(Continued).}
\label{fig:SED}
\end{figure}

\clearpage

\begin{figure}[tbph]
\begin{center}
\includegraphics[width=0.32\textwidth,angle=0]{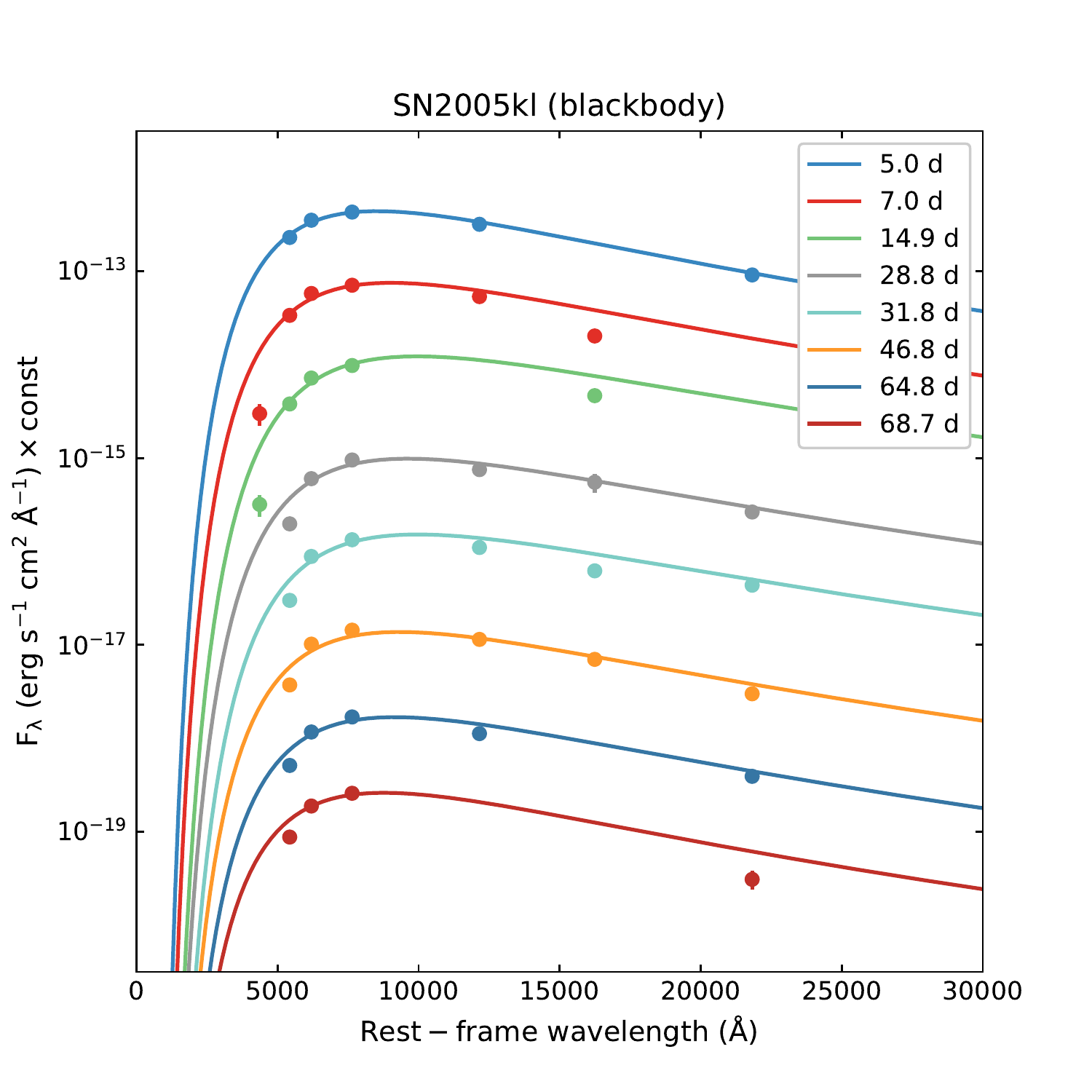}
\includegraphics[width=0.32\textwidth,angle=0]{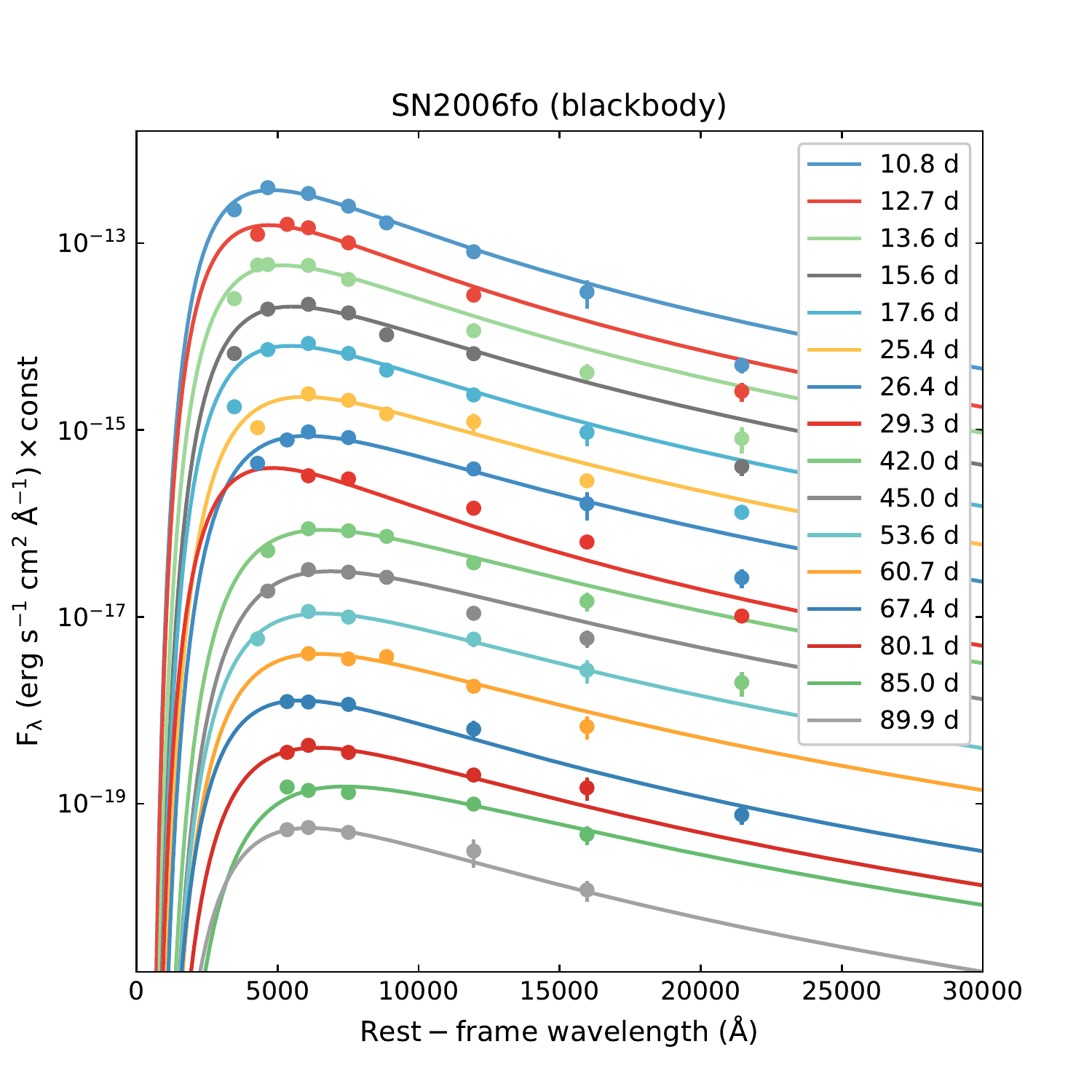}
\includegraphics[width=0.32\textwidth,angle=0]{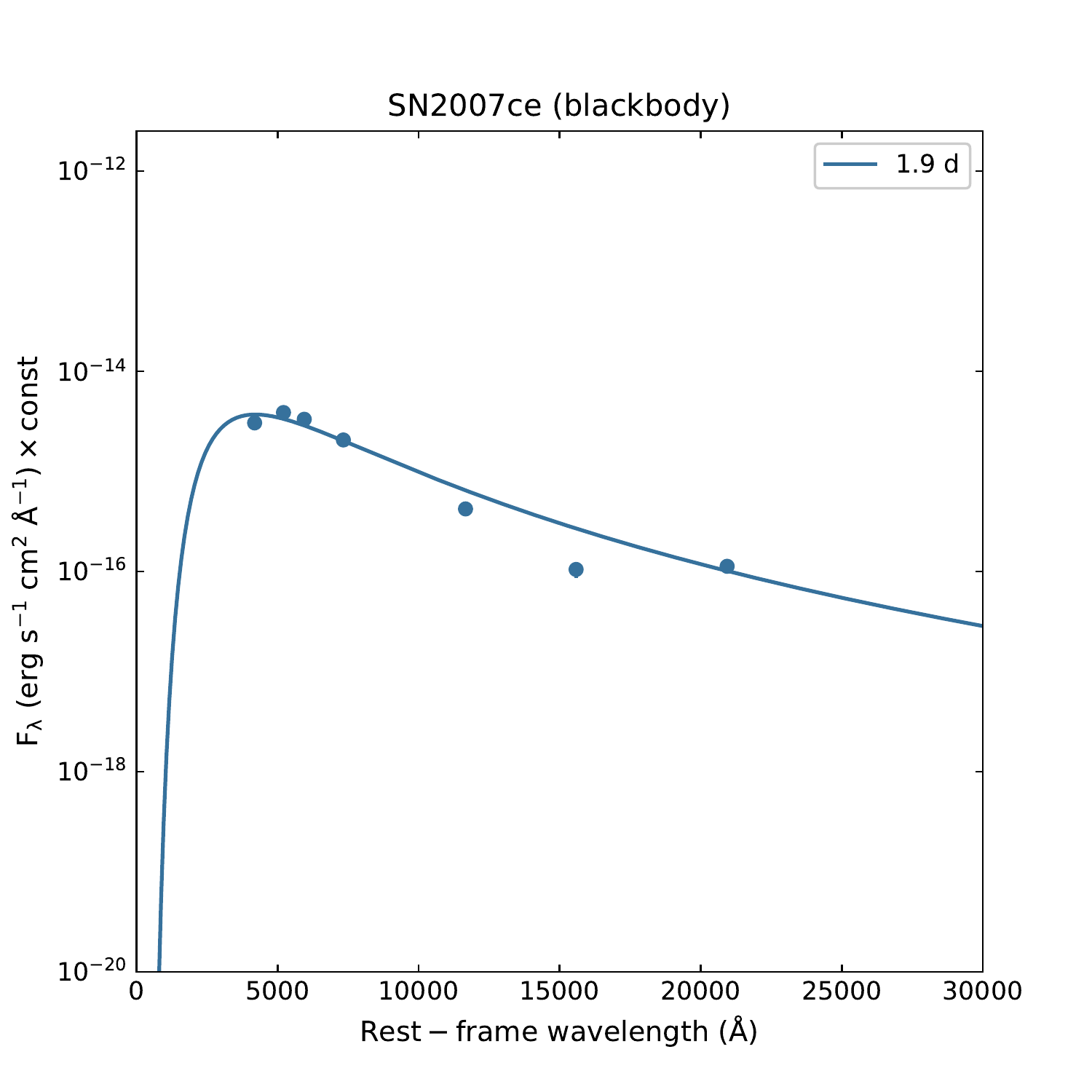}
\includegraphics[width=0.32\textwidth,angle=0]{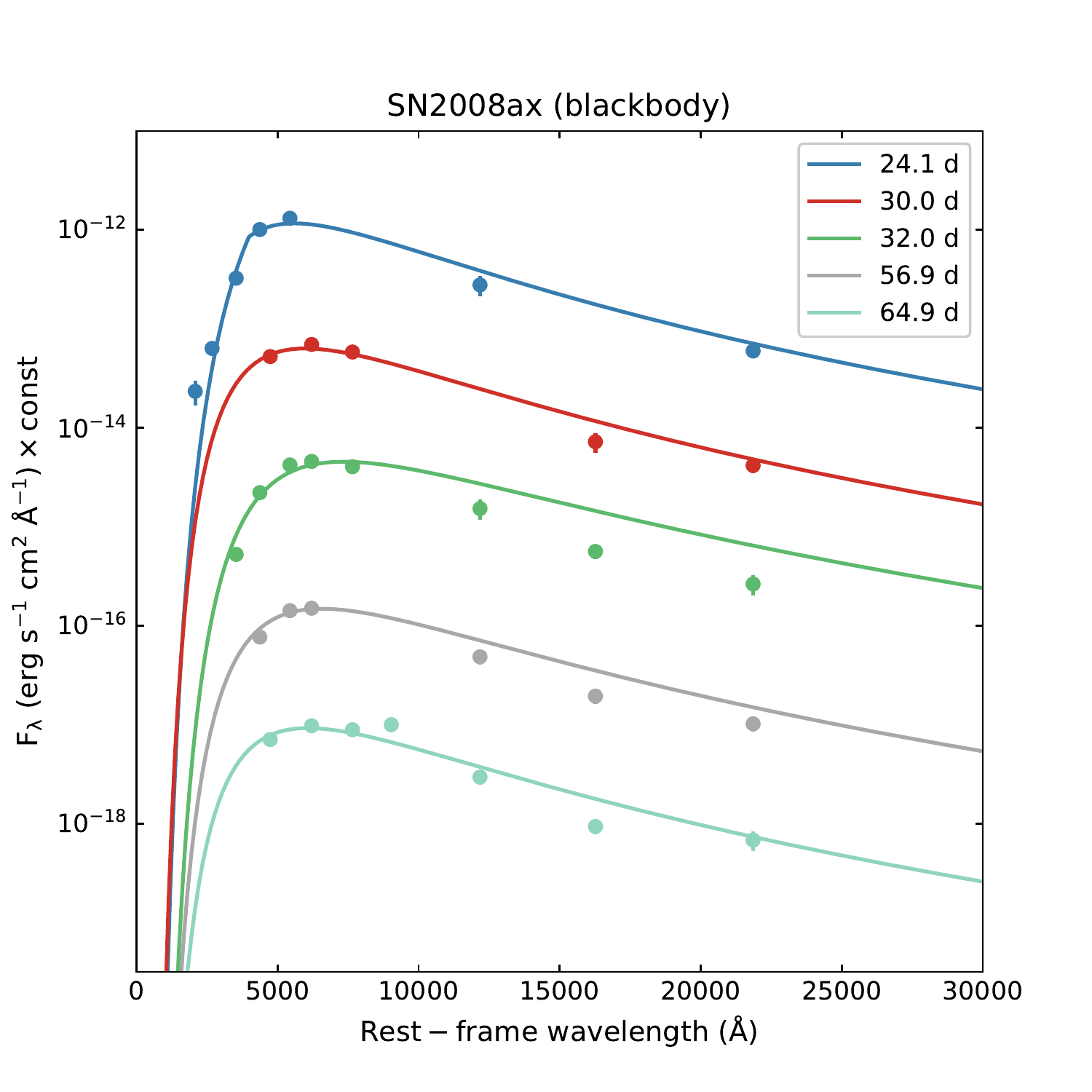}
\includegraphics[width=0.32\textwidth,angle=0]{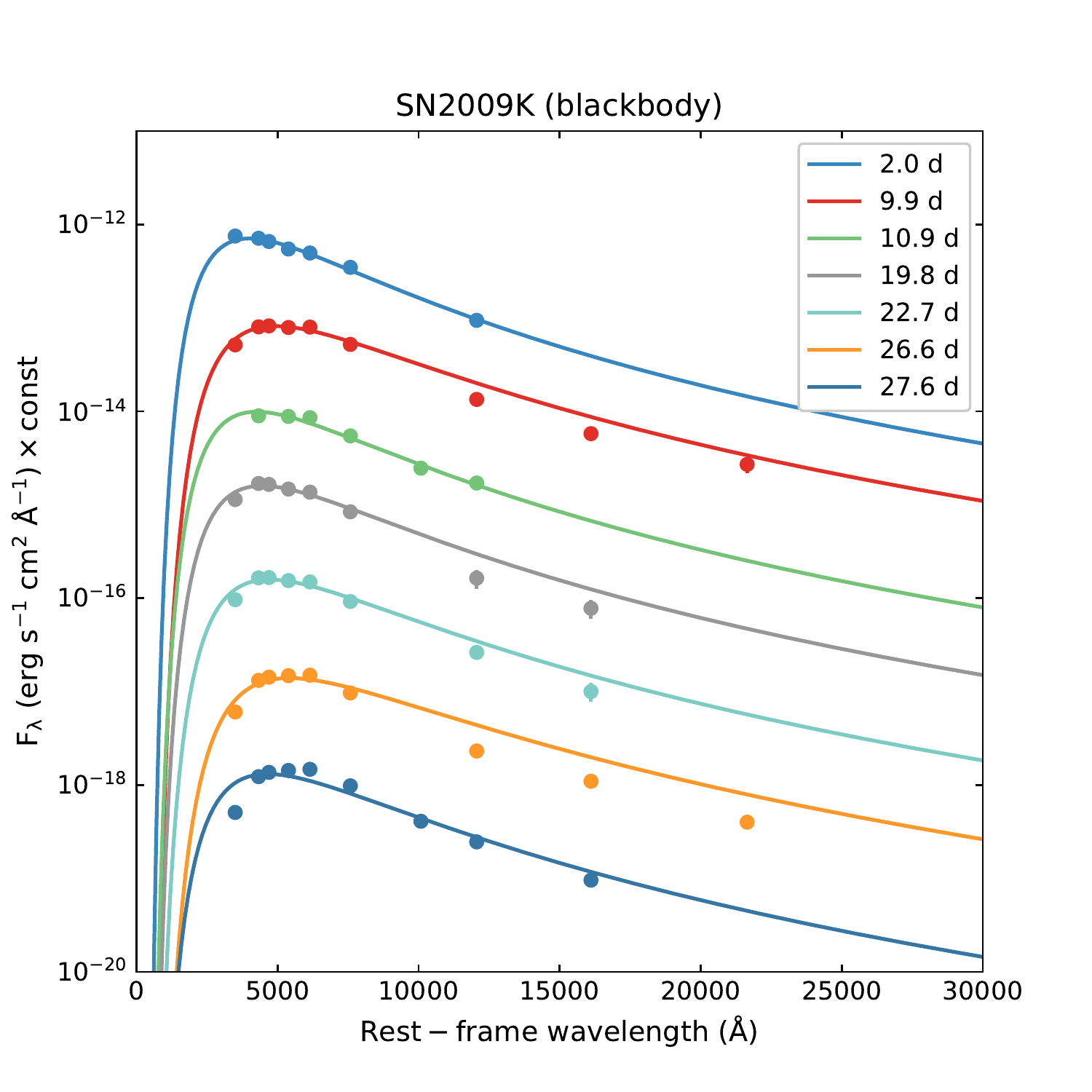}
\end{center}
\caption{The blackbody fits for the optical and NIR SEDs of SNe that cannot be
fitted by the blackbody model but don't show NIR excesses.
The data are from the references listed in Table \ref{table:details}. For clarity,
the flux at all epochs are shifted
by adding different constants.}
\label{fig:SED-BB-2}
\end{figure}

\clearpage

\begin{figure}[tbph]
\begin{center}
\includegraphics[width=0.32\textwidth,angle=0]{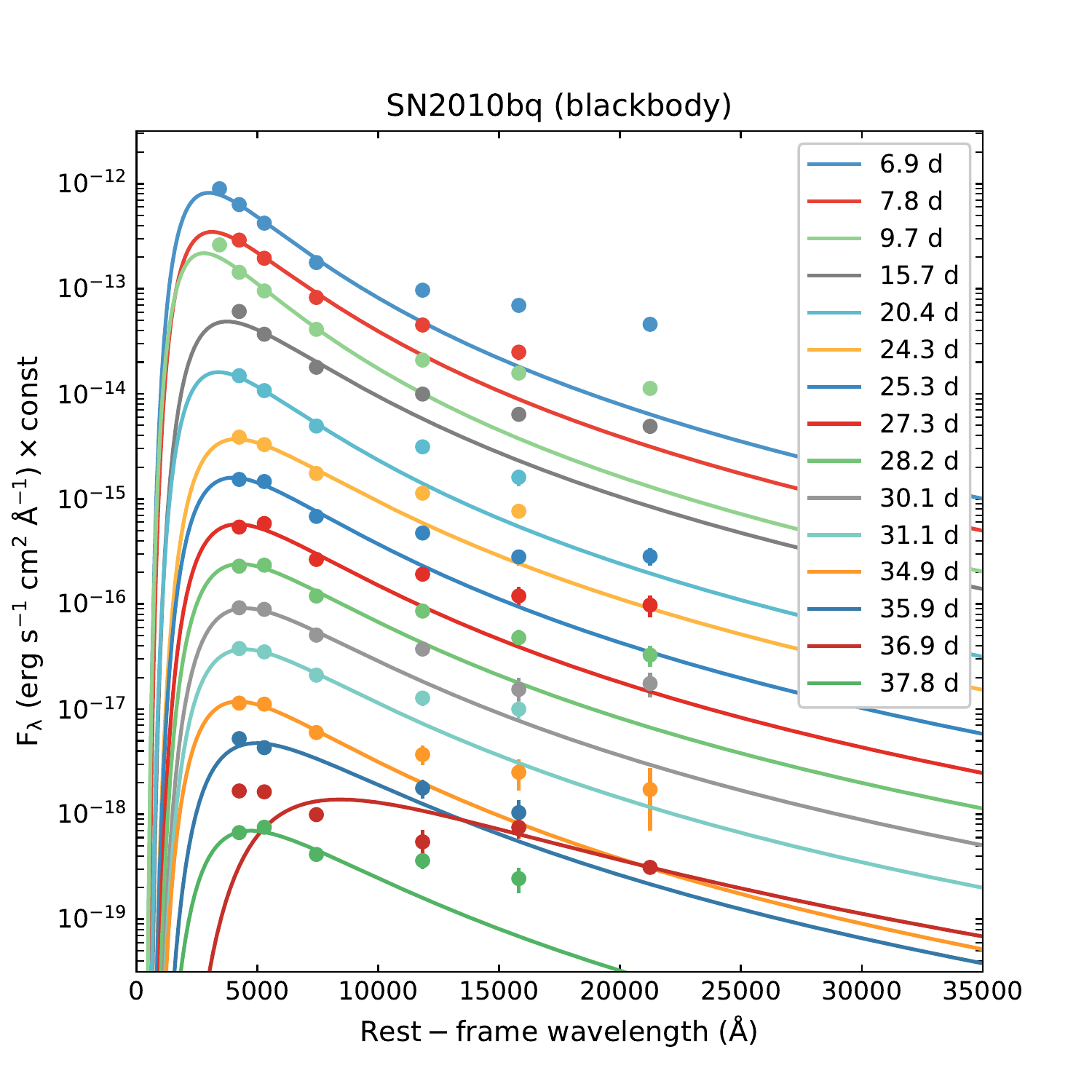}
\includegraphics[width=0.32\textwidth,angle=0]{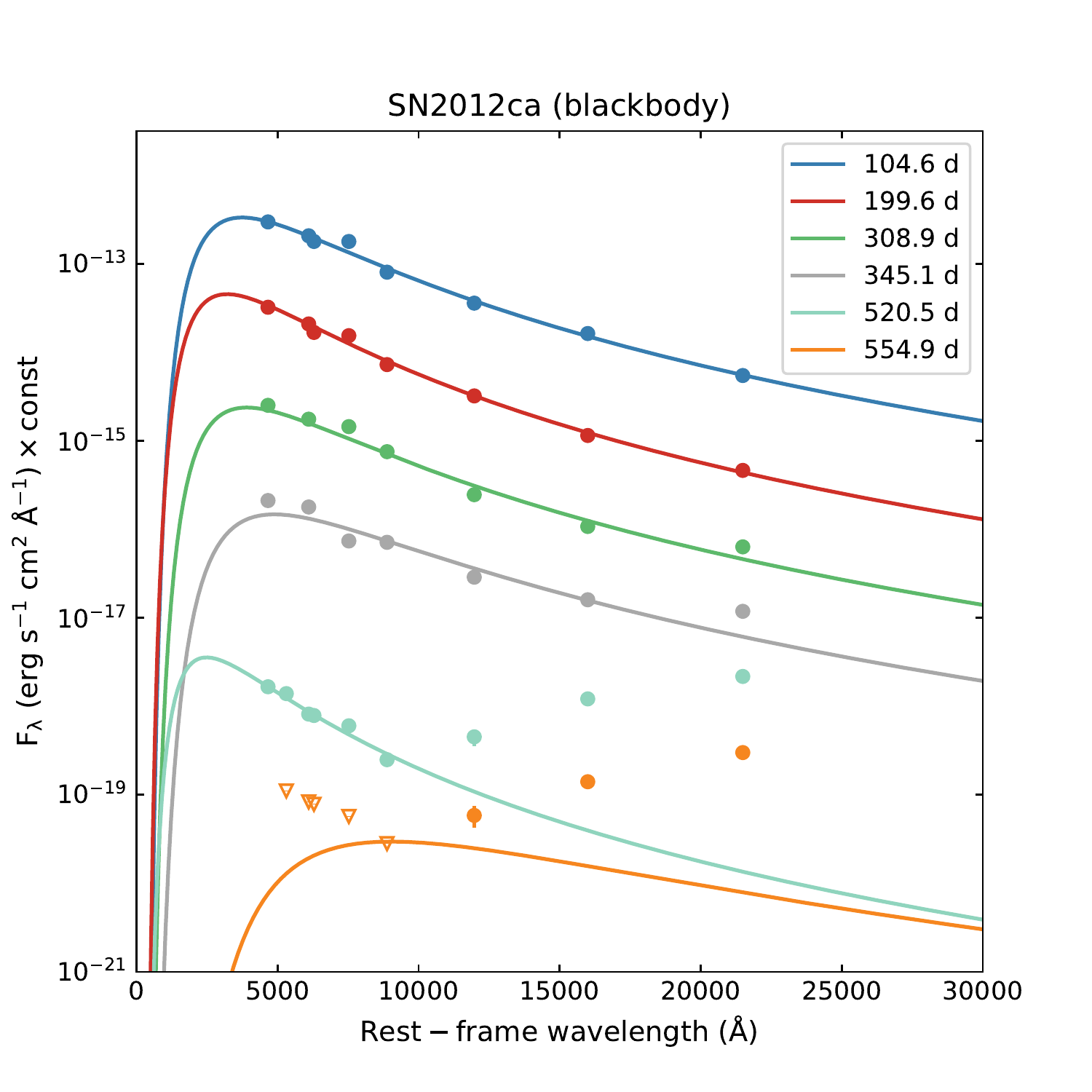}
\end{center}
\caption{The blackbody fits for the optical and NIR SEDs of SNe in our sample
(show evidence of IR excesses).
The data are from the references listed in Table \ref{table:details}. For clarity,
the flux at all epochs are shifted
by adding different constants.}
\label{fig:SED-BB-3}
\end{figure}

\clearpage

\begin{figure}[tbph]
\begin{center}
\includegraphics[width=0.40\textwidth,angle=0]{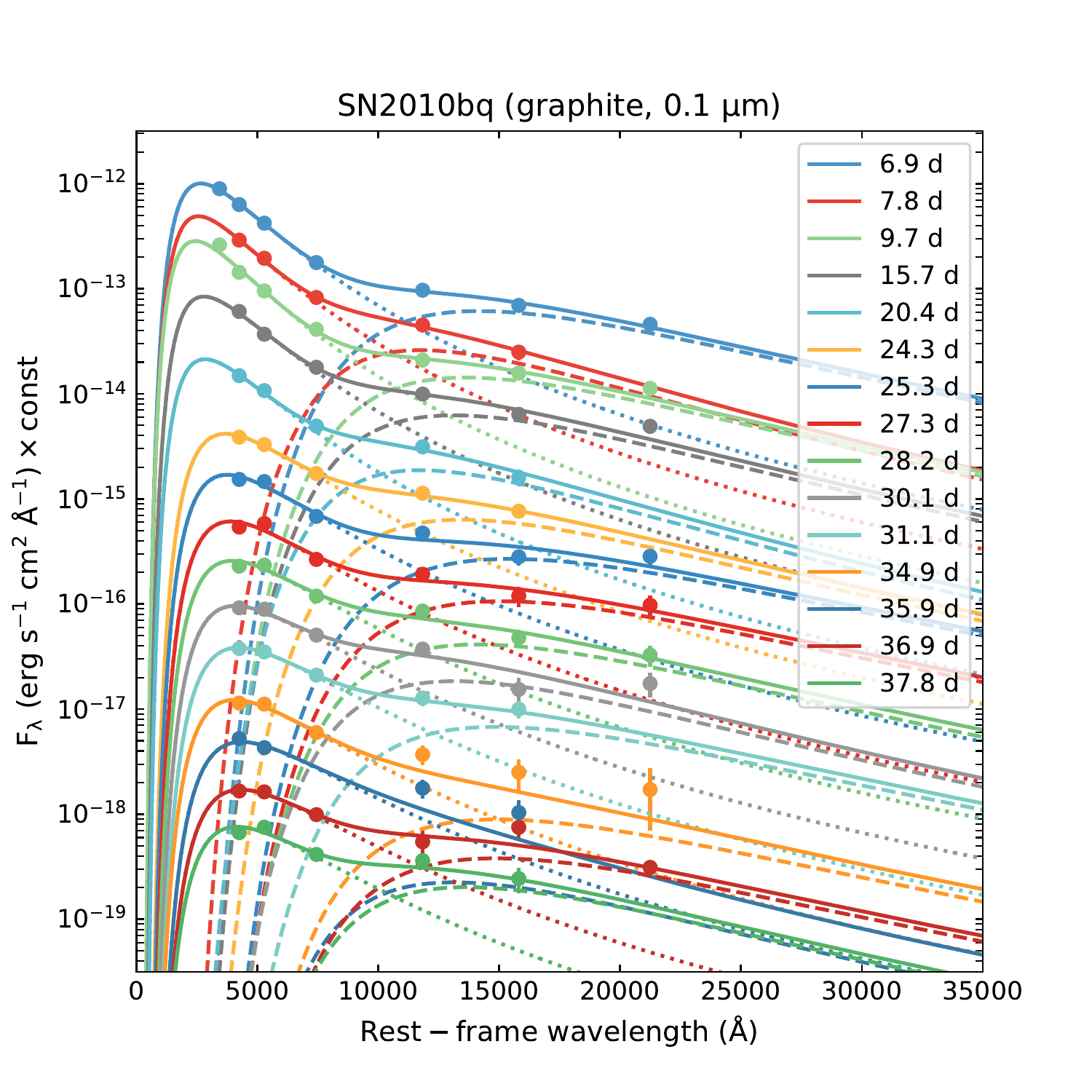}
\includegraphics[width=0.40\textwidth,angle=0]{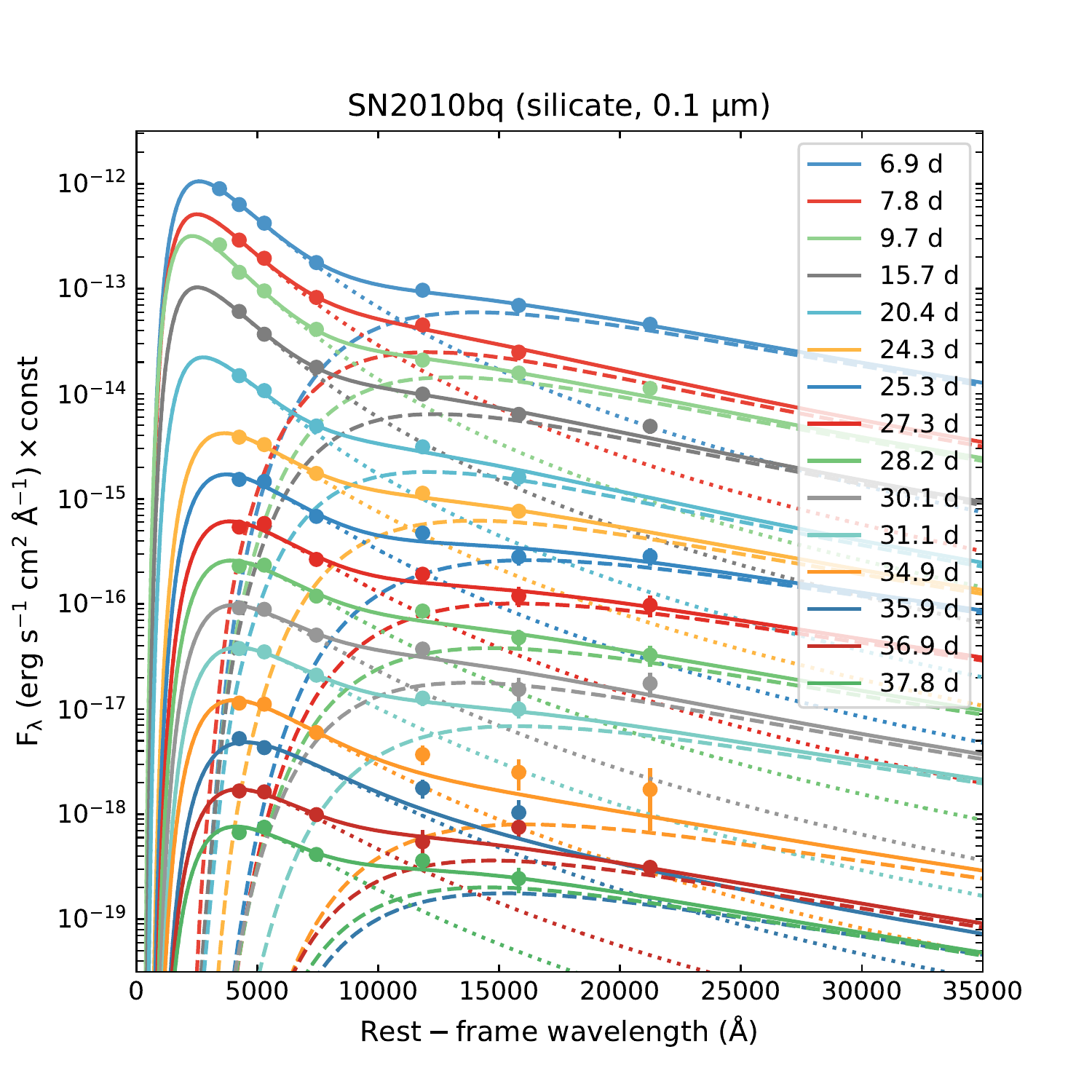}
\includegraphics[width=0.40\textwidth,angle=0]{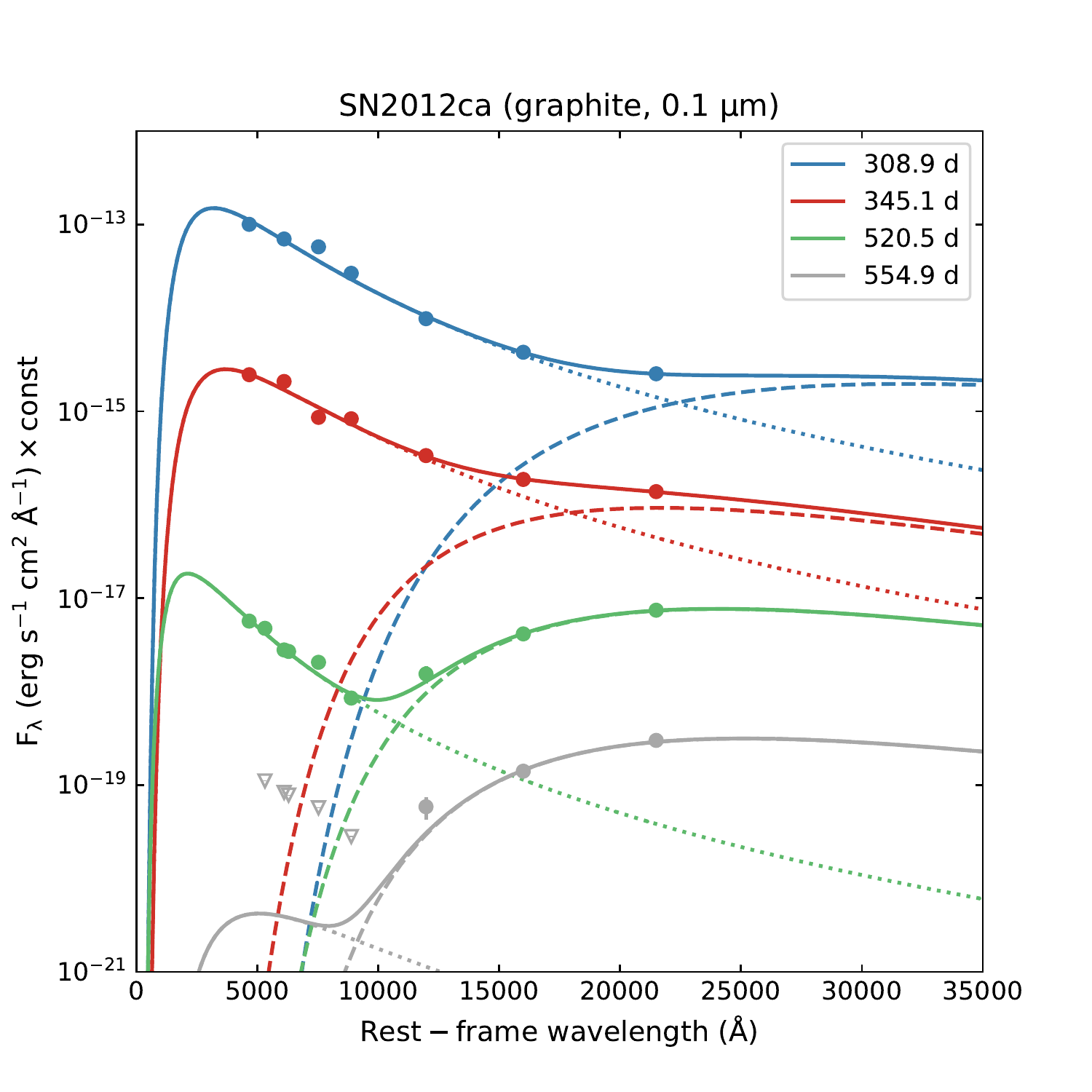}
\includegraphics[width=0.40\textwidth,angle=0]{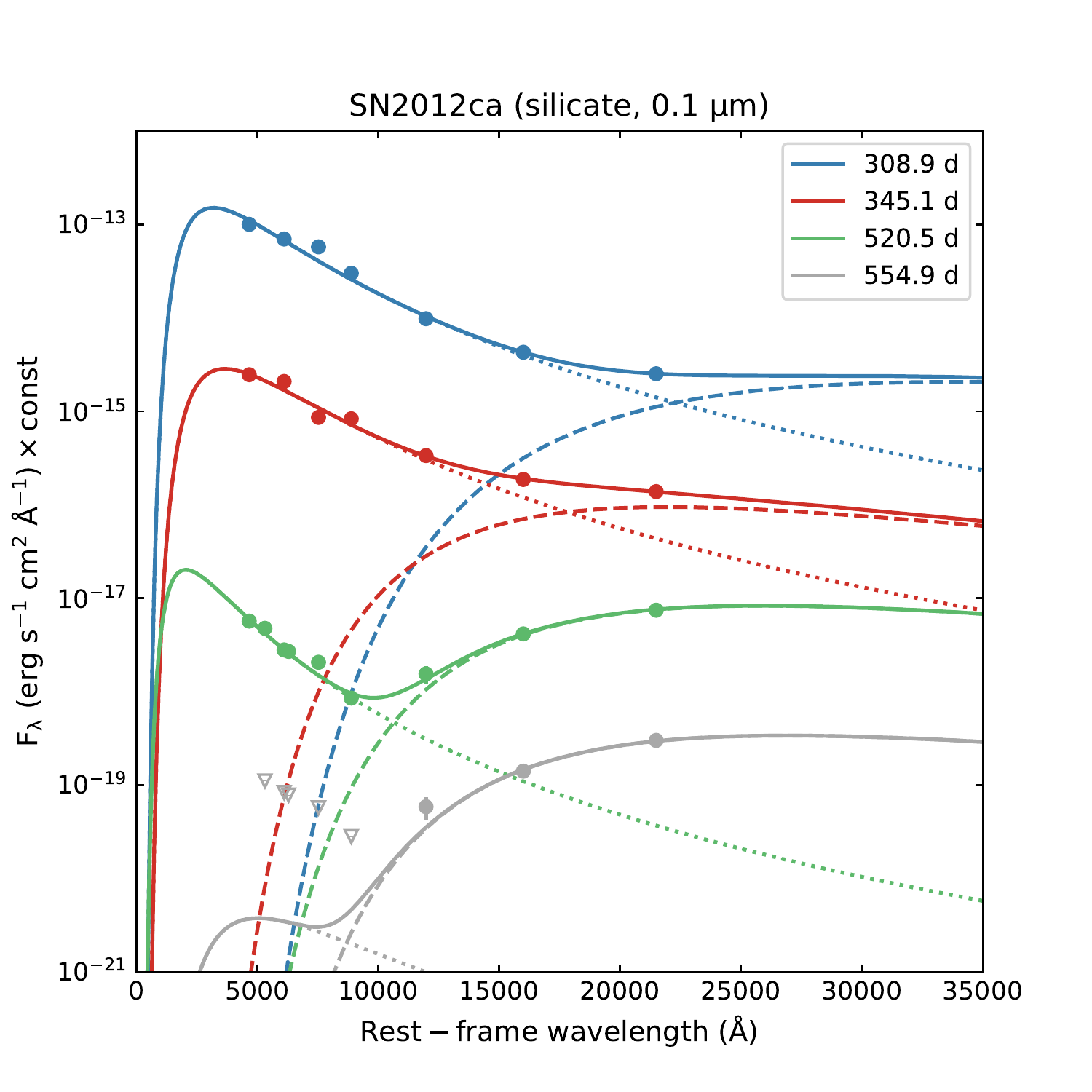}
\end{center}
\caption{The best-fitting SEDs of SNe~2010bq and 2012ca obtained by using the
blackbody plus dust {emission} model. The dotted, the dashed, and the solid lines present the flux of SN photospheres, dust,
and the sum of the two components, respectively.
The data are from the references listed in Table \ref{table:details}. For clarity, the plots at all epochs have been shifted
vertically.}
\label{fig:SED-double}
\end{figure}

\clearpage

\begin{figure}[tbph]
\begin{center}
\includegraphics[width=0.45\textwidth,angle=0]{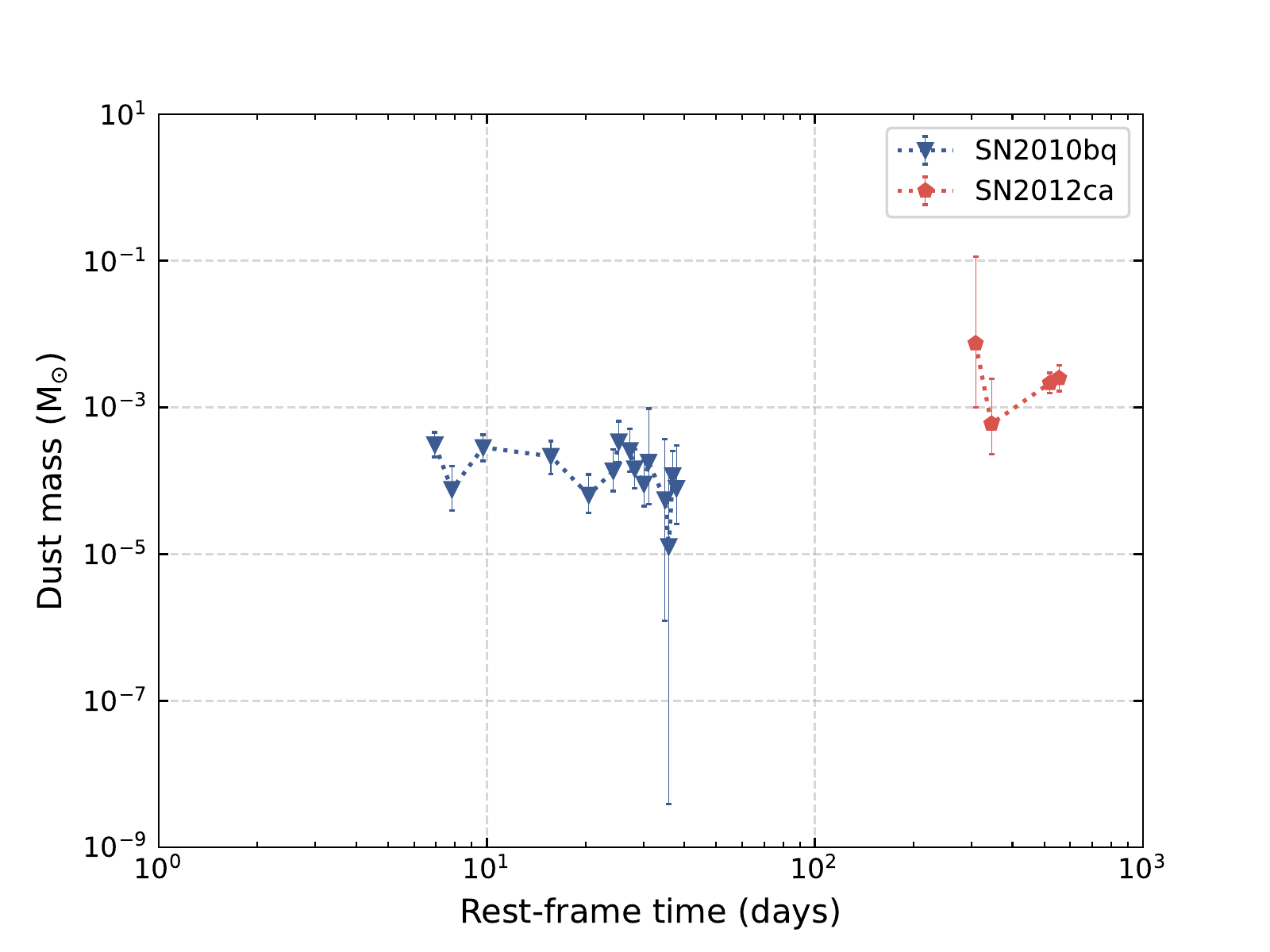}
\includegraphics[width=0.45\textwidth,angle=0]{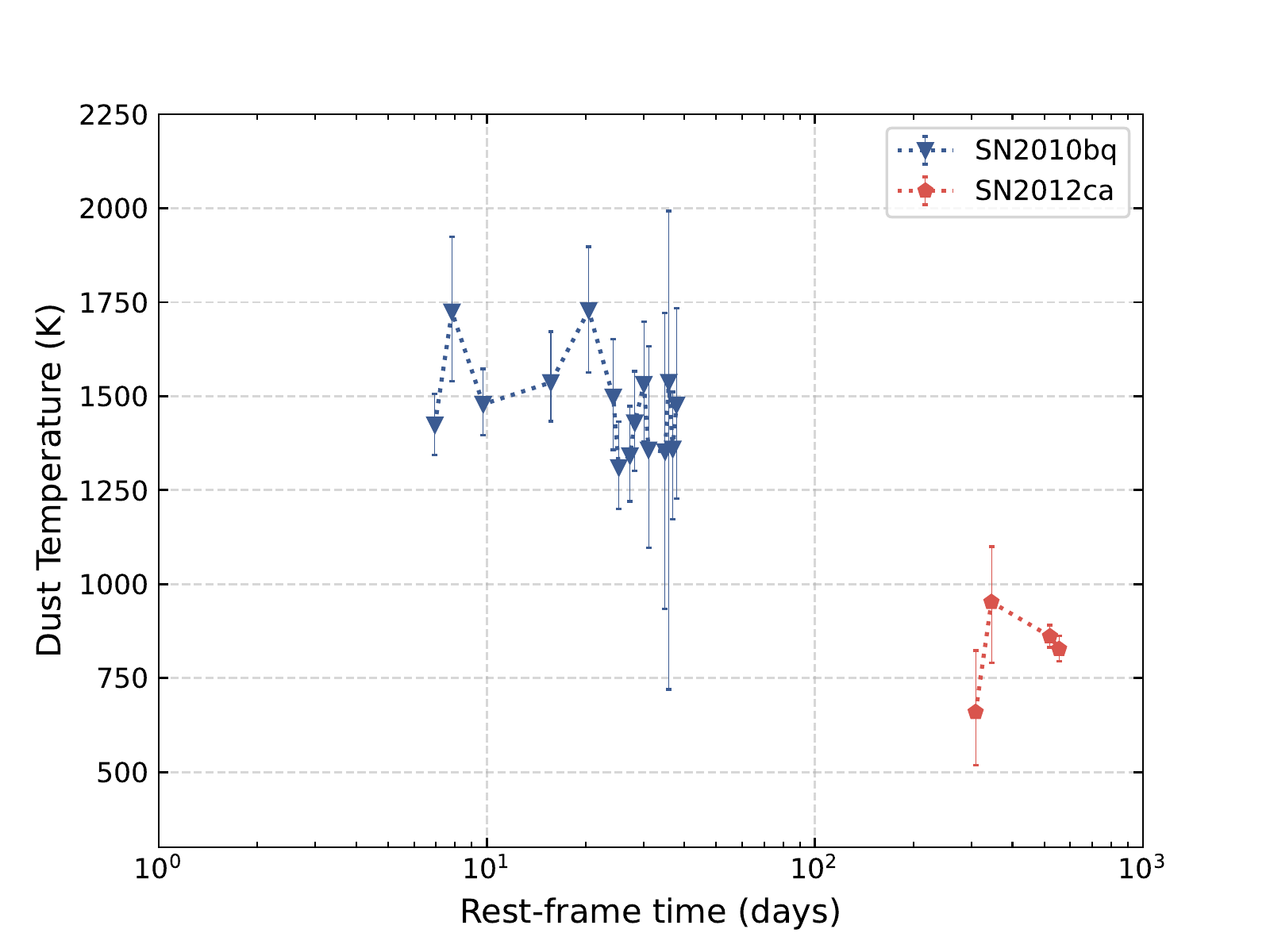}
\end{center}
\caption{The evolution of the masses (the left panel) and the temperatures (the right panel) of the dust shells
of SNe~2010bq and 2012ca.}
\label{fig:evo_T_M}
\end{figure}

\clearpage

\begin{figure}[tbph]
\begin{center}
\includegraphics[width=0.48\textwidth,angle=0]{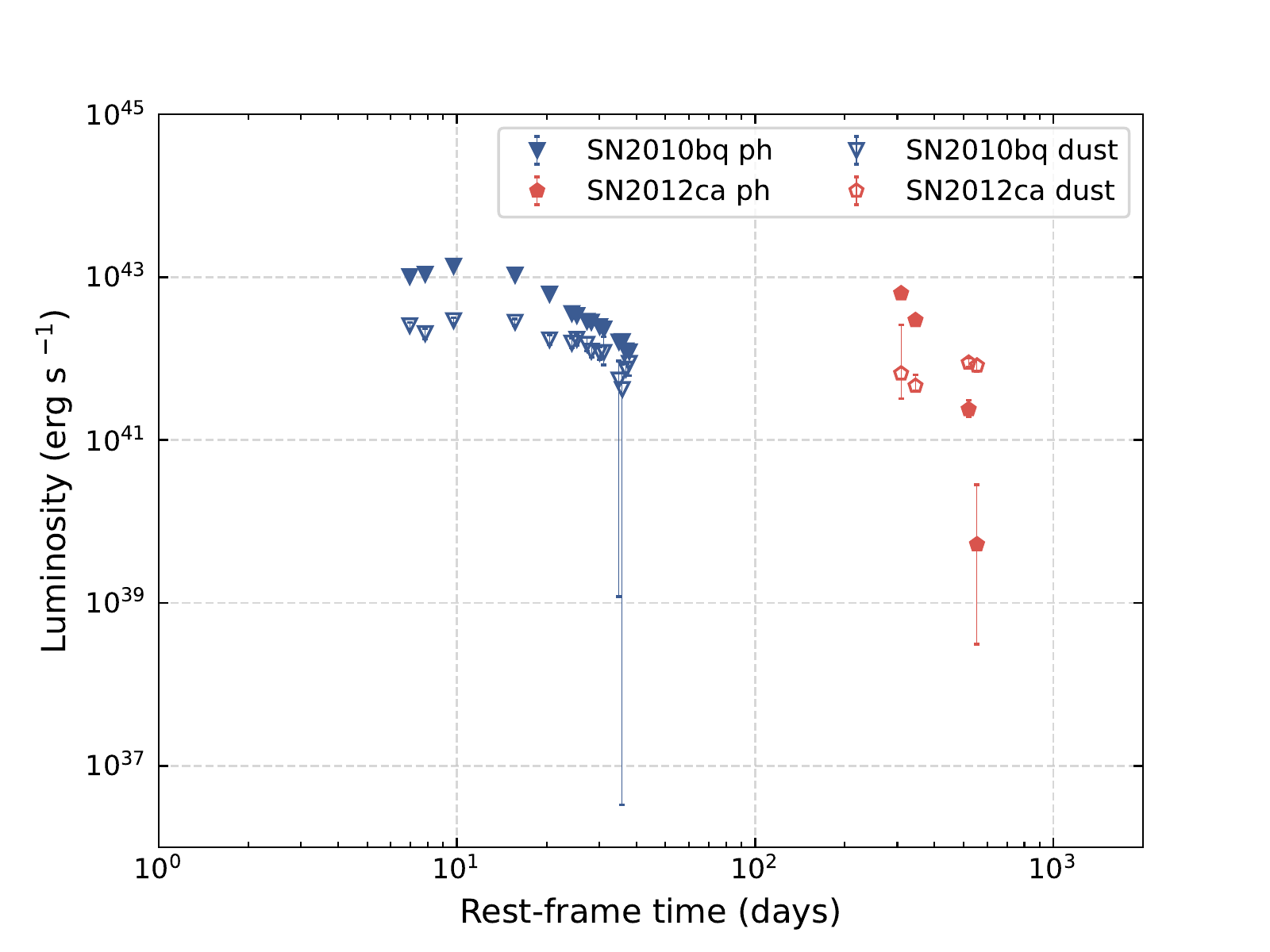}
\includegraphics[width=0.48\textwidth,angle=0]{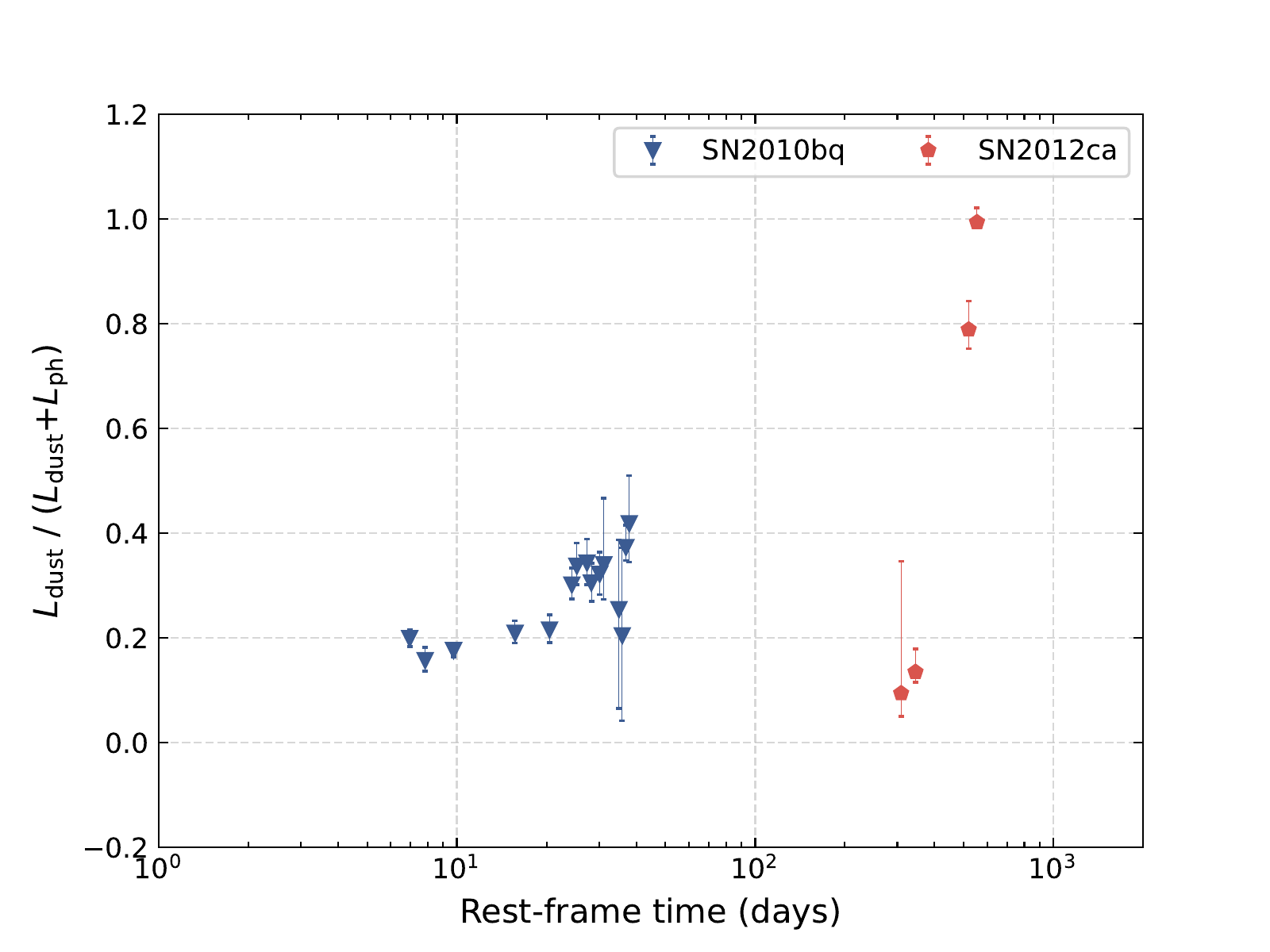}
\end{center}
\caption{The evolution of the luminosities
of the photospheres (filled symbols) and the dust shells (open symbols)
of SNe~2010bq and 2012ca {(the left panel) and the optical depth derived by using
$\tau = {L_{\rm d}}/({L_{\rm ph}+L_{\rm d}})$ (the right panel) at different epochs}.}
\label{fig:evo_L}
\end{figure}

\clearpage

\begin{figure}[tbph]
\begin{center}
\includegraphics[width=0.6\textwidth,angle=0]{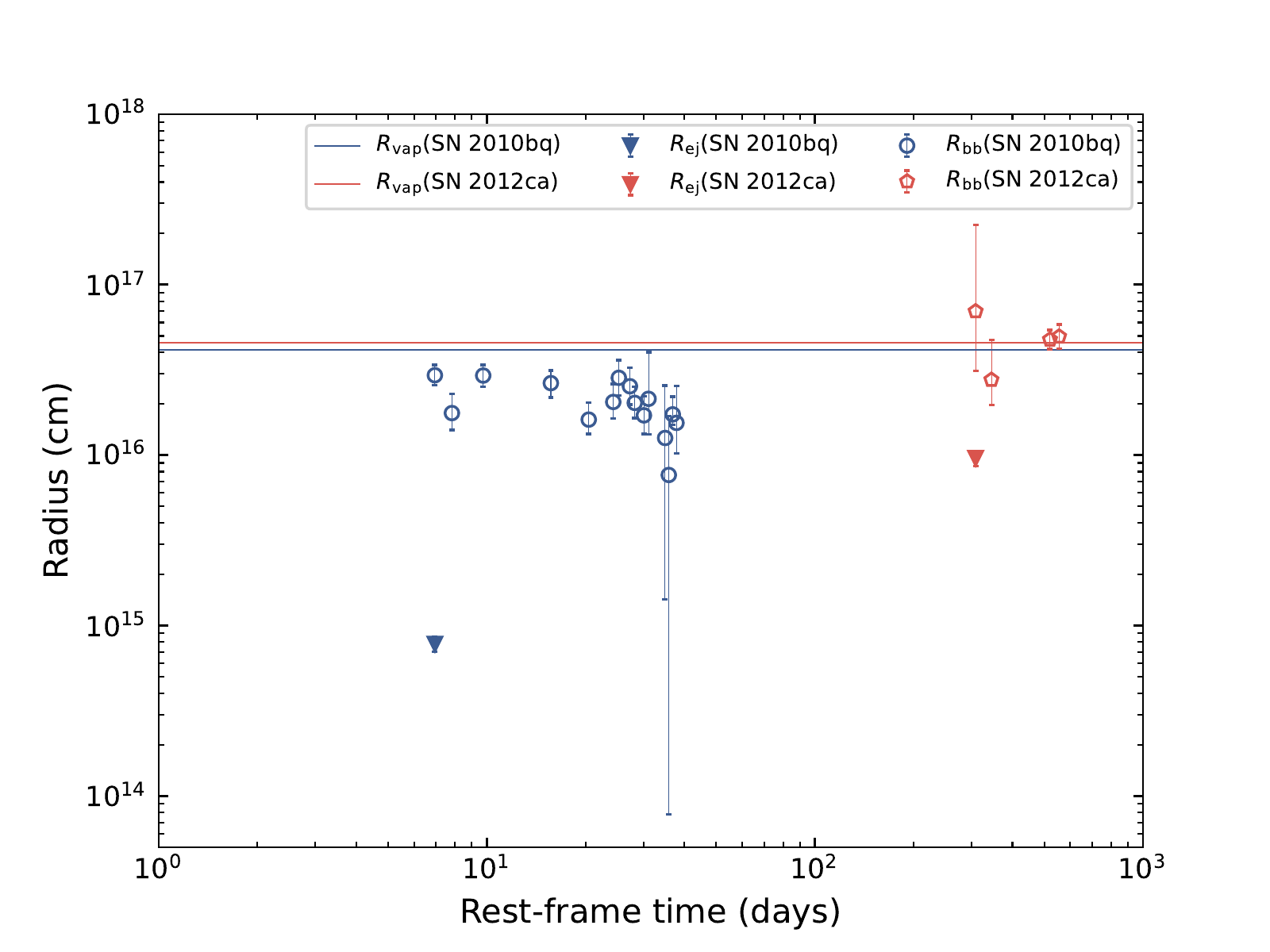}
\end{center}
\caption{The ejecta radii at the first epochs (filled symbols), the blackbody radii of the dust shells (open symbols),
and the evaporation radii (horizontal solid lines) of SNe~2010bq and 2012ca.}
\label{fig:evo_R}
\end{figure}

\clearpage

\begin{figure}[tbph]
\begin{center}
\includegraphics[width=0.45\textwidth,angle=0]{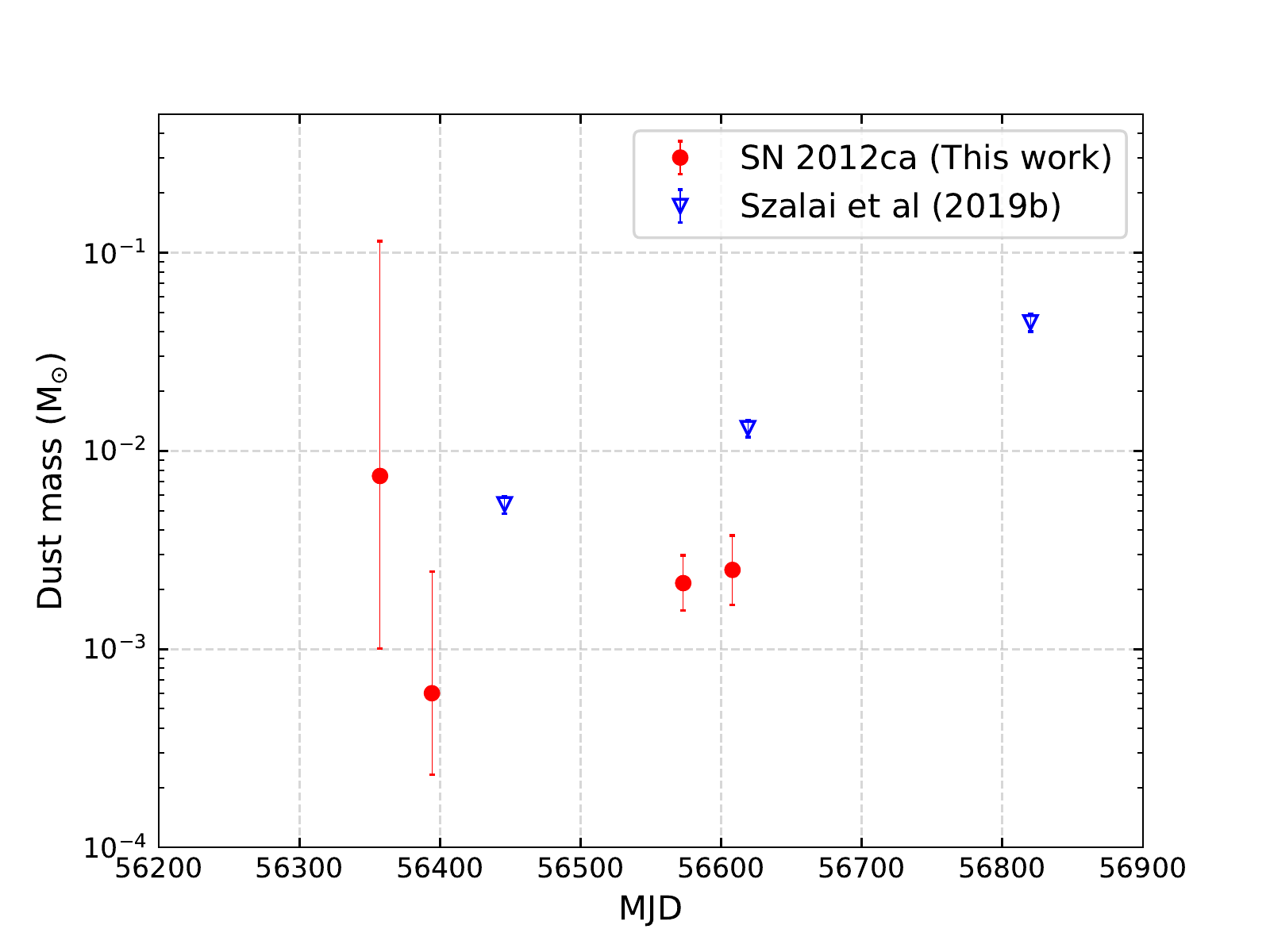}
\includegraphics[width=0.45\textwidth,angle=0]{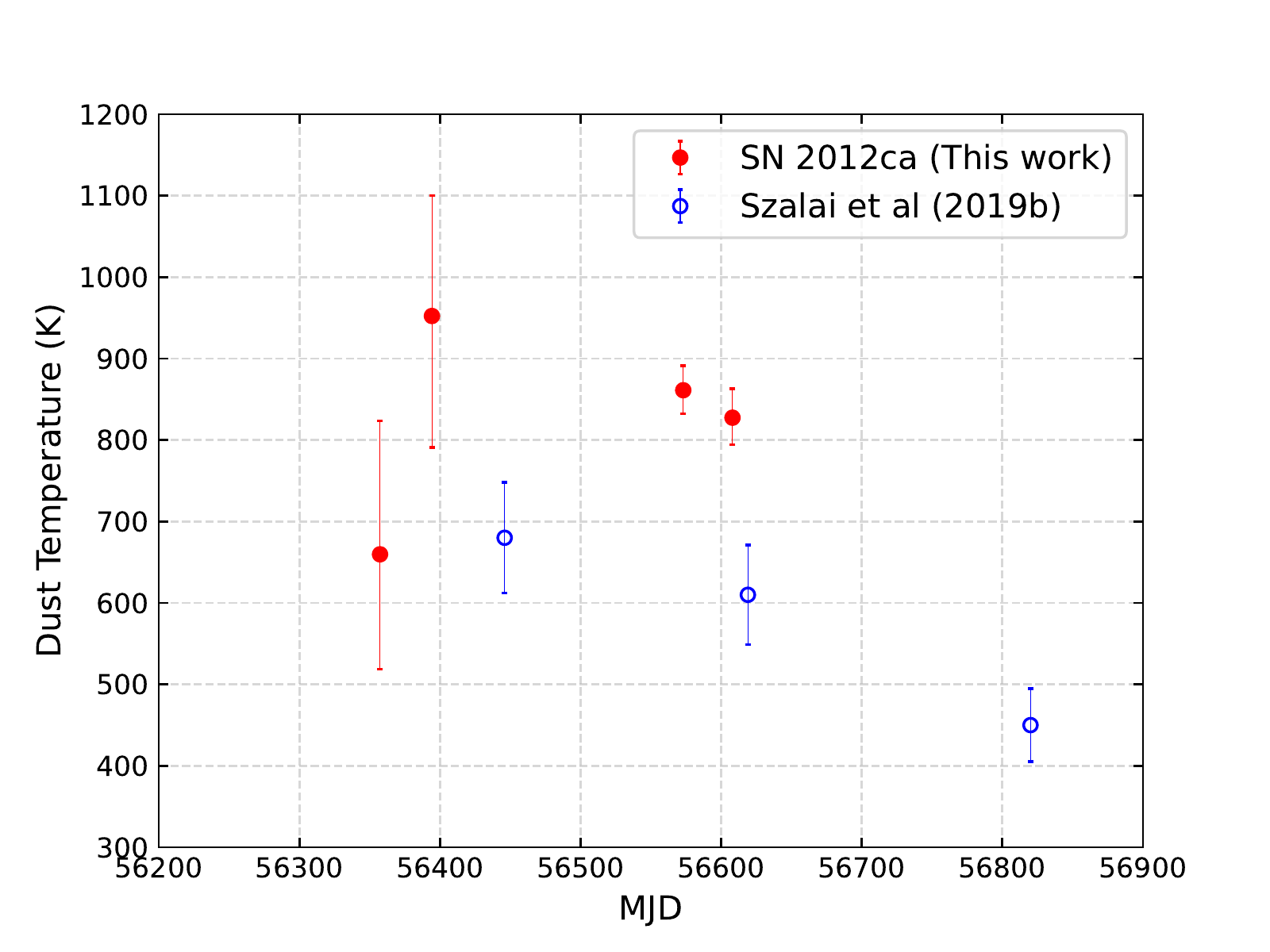}
\end{center}
\caption{The evolution of the dust mass (the left panel) and
the temperature (the right panel) derived by this work (filled symbols) and \cite{Szalai2019b} (open symbols)
of SN~2012ca. The parameters we derived are consistent with the parameters or upper limits derived by \cite{Szalai2019b} and pose more
 stringent constraints on the mass of the SN~2012ca.}
\label{fig:12ca_MT}
\end{figure}

\clearpage

\appendix
\setcounter{table}{0}
\setcounter{figure}{0}
\setcounter{equation}{0}
\renewcommand{\thetable}{A\arabic{table}}
\renewcommand{\thefigure}{A\arabic{figure}}
\renewcommand\theequation{A.\arabic{equation}}

Table \ref{table:SED_BB} lists the best-fitting parameters and the values of $\chi^2$/dof
for all fits using the blackbody model. Table \ref{table:SED_PARAM-double-uppperlimit}
presents {the fits for the SEDs} the SNe whose SEDs can be fitted by the blackbody model.

\begin{figure}[tbph]
\begin{center}
\includegraphics[width=0.8\textwidth,angle=0]{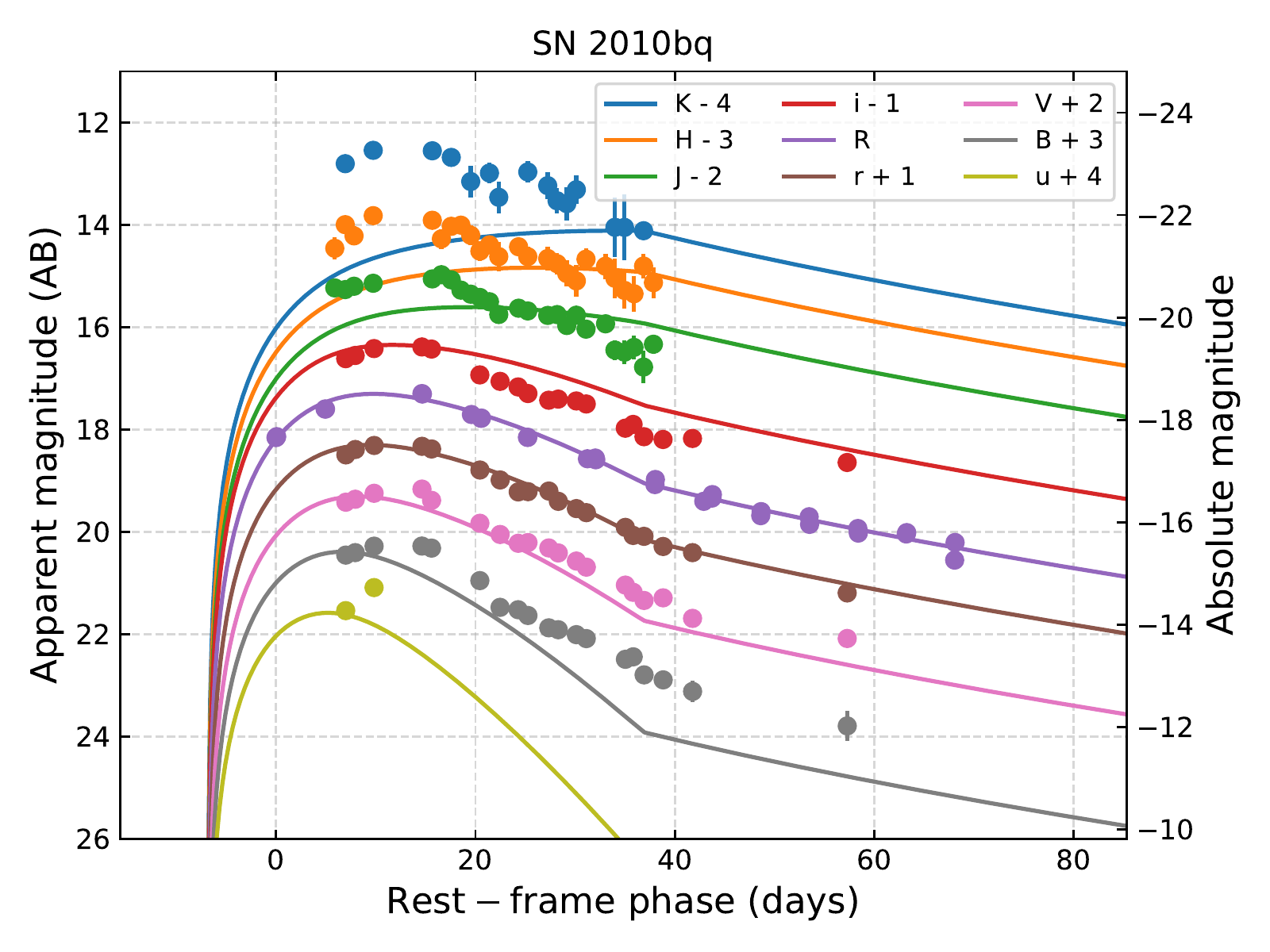}
\end{center}
\caption{{The fit of the multi-band light curves of SN~2010bq using the $^{56}$Ni model.}}
\label{fig:SN2010bq_lc}
\end{figure}

\begin{figure}[tbph]
\begin{center}
\includegraphics[width=0.8\textwidth,angle=0]{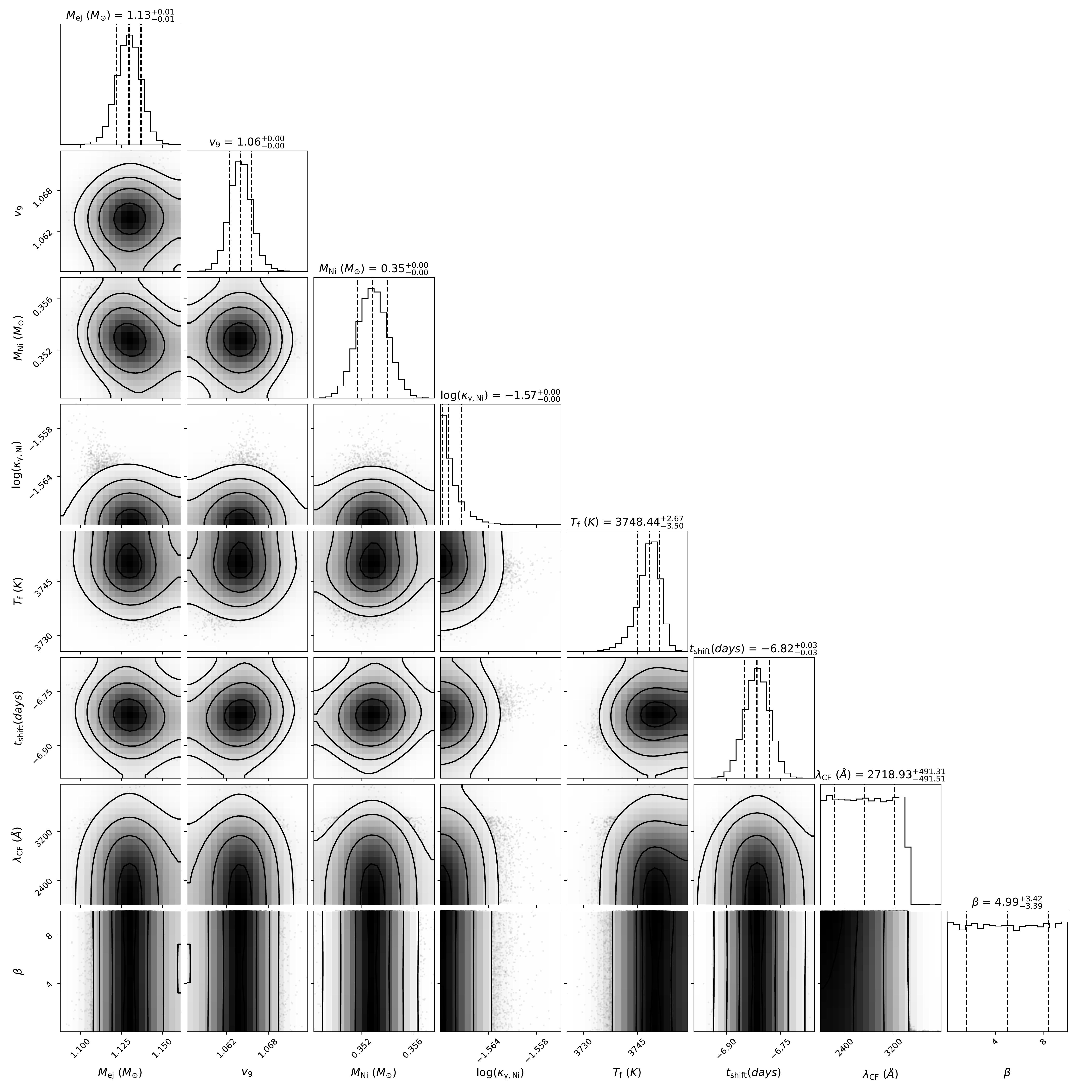}
\end{center}
\caption{{The corner plot of the $^{56}$Ni model for the multi-band light curves of SN~2010bq.}}
\label{fig:SN2010bq_corner}
\end{figure}

\begin{figure}[tbph]
\begin{center}
\includegraphics[width=0.8\textwidth,angle=0]{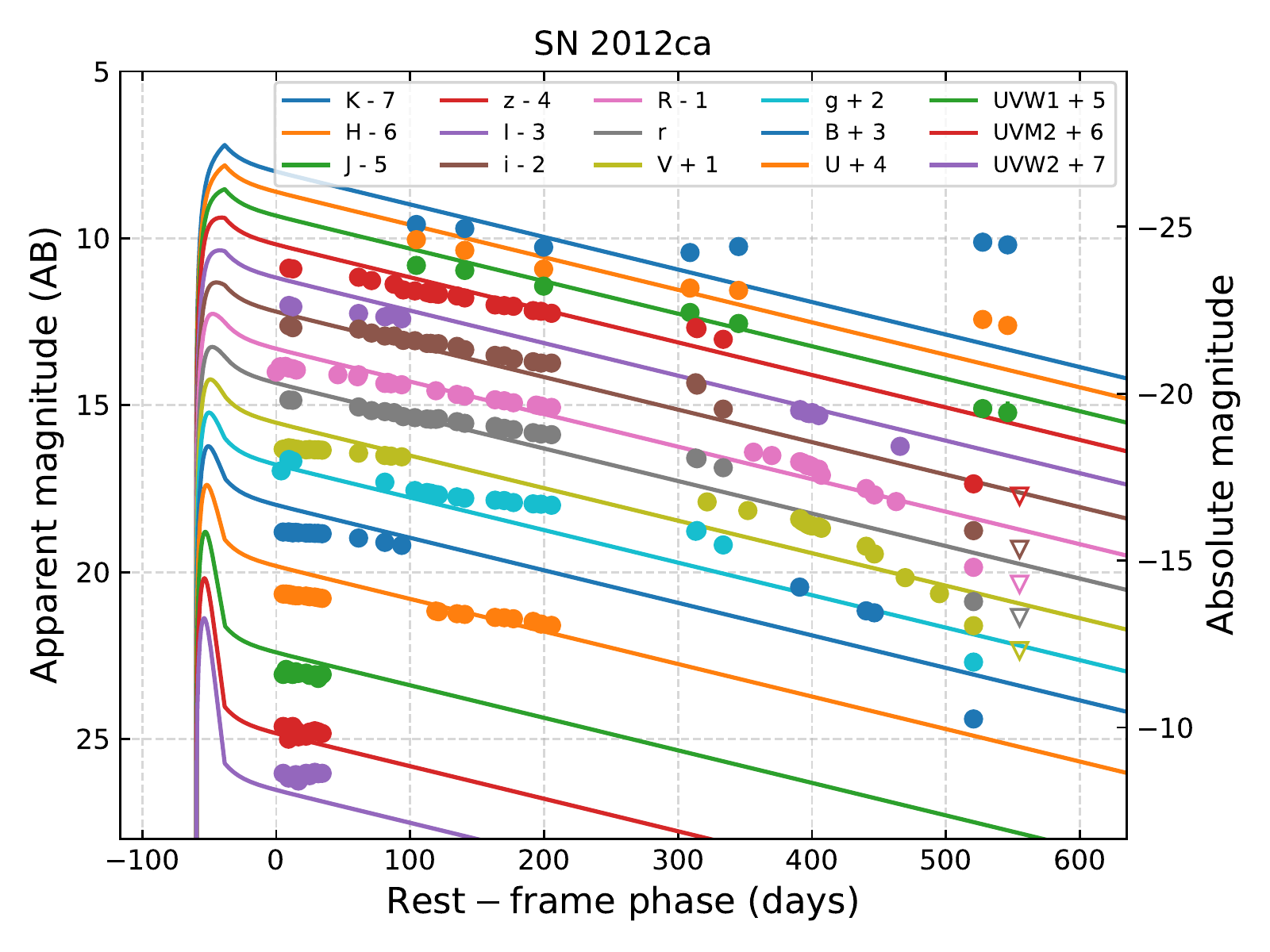}
\end{center}
\caption{{The fit of the multi-band light curves of SN~2012ca using the $^{56}$Ni model.}}
\label{fig:SN2012ca_lc}
\end{figure}

\begin{figure}[tbph]
\begin{center}
\includegraphics[width=0.8\textwidth,angle=0]{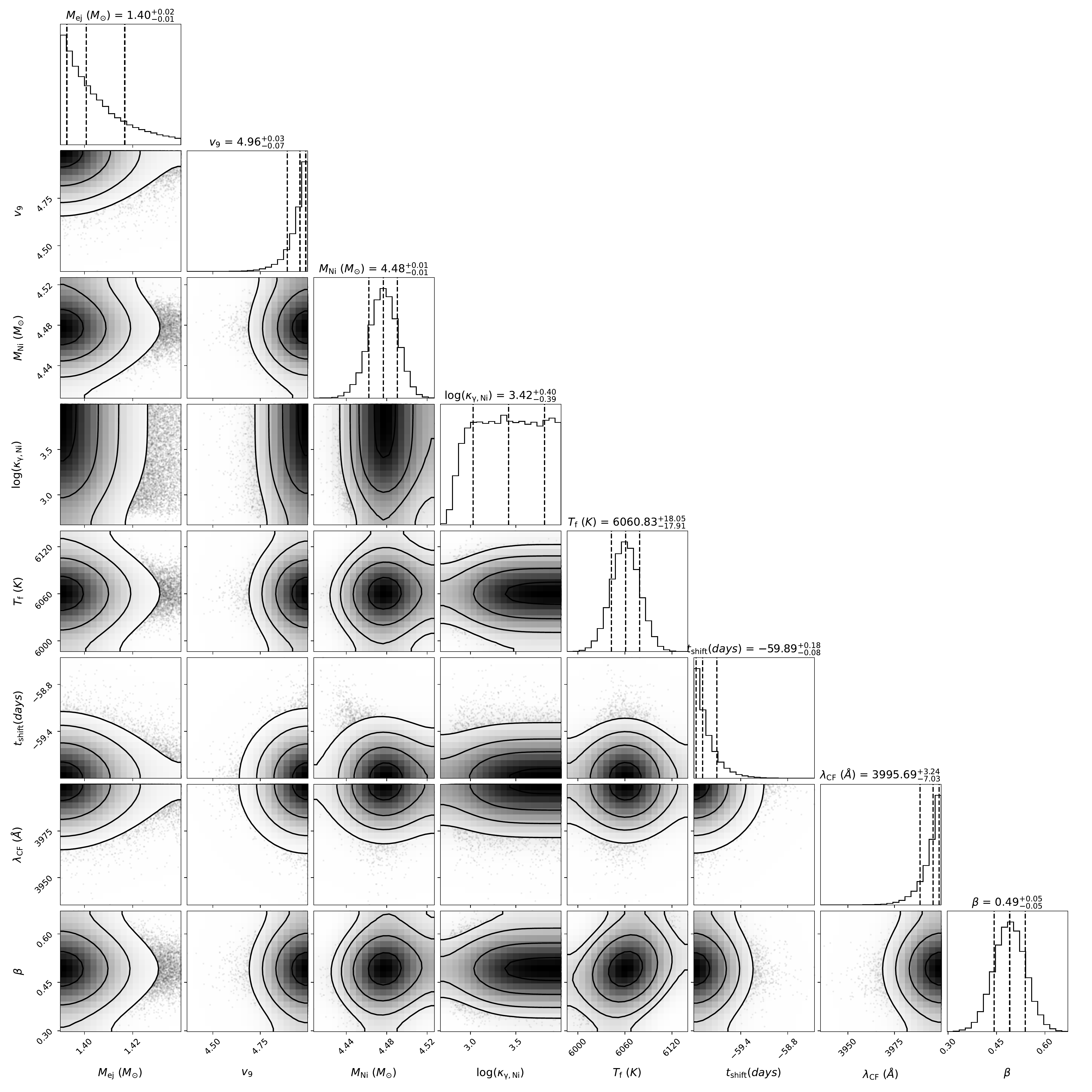}
\end{center}
\caption{{The corner plot of the $^{56}$Ni model for the multi-band light curves of SN~2010bq.}}
\label{fig:SN2012ca_corner}
\end{figure}

{Figures \ref{fig:SN2010bq_lc} and \ref{fig:SN2012ca_lc} show the fits using the $^{56}$Ni model
for the multi-band light curves of SNe~2010bq and 2012ca, respectively; figures \ref{fig:SN2010bq_corner}
and \ref{fig:SN2012ca_corner} show the corresponding corner plots of the two fits.}

\clearpage

\begin{center}
\setlength{\LTcapwidth}{\textwidth}
\setlength{\tabcolsep}{10pt}{
\begin{longtable}{c c c c c c c c c c}
\caption{The best-fitting parameters of the blackbody model for the SEDs of 66 SNe at the different rest-frame epochs. Here, $T_{\rm ph}$ is the temperature of the SN photosphere, $R_{\rm ph}$ is the radius of the SN photosphere.}
\label{table:SED_BB}
\\\hline\hline\noalign{\smallskip}

\colhead{SN} & \colhead{Phase$^a$} & \colhead{$T_{\rm ph}$} & \colhead{$R_{\rm ph}$} & \colhead{$\chi^{\rm 2}$/dof} \\
&(days)& (K) & (10$^{15}$ cm) & \\\noalign{\smallskip}\hline\noalign{\smallskip}
SN1996al & 6.0 d & $7744.61^{+115.9}_{-118.2}$ & $0.85^{+0.0}_{-0.0}$ & 4.16\\
& 130.0 d & $6225.11^{+121.3}_{-117.9}$ & $0.2^{+0.0}_{-0.0}$ & 18.11\\
& 418.0 d & $6878.74^{+333.7}_{-330.7}$ & $0.089^{+0.0}_{-0.0}$ & 7.94\\
& 419.0 d & $6870.88^{+378.0}_{-341.9}$ & $0.089^{+0.0}_{-0.0}$ & 8.44\\
& 527.3 d & $7027.07^{+426.9}_{-402.6}$ & $0.052^{+0.0}_{-0.0}$ & 8.47\\
\end{longtable}}
$^a$ \textbf{Phase}. All the phases are relative to the first data which are supposed to be respectively JD 2450287.79 (SN~1996al), JD 2451477.31 (SN~1999el), JD 2453293.62 (SN~2004ex), JD 2453310.8 (SN~2004ff), JD 2453339.0 (SN~2004gk), JD 2453352.79 (SN~2004gq), JD 2453356.86 (SN~2004gt), JD 2453356.64 (SN~2004gv), JD 2453456.88 (SN~2005aw), JD 2453458.45 (SN~2005ay), JD 2453465.8 (SN~2005az), JD 2453467.67 (SN~2005bf), JD 2453549.41 (SN~2005cs), JD 2453637.0 (SN~2005ek), JD 2453622.96 (SN~2005em), JD 2453639.94 (SN~2005gj), JD 2453670.6 (SN~2005hg), JD 2453692.0 (SN~2005kj), JD 2453699.98 (SN~2005kl), JD 2453733.0 (SN~2005mf), JD 2453767.72 (SN~2006T), JD 2453774.98 (SN~2006aa), JD 2453784.65 (SN~2006aj), JD 2453801.7 (SN~2006au), JD 2453820.68 (SN~2006ba), JD 2453823.75 (SN~2006bf), JD 2453830.65 (SN~2006bo), JD 2453979.78 (SN~2006ep), JD 2453994.93 (SN~2006fo), JD 2454016.55 (SN~2006ir), JD 2454029.7 (SN~2006lc), JD 2454036.77 (SN~2006ld), JD 2454109.85 (SN~2007C), JD 2454117.95 (SN~2007I), JD 2454150.52 (SN~2007Y), JD 2454152.85 (SN~2007aa), JD 2454168.64 (SN~2007ag), JD 2454182.76 (SN~2007av), JD 2454226.73 (SN~2007ce), JD 2454349.66 (SN~2007hn), JD 2454376.6 (SN~2007kj), JD 2454442.0 (SN~2007rz), JD 2454470.98 (SN~2007uy), JD 2454475.02 (SN~2008D), JD 2454490.3 (SN~2008S), JD 2454525.48 (SN~2008aq), JD 2454529.67 (SN~2008ax), JD 2454792.59 (SN~2008hh), JD 2454818.8 (SN~2008if), JD 2454830.84 (SN~2008in), JD 2454846.64 (SN~2009K), JD 2454914.01 (SN~2009ay), JD 2454914.5 (SN~2009bb), JD 2454927.86 (SN~2009ca), JD 2454956.92 (SN~2009dt), JD 2454980.96 (SN~2009er), JD 2455054.9 (SN~2009ib), JD 2455095.93 (SN~2009iz), JD 2455144.92 (SN~2009kr), JD 245296.76 (SN~2010bq), JD 2456043.08 (SN~2012ca), JD 2456154.72 (SN~2012ec), JD 2456510.5 (SN~2013dn), JD 2457470.84 (SN~2016bkv), JD 2456132.76 (PTF12gzk), and JD 2456369.61 (LSQ13zm).
\end{center}

\clearpage

\begin{center}
\setlength{\LTcapwidth}{\textwidth}
\setlength{\tabcolsep}{-1pt}{
\begin{longtable}{c c c c c c c c c c c c c}
\caption{The best-fitting parameters of the blackbody plus graphite (silicate) model {for the SEDs that can be fitted by the blackbody model}.
Here, $T_{\rm ph}$ is the temperature of the SN photosphere, $R_{\rm ph}$ the radius of the SN photosphere, $T_{\rm d}$ the temperature of the dust shell,
$M_{\rm d}$ the upper limits of mass of the dust shell.}
\label{table:SED_PARAM-double-uppperlimit}
\\\hline\hline\noalign{\smallskip}
&\multicolumn{5}{c}{{Blackbody plus Graphite}} & \multicolumn{5}{c}{{Blackbody plus Silicate}}\\\noalign{\smallskip}
\cmidrule(lr){2-6} \cmidrule(lr){7-11}
\colhead{Phase$^a$} & \colhead{$T_{\rm ph}$}&\colhead{$R_{\rm ph}$}&\colhead{$T_{\rm d}$}&\colhead{$M_{\rm d}$}&\colhead{$\chi^{\rm 2}$/dof}&\colhead{$T_{\rm ph}$}&\colhead{$R_{\rm ph}$}&\colhead{$T_{\rm d}$} & \colhead{$M_{\rm d}$}  & \colhead{$\chi^{\rm 2}$/dof} \\
& (K) & (10$^{15}$ cm) &  (K) & (10$\rm ^{-6}\ M_{\odot}$) &  & (K) & (10$^{15}$ cm) & (K) & (10$\rm^{-6}\ M_{\odot}$) & \\\noalign{\smallskip} \hline \noalign{\smallskip}
\multicolumn{11}{c}{{SN 1996al}}\\\noalign{\smallskip} \hline \noalign{\smallskip}
6.0 d & $7835.31^{+164.8}_{-152.5}$ & $0.83^{+0.0}_{-0.0}$ & $1090.09^{+652.1}_{-449.9}$ & $10.57^{+4997.8}_{-10.6}$ & 8.42 & $7835.17^{+171.4}_{-145.7}$ & $0.83^{+0.0}_{-0.0}$ & $1184.62^{+609.2}_{-459.3}$ & $123.28^{+12458.2}_{-123.3}$ & 8.03\\\noalign{\smallskip} \hline \noalign{\smallskip}
130.0 d & $6220.91^{+110.9}_{-107.5}$ & $0.2^{+0.0}_{-0.0}$ & $762.72^{+913.5}_{-487.3}$ & $0.0015^{+8.0}_{-0.0}$ & 36.21 & $6220.51^{+116.9}_{-114.3}$ & $0.2^{+0.0}_{-0.0}$ & $868.78^{+891.1}_{-557.0}$ & $0.004^{+11.7}_{-0.0}$ & 36.21\\\noalign{\smallskip} \hline \noalign{\smallskip}
418.0 d & $6934.57^{+402.9}_{-359.7}$ & $0.088^{+0.0}_{-0.0}$ & $892.37^{+800.4}_{-590.9}$ & $0.013^{+8.6}_{-0.0}$ & 15.89 & $6936.66^{+392.2}_{-348.4}$ & $0.088^{+0.0}_{-0.0}$ & $982.0^{+809.2}_{-636.4}$ & $0.099^{+23.0}_{-0.1}$ & 15.89\\
\end{longtable}
}

$^a$ \textbf{Phase}. All the phases are relative to the first data which are supposed to be respectively JD 2450287.79 (SN~1996al), JD 2451477.31 (SN~1999el), JD 2453293.62 (SN~2004ex), JD 2453310.8 (SN~2004ff), JD 2453339.0 (SN~2004gk), JD 2453352.79 (SN~2004gq), JD 2453356.86 (SN~2004gt), JD 2453356.64 (SN~2004gv), JD 2453456.88 (SN~2005aw), JD 2453458.45 (SN~2005ay), JD 2453465.8 (SN~2005az), JD 2453467.67 (SN~2005bf), JD 2453549.41 (SN~2005cs), JD 2453637.0 (SN~2005ek), JD 2453622.96 (SN~2005em), JD 2453639.94 (SN~2005gj), JD 2453670.6 (SN~2005hg), JD 2453692.0 (SN~2005kj), JD 2453733.0 (SN~2005mf), JD 2453767.72 (SN~2006T), JD 2453774.98 (SN~2006aa), JD 2453784.65 (SN~2006aj), JD 2453801.7 (SN~2006au), JD 2453820.68 (SN~2006ba), JD 2453823.75 (SN~2006bf), JD 2453830.65 (SN~2006bo), JD 2453979.78 (SN~2006ep), JD 2454016.55 (SN~2006ir), JD 2454029.7 (SN~2006lc), JD 2454036.77 (SN~2006ld), JD 2454109.85 (SN~2007C), JD 2454117.95 (SN~2007I), JD 2454150.52 (SN~2007Y), JD 2454152.85 (SN~2007aa), JD 2454168.64 (SN~2007ag), JD 2454182.76 (SN~2007av), JD 2454349.66 (SN~2007hn), JD 2454376.6 (SN~2007kj), JD 2454442.0 (SN~2007rz), JD 2454470.98 (SN~2007uy), JD 2454475.02 (SN~2008D), JD 2454490.3 (SN~2008S), JD 2454525.48 (SN~2008aq), JD 2454792.59 (SN~2008hh), JD 2454818.8 (SN~2008if), JD 2454830.84 (SN~2008in), JD 2454914.01 (SN~2009ay), JD 2454914.5 (SN~2009bb), JD 2454927.86 (SN~2009ca), JD 2454956.92 (SN~2009dt), JD 2454980.96 (SN~2009er), JD 2455054.9 (SN~2009ib), JD 2455095.93 (SN~2009iz), JD 2455144.92 (SN~2009kr), JD 2456043.08 (SN~2012ca), JD 2456154.72 (SN~2012ec), JD 2456510.5 (SN~2013dn), JD 2457470.84 (SN~2016bkv), JD 2456132.76 (PTF12gzk), and JD 2456369.61 (LSQ13zm).
\end{center}

\clearpage

\clearpage

\clearpage

\clearpage

\end{document}